\documentclass[aps,prd,eqsecnum,11pt,showpacs,preprintnumbers,nofootinbib]{revtex4}

\def\nbox#1#2{\vcenter{\hrule \hbox{\vrule height#2in
\kern#1in \vrule} \hrule}}
\def\sq{\,\raise.5pt\hbox{$\nbox{.09}{.09}$}\,}
\def\sqb{\,\raise.5pt\hbox{$\overline{\nbox{.09}{.09}}$}\,}

\newcommand{\bea}{\begin{eqnarray}}
\newcommand{\eea}{\end{eqnarray}}
\newcommand{\be}{\begin{equation}}
\newcommand{\ee}{\end{equation}}
\newcommand{\bes}{\begin{subequations}}
\newcommand{\ees}{\end{subequations}}
\def\lag{\langle}
\def\rag{\rangle}

\def\nn{\nonumber \\}

\usepackage{amssymb,amsmath}
\usepackage{graphicx}
\usepackage{graphics}

\begin{document}
\
\preprint{\rm LA-UR 09-01895}

\title{Cosmological Horizon Modes and Linear Response in de Sitter Spacetime}
\author{Paul R. Anderson}
\affiliation{Department of Physics,\\
Wake Forest University, \\
Winston-Salem, NC 27109 USA\\
and Departamento de F\'{\i}sica Te\'orica and IFIC, \\
Universidad de Valencia-CSIC,\\ C. Dr. Moliner 50, Burjassot-46100,
Valencia, Spain}
\email{anderson@wfu.edu}

\author{Carmen Molina-Par\'{\i}s}
\affiliation{Department of Applied Mathematics,\\
University of Leeds,\\
Leeds LS2 9JT, UK}
\email{carmen@maths.leeds.ac.uk}

\author{Emil Mottola}
\affiliation{Theoretical Division \\
Los Alamos National Laboratory \\
Los Alamos, NM 87545 USA}
\email{emil@lanl.gov}

\begin{abstract}
\vskip .3cm

Linearized fluctuations of quantized matter fields and the spacetime geometry
around de Sitter space are considered in the case that the matter fields are
conformally invariant. Taking the unperturbed state of the matter to be the
de Sitter invariant Bunch-Davies state, the linear variation of the stress
tensor about its self-consistent mean value serves as a source for fluctuations
in the geometry through the semi-classical Einstein equations. This linear
response framework is used to investigate both the importance of quantum
backreaction and the validity of the semi-classical approximation in cosmology.
The full variation of the stress tensor $\delta \langle T^a_{\ b}\rangle$
contains two kinds of terms: (1) those that depend explicitly upon the linearized
metric variation $\delta g_{cd}$ through the $\langle [T^a_{\ b}, T^{cd}] \rangle$
causal response function; and (2) state dependent variations, independent of
$\delta g_{cd}$. For perturbations of the first kind, the criterion for the
validity of the semi-classical approximation in de Sitter space is satisfied
for fluctuations on all scales well below the Planck scale. The perturbations
of the second kind contain additional massless scalar degrees of freedom
associated with changes of state of the fields on the cosmological horizon
scale. These scalar degrees of freedom arise necessarily from the local
auxiliary field form of the effective action associated with the trace anomaly,
are potentially large on the horizon scale, and therefore can lead to
substantial non-linear quantum backreaction effects in cosmology.

\end{abstract}


\maketitle

\vskip 1cm

\section{Introduction}

Although matter is undeniably quantum, in General Relativity its
gravitational effects are treated classically. In order to include the
quantum effects of matter, a natural next step from the purely
classical theory is a semi-classical treatment, in which the classical
matter stress tensor $T^a_{\ b}$ in Einstein's equations is replaced
by its expectation or average value $\lag T^a_{\ b}\rag$, but the
spacetime geometry is still treated classically. This is clearly an
approximation to a more complete treatment, whose validity depends
upon a number of assumptions, chief among them that the quantum
fluctuations of the matter energy-momentum-stress tensor $T^a_{\ b}$
about its average value are ``small," at least on macroscopic distance
scales.

This heuristic condition can be given a precise meaning in the
semi-classical or mean field limit of an effective field theory
approach to gravity. In the limit, $\hbar \rightarrow 0$ but the
number of quantum fields $N \rightarrow \infty$ with $\hbar N$ fixed,
it becomes permissible to replace the conserved $T^a_{\ b}$ of a
quantum theory by its expectation value $\lag T^a_{\ b}\rag$, as the
source term for the semi-classical Einstein equations,
\be R^a_{\ b} -
\frac{R}{2}\, \delta^a_{\ b} + \Lambda\, \delta^a_{\ b} = 8\pi G_{_N}
\lag T^a_{\ b}\rag_{_R}\,.
\label{scE}
\ee
The corrections to the renormalized expectation value $\lag
T^a_{\ b}\rag_{_R}$ are suppressed by $1/N$, and negligible in the
large $N$ limit \cite{Tomb}. In this way and in this limit the
treatment of the metric geometry of spacetime as classical, sharply
peaked about its mean value, can be obtained from an underlying quantum
theory of matter.

Replacing the classical source of Einstein's equations by a quantum
expectation value has a number of important implications. Since
$T^a_{\ b}(x)$ is an operator which is at least bi-linear in local
quantum fields at the same spacetime point, any direct evaluation of
its expectation value is divergent. The expectation value $\lag
T^a_{\ b}\rag$ must be renormalized, {\it i.e.} its divergent parts
identified and removed by means of suitable local counterterms in the
effective action of low energy gravity.

It is essential that the renormalization procedure maintain the
covariant conservation of $\lag T^a_{\ b}\rag_{_R}$, for it is to act
as a source of the gravitational field through the semi-classical
Einstein equations (\ref{scE}); otherwise (\ref{scE}) would be
inconsistent. Covariant methods for identifying and removing the
ultraviolet divergences $\lag T^a_{\ b}\rag$ by point-splitting, heat
kernel expansions, or dimensional regularization which satisfy the
requirements of general covariance have been developed, and lead to
consistent well-defined finite results for $\lag T^a_{\ b}\rag_{_R}$
\cite{DeW,BirDav}.

Eqs. (\ref{scE}) also provide consistency conditions for the finite
terms in the renormalized expectation value $\lag T^a_{\ b}\rag_{_R}$.
To evaluate this quantity in the usual Minkowski vacuum state
of flat spacetime, one is {\it required} to define the renormalized
expectation value of $T^a_{\ b}$ to vanish identically, when $\Lambda
= 0$, or (\ref{scE}) will not be satisfied for $R^a_{\ b}=0$.
Equivalently, by transposing the cosmological term $\Lambda$ to the
right side of (\ref{scE}), the existence of a flat spacetime solution
to the semi-classical Einstein equations requires that the difference,
$\lag T^a_{\ b}\rag_{_R} - \frac{\Lambda}{8\pi G_{_N}}\, \delta^a_{\ b}$
vanish for $R^a_{\ b}=0$, so that any cosmological term is exactly
canceled by a finite renormalization of $\lag T^a_{\ b}\rag_{_R}$ in
flat space. Thus, neither the vacuum expectation value $\lag T^a_{\ b}\rag_{_R}$
nor the cosmological term $\Lambda \,\delta^a_{\ b}$ alone have direct
physical significance, but only the combination, $\lag T^a_{\ b}\rag_{_R} -
\frac{\Lambda}{8\pi G_{_N}}\, \delta^a_{\ b}$, which acts as the
source of curvature.

The condition that flat spacetime be a solution to the semi-classical
Eqs.  (\ref{scE}) with all matter in its Lorentz invariant Minkowski
vacuum state is a renormalization condition on the low energy effective
theory of gravity. This condition serves to define the effective theory
by fixing the subtractions of infinite and finite vacuum terms, and
makes the semi-classical framework a predictive one in spaces
other than flat space and/or in states other than the vacuum state.
At present this renormalization condition cannot be justified by any
fundamental theory, but solely by appeal to the experimental fact that
the very high frequency fluctuations responsible for the divergences
in $\lag T^a_{\ b}\rag$ do not produce any appreciable mean curvature of
space. The semi-classical theory described by (\ref{scE}) must be regarded
then as a low energy effective theory \cite{Dono}, on a par with pion
effective theories of nuclear forces or the Fermi theory of the weak
interactions, whose UV divergences are subsumed into effective low energy
parameters, and whose predictions are reliable only at sufficiently low
energies or long wavelengths, pending a more complete, but unspecified
theory at high energies and short wavelengths.

Classically, with $\Lambda > 0$ and in the absence of any other sources,
there is a maximally symmetric non-flat solution to Einstein's equations,
namely de Sitter spacetime with $R^a_{\ b} =\Lambda\, \delta^a_{\ b}$.
In the classical theory $\Lambda$ is a fixed constant with absolutely
no dynamics, and de Sitter space is a stable solution to Einstein's
equations with a positive cosmological constant. On the other hand,
in the semi-classical framework of (\ref{scE}), as just observed,
the cosmological constant term cannot be divorced from the definition
of $\lag T^a_{\ b}\rag_{_R}$ itself, as $\Lambda$ can be reabsorbed
into a finite redefinition of $\lag T^a_{\ b}\rag_{_R}$. Since
$\lag T^a_{\ b}\rag_{_R}$ depends upon the state in which it is
evaluated, the energy of the vacuum is no longer a fixed non-dynamical
constant of the Lagrangian, as in the classical theory, but instead
depends upon the infrared choice of ``vacuum" state in curved space,
in which the expectation value $\lag T^a_{\ b}\rag_{_R}$ is defined.
Which infrared state and which spacetime is preferred then becomes
a dynamical question, which can be investigated by probing the response
of the system to perturbations of the state of the fields and the spacetime
geometry together on length scales much greater than the Planck scale.
In the semi-classical theory (\ref{scE}) the stability of de Sitter
space to these infrared perturbations depends upon the behavior
of the quantum fluctuations of the matter sources on cosmological
length scales of the order of $H^{-1} = \sqrt{3/\Lambda}$, and the
dynamical question can be studied in a systematic way with the same
effective action that gives rise to (\ref{scE}) by the method of linear
response.

The linear response of a system to a small perturbation is familiar
from other branches of physics, and has been formulated for
semi-classical gravity in Refs. \cite{Mo86,AndMolMot}.
The fluctuations of the matter fields lead to fluctuations in the
expectation value $\lag T^a_{\ b}\rag_{_R}$, which is probed by
considering the response of $\lag T^a_{\ b}\rag_{_R}$ to a small
perturbation of the geometry $g_{ab} \rightarrow g_{ab} + \delta
g_{ab}$. The quantum action principle tells us that this variation is
given formally by
\bea
&&\delta \lag T^a_{\ b}(x)\rag = \frac{1}{2}\int d^4 x'\sqrt{-g(x')}\,
\Pi^{a\ cd({\rm ret})}_{\ b}(x, x')\,\delta g_{cd}(x') \nonumber\\
&& \qquad =  -\frac{i\hbar}{2} \int d^4 x'\, \sqrt{-g(x')}\,\theta (t,t')  \,
\lag in \vert [T^a_{\ b}(x), T^{cd}(x')]\vert in \rag \ \delta g_{cd}(x')\, ,
\label{polten}
\eea
up to contact terms proportional to delta functions and derivatives
thereof with support at $x=x'$ \cite{Mo86,AndMolMot}. The local contact terms
are determined by the renormalization of the retarded polarization
$\Pi^{a\ cd({\rm ret})}_{\ b}(x, x')$ of the matter field(s), evaluated
in the same state as the self-consistent background solution of (\ref{scE}).

The linear variation (\ref{polten}) is essentially equivalent to first
order perturbation theory in the metric perturbation $\delta g_{ab} \equiv h_{ab}$,
and assumes that the entire change in the quantum state and the expectation value,
$\lag T^a_{\ b}\rag_{_R}$ comes about directly from the change in the
geometry. These variations, linear in the metric variation $h_{ab}$,
we term variations of the first kind. This first class of linear perturbations
are the ones upon which attention has been focused in the earlier literature
on this subject \cite{HarHu,Hor,HorWal,Starob,CalHu,CamVer}. However it is also possible
to consider variations of the state that lead to variations in $\lag T^a_{\ b}\rag_{_R}$
which are {\it not} directly driven by local variations in the metric. For example
one might consider variations of the temperature of a system coupled to a heat
reservoir, or other variations of the boundary conditions imposed on the system.
These variations may depend upon dynamical degrees of freedom contributing
to the stress tensor other than the local metric geometry, and hence are not
proportional to $\delta g_{cd}$. The corresponding state dependent variations
in $\lag T^a_{\ b}\rag_{_R}$ which do not depend upon the local $\delta g_{cd}$,
we term variations of the second kind. We shall see that the trace anomaly
parameterizes certain specific state dependent variations of the stress tensor
which are of this second kind, and that they can have significant effects
in de Sitter space.

The dynamical linear response equation which probes the self-consistent
solution of (\ref{scE}) to small fluctuations of both kinds is the linear
integro-differential equation for the self-consistent metric perturbation $h_{ab}$,
\be \delta
\left\{R^a_{\ b} - \frac{R}{2} \delta^a_{\ b} + \Lambda
\delta^a_{\ b}\right\} = 8\pi G_{_N} \delta\, \lag T^a_{\ b}\rag_{_R}
\, ,
\label{linresgen}
\ee
obtained when the local linearized variation of the left side of
(\ref{scE}) is set equal to the properly renormalized right side.
Once the ultraviolet divergences of this linear response equation are
isolated and removed, (\ref{linresgen}) can be used to study the infrared
fluctuations of matter and geometry of de Sitter space in a self-consistent
and reliable way. This is our main purpose in this paper.

The most direct way of deriving (\ref{linresgen}) and determining the
correct renormalization counterterms necessary to properly define the
formal expressions for $\lag T^a_{\ b}\rag$ and $\delta \lag
T^a_{\ b}\rag$ is to integrate out the matter fields, obtaining their
one-loop effective action in a general gravitational background.
In the large $N$ limit, the contributions of higher order
gravitational metric fluctuations themselves to $\delta \lag
T^a_{\ b}\rag$ are suppressed by $1/N$ and may be neglected at leading
order. In this limit the metric may be treated as a (semi-)classical
quantity. In particular, there are no stochastic or ``noise" terms in
the linear response equations, (\ref{linresgen}) at leading order in the
$1/N$ expansion. These can arise only at higher orders and are
outside the domain of the semi-classical theory.

Renormalizing the one-loop matter effective action
and taking its first variation yields the semi-classical Einstein
equations (\ref{scE}). In addition to fixing the background geometry,
the solution of (\ref{scE}) involves also fixing the quantum state
(or density matrix) $\vert in\rag \lag in \vert$ of the fields, with
respect to which all expectation values are to be computed. Then a
second variation of the effective action, or first variation of
(\ref{scE}) is performed around this background metric to yield the
linear response equation (\ref{linresgen}) for
$h_{ab}$. The kernel of the polarization integral in (\ref{polten}) is
computed using the background geometry and the Green's functions for
the quantum field in the specific quantum state (density matrix) fixed
in the previous step.  The form of the counterterms used to define the
renormalized expectation value $\lag T^a_{\ b}\rag_{_R}$ are the same
used to renormalize its linear variation $\delta
\lag T^a_{\ b}\rag_{_R}$ \cite{AndMolMot}.

In \cite{AndMolMot} we proposed a criterion for the validity of the
semi-classical Einstein equations (\ref{scE}), namely, that the solutions
of the linearized equations (\ref{linresgen}) should lead to linearized
gauge invariant amplitudes which remain bounded for all times. The
criterion is satisfied in flat spacetime. The present work extends the
analysis and tests the criterion in de Sitter spacetime for conformal
matter, where the variation of the stress tensor in (\ref{linresgen})
also can be evaluated in closed form.

The interest of applying this linear response method to de Sitter
spacetime arises from several different considerations. First de Sitter
spacetime serves as a fundamental testing ground for semi-classical
effects of quantum vacuum fluctuations in a curved space with a
cosmological constant, in which analytic results can be obtained.
Secondly, current cosmological models of structure formation in the
universe assume that this structure grew by gravitational instability
from small quantum inhomogeneities in an otherwise featureless,
primordial de Sitter-like inflationary phase. Finally, observations
of Type Ia supernovae suggest that the universe may be in a
de Sitter-like accelerating phase in the present epoch, with some
$70$-$75$\% of the energy density of the universe today being of a
hitherto undetected dark variety, with negative pressure,
$p \approx - \rho$ \cite{DE}. Since this is exactly the equation of
state of a cosmological term or that of the quantum ``vacuum," the
study of quantum fluctuations and their stability in de Sitter spacetime
is of possible direct relevance to the cosmology of the early universe,
structure formation, and the cosmic acceleration of the present universe.

In this paper we study the solutions of this dynamical linear response
equation (\ref{linresgen}) under small fluctuations in de Sitter
spacetime for the case of conformal matter fields, {\it i.e.} those
which are classically conformally invariant, up to the quantum
conformal or trace anomaly. The study of metric perturbations
in cosmology due to quantum conformal matter in cosmology was
initiated in Ref. \cite{HarHu}, whose authors made use of an
in-out effective action formalism, rather than the retarded
boundary conditions appropriate for causal linear response.
The in-in formalism for the effective action in cosmology
was developed in Ref. \cite{CalHu}, and extended and elaborated in
Ref. \cite{CamVer}. The general form of the perturbed stress tensor
in an arbitrary FRW cosmological spacetime obtained in \cite{CamVer},
agrees with the earlier results of Refs. \cite{Hor,HorWal,Starob}.
We study the solution of the resulting causal linear response
Eqs.  (\ref{linresgen}) with this general stress tensor source for
inhomogeneous and anisotropic perturbations away from de Sitter
space in Sec. VII A. Previous discussions of linearized
perturbations of de Sitter space have focused instead on
spatially homogeneous, isotropic perturbations, earlier in
the functional Schr\"odinger initial state formalism \cite{IsaRog},
and more recently for a minimally coupled scalar field in the in-in
effective action formalism \cite{NadRouVer}.

For the more general case of spatially inhomogeneous and anisotropic
perturbations, we find the analogous short distance ({\it i.e.} Planck
scale) quantum fluctuations expected from the corresponding analysis
in flat spacetime, which should be discarded as outside the range
of validity of the semi-classical theory. Going further, when general
inhomogeneous and anisotropic state dependent variations of the stress
tensor are considered ({\it i.e.} variations of the second kind), we
find additional modes arising from coherent macroscopic changes of state
on every scale, characterized by a spatial wave vector $\vec k$, including
the scale of the de Sitter horizon itself. These cosmological horizon
modes are associated with the effective action of the trace anomaly,
not variations of the local metric geometry, and as such are different
than anything in the purely classical theory, or encountered in previous
studies of restricted classes of perturbations. The existence of this
new set of scalar cosmological horizon modes arising from the quantum
trace anomaly is a principal result of this paper.

To keep this result in perspective, and when considering quantum effects
in cosmology generally, it is important to recognize that the intrinsically
quantum phenomenon of phase coherence and consequently non-classical effects
may be present at any scale. Indeed macroscopic quantum states are encountered
at low enough temperatures in virtually all branches of physics on a very
wide variety of scales. In addition to low temperature and condensed matter
laboratory systems, the chiral symmetry breaking expectation value of quark
bilinears $\lag\bar q q\rag$ in QCD and the Higgs field $\lag
\Phi\rag$ in electroweak theory are examples of macroscopic vacuum
condensates, which extend over arbitrarily large distances. It is
these condensates which provide the connection between a more
microscopic approach to both the strong and electroweak interactions
and the low energy effective theories which preceded them. Such
coherence effects due to the quantum wave-like properties of matter, and
its propensity to form phase correlated states over macroscopic
distance scales cannot be treated in a purely classical description of
gravity. However, the semi-classical treatment of the effective
stress-energy tensor source $\lag T^a_{\ b}\rag_{_R}$ for Einstein's
equations already allows for such effects, in a mean field
description \cite{MotVau}.

A significant part of the effective action for conformally invariant fields
is determined by the conformal or trace anomaly. Although intrinsically
non-local in terms of the spacetime geometry, the effective action
determined by the anomaly can be cast into a covariant local form
by the introduction of one or more scalar auxiliary degree(s) of
freedom~\cite{MotVau,Rie,FraTse,ShaJac,BFS,AMMDE}. Since there are two
distinct cocycles in the non-trivial cohomology of the Weyl group in
four dimensions \cite{MazMot}, the most general representation of the
anomaly action is in terms of two auxiliary scalar degrees of freedom,
each satisfying fourth order linear differential equations of motion
(\ref{EFtraces}). The resulting stress tensor in terms of these scalar
auxiliary fields has been used to study the non-local macroscopic
quantum coherence effects contained in the semi-classical effective
theory (\ref{scE}) \cite{MotVau,RN}. The scalar auxiliary fields are
related to the freedom to change the macroscopic quantum ``vacuum "
state, and supply additional infrared relevant terms in the low energy
effective theory of gravity \cite{MazMot}. In perturbations
about de Sitter spacetime we shall show they give rise to new
cosmological horizon modes, not contained in classical perturbation
theory, that can lead to large backreaction effects at the
very largest Hubble scale of a de Sitter universe. Thus scalar
degrees of freedom in cosmology arise naturally from the
effective action of the conformal anomaly of massless quantum fields
in the Standard Model, without the {\it ad hoc} introduction of an
inflaton field.

The paper is organized as follows. In order to fix ideas and notations
we review in the next section the linear response fluctuation analysis
in flat spacetime with a vanishing cosmological constant. In Section III,
we consider linear response for conformal matter fields in conformally flat
spacetimes, where the stress tensor variation of the first kind (\ref{polten})
can be expressed in closed form. We show that for long wavelength solutions
of the equations within the range of validity of the semi-classical theory, it
is sufficient to study only the time-time component of the linear response
equations, in a general Friedman-Robertson-Walker (FRW) spacetime. In Section IV
we consider the stress tensor and linear response equations that follow from
the trace anomaly part of the effective action only, in the auxiliary
field description, comparing the result to the previous general approach in
conformally flat spacetimes. In Section V we derive the linear response
equations in both approaches for fluctuations in de Sitter spacetime, and
show how the auxiliary field degrees of freedom in the anomaly action
describe cosmological horizon scale modes. In Section VI we discuss coordinate
transformations and recast our linear response equations in a completely gauge
invariant form, showing in particular that the new cosmological horizon modes
are gauge invariant.  We give the quadratic gauge invariant action corresponding to the linear response
equations in these variables. In Section VII we give a qualitative and quantitative
discussion of the solutions of the linear response equations in de Sitter
space, showing that the auxiliary field anomaly action captures all the
infrared quantum effects correctly. In Section VIII the cosmological horizon
modes, their interpretation as fluctuations in the Hawking-de Sitter
temperature, and their possible implications for the the stability of
de Sitter space are discussed. Section IX contains a full summary and
discussion of our results. There are two appendices, the first containing
an in-depth discussion of the properties of the non-local kernel $K$
which arises in the linear response equations in conformally flat spaces,
and the second, some of the details of the second variation of the anomaly
action. Except where indicated the units used are $\hbar = c  = 1$ and
the metric and curvature conventions are the same as those of Misner, Thorne,
and Wheeler \cite{MTW}.

\section{Linear Response in Flat Spacetime}

In this section the derivation and solutions to the linear response equations
in a flat space background are reviewed. In flat spacetime Lorentz invariance
permits the full decomposition of perturbations into scalar and tensor components,
with respect to the background Minkowski four-metric $\eta_{ab} =$ diag $(-+++)$.
In Ref. \cite{AndMolMot} explicit results for these components were obtained for
a quantum scalar field with general mass $m$ and curvature coupling $\xi$. Making
use of those results as a specific example, we generalize the discussion here
to any set of underlying quantum fields. We emphasize that the four dimensional
tensor decomposition in Minkowski spacetime is distinct from the decomposition
of the perturbations into scalar, vector, and tensor components with respect
to the spatial three-metric $g_{ij}$ appropriate for general FRW spacetimes,
and employed in the succeeding sections.

The starting point of a linear response analysis of quantum
fluctuations is a self-consistent solution to the semi-classical
Einstein equations, (\ref{scE}).  In flat spacetime this requires that
$\lag T^a_{\ b}\rag_{_R} = 0$ in the usual Minkowski vacuum state with
$\Lambda = 0$. Since it is a dimension four operator, the
renormalization of $\lag T^a_{\ b}\rag$ requires fourth order
counterterms in the effective action. These gives rise to additional
geometric terms in equations (\ref{scE}), fourth order in derivatives,
which are usually displayed on the left side of the semi-classical
equations (\ref{scE}).  We shall instead adopt the convention of
including such geometric terms in the definition of the renormalized
expectation value, $\lag T^a_{\ b}\rag_{_R}$ itself: {\it cf.} equations
(\ref{linresflat}) and (\ref{Hdef}) below. Viewing the semi-classical
theory as an effective field theory, these local higher derivative
terms multiplied by finite coefficients are in any case negligibly
small at macroscopic length scales much greater than the extreme
microscopic Planck scale, $L_{Pl} = \sqrt{\hbar G_N/c^3} \simeq
1.616 \times 10^{-33}$ cm., or at energy scales much less than the
Planck energy $M_{Pl}c^2 = \sqrt{\hbar c^5/G_N} \simeq 1.223 \times
10^{19}$ GeV.

In flat spacetime, the non-local integro-differential equation
(\ref{polten}) in coordinate space is easily handled by Fourier
transforming to momentum space. With the self-consistent solution
of (\ref{scE}) for $\Lambda = 0$ just the usual Minkowski vacuum
state, the linear response equations around flat spacetime take
the form~\cite{AndMolMot}
\bea
&&\delta G_{ab}  =
8\pi G_{_N}\,\delta \langle T_{ab}\rangle_{_R} \nonumber\\
&& \qquad = 8\pi G_{_N}\,\left\{\left( \alpha_{_R} + \frac{F^{(T)}}{2}\right)\,
\delta\ ^{(C)\!\!}H_{ab} +
\left( \beta_{_R} + \frac{F^{(S)}}{12}\right)\,
\delta\ ^{(1)\!\!}H_{ab}\right\}\,.
\label{linresflat}
\eea
Here the two local, conserved geometric tensors, $^{(C)\hspace{-.1cm}}H_{ab}$
and $^{(1)\hspace{-.1cm}}H_{ab}$, also denoted as $-A_{ab}$ and $-B_{ab}$
respectively are
\bes\bea
&&\hspace{-2cm} ^{(C)\hspace{-.1cm}}H_{ab} \equiv \frac{1}{\sqrt{-g}}
\frac{\delta}{\delta g^{ab}}
\int\, d^4\,x\,\sqrt{-g}\, C_{abcd}C^{abcd} =
4\nabla_c \nabla_d\,C_{(a\ b)}^{\ \ c\ \,d}
\,+\, 2C_{a\ b}^{\ c\ d}R_{cd} = -A_{ab}\,, \label{Aabdef}\\
&&\hspace{-2cm} ^{(1)\hspace{-.1cm}}H_{ab} \equiv \frac{1}{\sqrt{-g}}
\frac{\delta}{\delta g^{ab}}
\int\, d^4\,x\,\sqrt{-g}\, R^2 = 2 g_{ab}\sq R  - 2\nabla_a\nabla_b R
+ 2 R R_{ab} - \frac{g_{ab}}{2}R^2 = -B_{ab} . \label{H1def}
\eea
\label{Hdef}
\ees
\vspace{-.6cm}

\noindent
These tensors, which are derived from the two dimension-four local invariants,
$C_{abcd}C^{abcd}$ and $R^2$ respectively, are required for renormalization of the
expectation value $\lag T_{ab}\rag$. The parameters $\alpha_{_R}$, and $\beta_{_R}$
are the finite, renormalized coefficients of these local terms. The causal
response functions,
\be
F^{(T,S)} (K^2) \equiv K^2 \int_0^{\infty} \frac{d s}{s^3} \
\frac{\rho^{(T,S)}(s)}{s + K^2 - i\epsilon\ {\rm sgn}\,\omega}\,,
\label{FTSdef}
\ee
in (\ref{linresflat}) have been renormalized by subtracting the dispersion
integral three times at $s=0$.\footnote{In this section only we employ the
notation $K^a$ for a four-vector in flat spacetime, and
$K^2 \equiv \eta_{ab} K^a K^b = -\omega^2 + k^2$ to distinguish it
from $k \equiv |\vec k|$, the magnitude of its three spatial components
used throughout the paper.}
They are the non-local part of $\delta \lag T_{ab}\rag_{_R}$ arising from
the matter fluctuations. The renormalization effected by subtracting the
Taylor series expansion in $K^2$ of the unrenormalized response function
up to order $(K^2)^2$ fixes the value of the cosmological constant term,
Newtonian constant (order $K^2$) and fourth order terms $A_{ab}$ and $B_{ab}$
at order $(K^2)^2$ to be their prescribed renormalized values at $K^2 =0$.
The renormalization prescription and local counterterms of the retarded
polarization tensor $\Pi^{a\ cd({\rm ret})}_{\ b}(x, x')$ at $x=x'$
are defined in momentum space by this procedure at $K^2 = 0$. The
modification of the renormalization procedure necessary for massless
quantum fields is discussed below.

\begin{figure}
\includegraphics*[viewport=200 530 450 650,angle=0,width = 0.5\textwidth,clip]{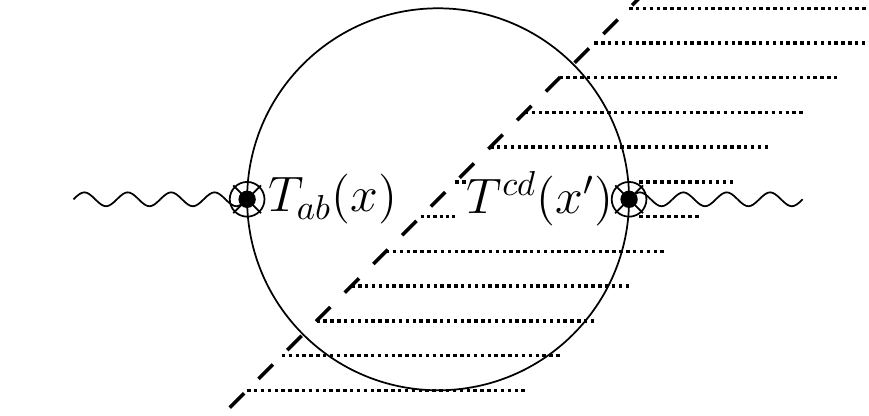}
\caption{The one-loop vacuum polarization $\Pi_{ab}^{\ \ cd}(x,x')$.
Its imaginary part, obtained by cutting the internal lines as illustrated
and placing them on-shell gives the spectral functions $\rho^{(T,S)}$.}
\label{Fig:polcut}
\end{figure}

The superscripts $T$ and $S$ in (\ref{FTSdef}) denote quantities associated
with tensor (spin-$2$) and scalar (spin-$0$) perturbations which are defined
with respect to the four dimensional background Minkowski metric by
Eqs. (\ref{projs}) below. As discussed in Ref. \cite{AndMolMot} these
causal response functions are completely determined by their imaginary
parts in terms of the spectral functions $\rho^{(T,S)}$ of the matter fields.
The spectral functions $\rho^{(T,S)}$ are finite and may be computed
directly from the one-loop cut diagram of Fig. \ref{Fig:polcut} in flat
spacetime.

It is essential that the linear response equation (\ref{linresflat}) is
causal, {\it i.e.} that the stress tensor response at $(t, \vec x)$ is
sensitive to variations of the geometry only within the past light
cone of $(t, \vec x)$. The $ - i\epsilon\ {\rm sgn}\, \omega$
prescription (where $K^2 = -\omega^2 + k^2$, $k \equiv |\vec k|$ and
$\omega \equiv K^0$) enforces retarded boundary in-in conditions,
appropriate for this causal linear response.

Note also that both terms on the right side of (\ref{linresflat})
are geometrical, involving the linearized metric fluctuation $h_{ab}$
itself, {\it i.e.} they are homogeneous variations of the first kind
only. It is clear that non-geometrical terms could always be added to
$\delta \lag T^a_{\ b}\rag_{_R}$, by allowing states other than the
Minkowski vacuum to contribute to the variation, even in the absence
of any metric variation. These {\it state dependent} contributions
will be considered in the next section.

The expressions (\ref{linresflat})-(\ref{FTSdef}) are finite and
general for the linear response of a quantum field theory in the usual
Minkowski vacuum perturbed around flat spacetime. To linear order all
indices are raised and lowered with the flat space metric
$\eta_{ab}$. The dependence on the particular quantum matter theory
enters only through the spectral functions $\rho^{(T,S)}$, which must
be positive definite on general grounds of unitarity. Explicit
calculations of the spectral functions and causal response functions
of a free scalar field theory with mass $m$ and curvature coupling
$\xi$ are given in Ref. \cite{AndMolMot}. The results are:
\bes\bea
\rho^{\rm (S)}(s)&=&
\frac{\theta(s-4m^2)} {24 \pi^2} \, \sqrt{1-\frac{4m^2}{s}}
\,\left[m^2 +\frac{(1-6 \xi)s}{2}\right]^2\, .
\label{rhoS}\\
\rho^{(\rm T)}(s)&=&
\frac{\theta(s-4m^2)}{60 \pi^2}\,  \sqrt{1-\frac{4m^2}{s}}\,
\left(\frac{s}{4}-m^2\right)^2\, .
\label{rhoT}
\eea\ees
For conformal scalars, $m=0$ and $\xi = \frac{1}{6}$, the scalar
spectral function $\rho^{\rm (S)}$ vanishes for all $s>0$. However,
the expression $F^{(S)}$ in (\ref{FTSdef}) is finite in the limit
$m\rightarrow 0$ limit. The two response functions in this limit are
\bes\bea
&&F^{(S)} =  \frac{1}{1440\pi^2} = \frac{4b}{3}\,,\label{FS}\\
&&F^{(T)} \rightarrow  \frac{1}{960\pi^2} \,\ln
\left(\frac{K^2- i\epsilon\ {\rm sgn}\, \omega} {m^2}\right)
= 2\,b\, \ln \left(\frac{K^2- i\epsilon\ {\rm sgn}\, \omega} {m^2}\right)
\,,\label{FT}
\eea
\label{Fconf}
\ees
\vspace{-.6cm}

\noindent
where $b = \frac{1}{(4\pi)^2}\frac{1}{120}$ is the coefficient of the
$C_{abcd}C^{abcd}$ term of the trace anomaly for the conformal scalar field
in an arbitrary curved background. The logarithmic divergence in (\ref{FT})
in the limit of zero mass is the result of the renormalization of
$\alpha_{_R}$ at $K^2=0$. If the renormalization is performed instead
at an arbitrary but non-zero mass scale $\mu^2$, and we define
\be
\alpha_{_R} (\mu^2) \equiv \alpha_{_R} + b \ln \left( \frac{\mu^2}{m^2}\right)\,,
\label{alpmu}
\ee
then the combinations appearing in the linear response equation (\ref{linresflat}) are
\bes\bea
&&\alpha_{_R} + \frac{F^{(T)}}{2}= \alpha_{_R} (\mu^2) + b \,\ln
\left(\frac{K^2- i\epsilon\ {\rm sgn}\, \omega} {\mu^2}\right)\,,\\
&&\beta_{_R} + \frac{F^{(S)}}{12} = \beta_{_R} + \frac{b}{9}\,.\label{betaRb}
\eea\label{combin}\ees
These are finite in the conformal limit $m\rightarrow 0$, provided
$\alpha_{_R} (\mu^2)$ remains finite in that limit.  Note also that since
\be
\mu^2\frac{d}{d\mu^2} \left[ \alpha_R (\mu^2) + b \ln
\left(\frac{K^2- i\epsilon\ {\rm sgn}\, \omega}{\mu^2}\right) \right] = 0\,,
\label{muindep}
\ee
the physical linear response does not depend on the arbitrary renormalization
scale $\mu^2$, {\it i.e.} this combination is renormalization group invariant.
For the scalar response function $F^{(S)}$ of a conformal matter field, there
is no logarithmic term and no $\mu^2$ dependence at one-loop order, and the
last term in (\ref{betaRb}) is simply an additional finite renormalization
of the coefficient $\beta_{_R}$ of the $R^2$ term in the effective action.

Eqs. (\ref{Fconf}) and (\ref{combin}) are valid in fact for any conformal
matter field, provided one uses the appropriate anomaly $b$ coefficient
for that field. This may be seen from the general form of the trace anomaly
for conformal matter fields in an arbitrary curved spacetime, {\it viz.},
\bea
&& \lag T^a_{\ a} \rag_{_R}
= b F + b' \left(E - \frac{2}{3}\sq R\right) + b'' \sq R \nonumber\\
&& \qquad = b\left(F + \frac{2}{3}\sq R\right)
+ b' E + \left( b'' -\frac{2b+ 2b'}{3}\right) \sq R\,,
\label{tranom}
\eea
where
\bes\bea
&&E \equiv ^*\hskip-.2cmR_{abcd}\,^*\hskip-.1cm R^{abcd} =
R_{abcd}R^{abcd}-4R_{ab}R^{ab} + R^2
\,,\qquad {\rm and} \label{EFdef}\\
&&F \equiv C_{abcd}C^{abcd} = R_{abcd}R^{abcd}
-2 R_{ab}R^{ab}  + \frac{R^2}{3}\,.
\eea\ees
\noindent  with $R_{abcd}$ the Riemann curvature tensor, $^*\hskip-.1cmR_{abcd}
= \frac{1}{2}\varepsilon_{abef}R^{ef}_{\ \ cd}$ its dual, and $C_{abcd}$ the Weyl
conformal tensor. The coefficients $b$, $b'$, and $b''$ are computed at
one-loop order, proportional to $\hbar$, and determined by the number of
massless conformal fields and their spin via
\bes\bea
b &=& \frac{\hbar}{120 (4 \pi)^2}\, (N_S + 6 N_F + 12 N_V)\,, \label{bdef}\\
b'&=& -\frac{\hbar}{360 (4 \pi)^2}\, (N_S + 11 N_F + 62 N_V)\,,
\eea
\label{bbprime}
\ees
\vskip -.3cm
\noindent with $N_S$ the number of spin $0$ fields, $N_F$ the number of spin
$\frac{1}{2}$ Dirac fields, and $N_V$ the number of spin $1$ fields \cite{BirDav}.
The $\hbar$ in (\ref{bbprime}) emphasizes that these coefficients appear at
one-loop order, and that the anomaly terms are present in the leading order
semi-classical limit, $\hbar \rightarrow 0, N \rightarrow \infty$, with
$\hbar N$ finite (where $N$ is any of $N_S, N_F, N_V$). With this understood,
we shall henceforth set $\hbar = 1$.

The $b$ coefficient is also the one that determines the logarithmic scale
dependence of the $C_{abcd}C^{abcd}$ term in the effective action. The
coefficient $b''$ is scheme dependent, corresponding as it does to the
trace of the tensor $^{(1)}H_{ab} = -B_{ab}$, which is derived from a local
action, whose coefficient can be shifted by a finite shift of $\beta_{_R}$.
For classically conformal invariant theories there is no logarithmic scale dependence
in the coefficient of the $R^2$ term in the effective action at one-loop
order and one finds $b'' = 2(b+ b')/3$ in covariant regularization
schemes, such as dimensional regularization, so that the last term in
(\ref{tranom}) vanishes \cite{Duff}.

From these considerations we deduce that the renormalized stress tensor
variation in (\ref{linresflat}) in Fourier space for conformal matter in
flat spacetime may be written in the form,
\be
\delta  \lag T_{ab}\rag_{_{R, \, conf.}} = -\left[\alpha_{_R}(\mu^2) + b\,
\ln \left(\frac{K^2- i\epsilon\ {\rm sgn}\, \omega} {\mu^2}\right)\right]\,
\delta A_{ab} - \left(\beta_{_R} + \frac{b}{9}\right) \delta B_{ab}\,,
\label{varflat}
\ee
which is obtained by substituting Eq. (\ref{combin}) into Eq. (\ref{linresflat}).
Since both the invariants $E$ and $F$ are quadratic in the curvature tensor,
their linear variations away from flat spacetime vanish, and only the
$\frac{2}{3} b\sq \delta R$ term survives in the variation of the trace,
\be
\delta  \langle T^a_{\ a}\rangle_{_{R, \, conf.}} =
\left(6\beta_{_R} + \frac{2b}{3}\right)\sq (\delta R)\,.
\label{vartracflat}
\ee
All of the dependence of the linear response on the conformal matter theory
in flat spacetime is contained therefore in the single parameter $b$, determined
by the trace anomaly (\ref{tranom}) and given by (\ref{bdef}) for free conformal
matter fields of any spin. The $b'$ anomaly coefficient of the
Euler-Gauss-Bonnet density $E$ does not enter the linear variation of
the stress tensor around flat spacetime.

The tensor and scalar components in (\ref{linresflat}) or (\ref{varflat})
are conveniently separated with the aid of the set of orthogonal projectors,
\bes\bea
P^{(T) \, cd}_{ab} &=& \frac{1}{2} \left( \theta_a^{\ c} \theta_b^{\ d}
+  \theta_a^{\ d} \theta_b^{\ c} \right) -\frac{1}{3}\, \theta_{ab}\, \theta^{cd}\,,\\
P^{(V) \, cd}_{ab} &=& \frac{1}{2}\left( \delta_a^{\ c} \delta_b^{\ d}
+ \delta_a^{\ d} \delta_b^{\ c} - \theta_a^{\ c} \theta_b^{\ d}
- \theta_a^{\ d}\, \theta_b^{\ c} \right) \,, \label{projV}\\
P^{(S) \, cd}_{ab} &=& \frac{1}{3} \, \theta_{ab}\, \theta^{cd}\,,\\
\theta_{ab} &\equiv & \eta_{ab} - \partial_a \frac{1}{\sq}\partial_b
\rightarrow \eta_{ab} - \frac{K_a K_b}{K^2}
\eea
\label{projs}
\ees

\vspace{-.6cm}
\noindent
in momentum space, so that the linearized tensor and scalar metric
perturbations around flat spacetime are
\be
h_{ab}^{(T, S)} \equiv P^{(T, S) \, cd}_{ab}h_{cd}\,,
\ee
respectively. Both perturbations are transverse in the four dimensional sense,
\be
\partial^b h_{ab}^{(T,S)} =  0 = K^b h_{ab}^{(T,S)}\,,
\ee
while the tensor perturbation is also tracefree in the four dimensional sense,
\be
\eta^{ab}  h_{ab}^{(T)} = 0\,.
\ee
The linear variations around flat spacetime of the three local
geometric tensors appearing in (\ref{linresflat}) may be expressed then as
\bes\bea
\delta G_{ab} &=& \delta G_{ab}^{(T)} + \delta G_{ab}^{(S)} =
- \frac{1}{2} \sq h_{ab}^{(T)} + \sq h_{ab}^{(S)} \rightarrow
\frac{K^2}{2}\, h_{ab}^{(T)} - K^2\, h_{ab}^{(S)}\,,\\
\delta A_{ab} &=& \sq^2 h^{(T)}_{ab} \rightarrow (K^2)^2 \, h_{ab}^{(T)}\,,\\
\delta B_{ab} &=& 6 \sq^2 h^{(S)}_{ab} \rightarrow 6\, (K^2)^2\, h_{ab}^{(S)}\,,
\eea
\label{wre}
\ees

\vspace{-.7cm}
\noindent in terms of this decomposition into tensor and scalar metric
components in momentum space. The remaining projector onto linearized
vector perturbations (\ref{projV}) is not needed because
$h_{ab}^{(V)} = P^{(V) \, cd}_{ab}\, h_{cd}$ is gauge dependent and
non-transverse, and hence cannot appear in the linear
variations of the conserved tensors in (\ref{linresflat}).

Making use of the projection operators (\ref{projs}), we find that the
linear response equation around flat spacetime (\ref{linresflat})
separates finally into two independent, orthogonal components, {\it viz.},
\bes\bea
&&\delta G_{ab}^{(T)} = \frac{K^2}{2}\, h_{ab}^{(T)}  =
- 8\pi G_{_N}\,\left\{ \alpha_{_R} (\mu^2) + b \,\ln
\left(\frac{K^2- i\epsilon\, {\rm sgn}\, \omega}
{\mu^2}\right)\right\}\,(K^2)^2\, h_{ab}^{(T)}\,,
\label{lintena}\\
&&\delta G_{ab}^{(S)} = -\frac{1}{3} \theta_{ab} \delta R = -K^2\, h_{ab}^{(S)}
= - 8\pi G_{_N}\, \left(6\, \beta_{_R} + \frac{2b}{3}\right)\,
(K^2)^2\, h_{ab}^{(S)} \,,
\label{linscal}
\eea\label{linflat}
\ees
The solutions of the linear response equations in flat spacetime were
discussed in ref. \cite{AndMolMot} for scalar fields with arbitrary mass and
curvature coupling $\xi$.  In Fourier space it is straightforward to show
that there are no low energy solutions to (\ref{linflat}) other than
the usual propagating transverse, traceless, spin-2 gravitational wave
modes at $K^2 = 0$ of the pure Einstein theory with no sources.  Since
these transverse degrees of freedom are massless, there are only two
helicity states which propagate, the other three spin-2 states and the
one apparent scalar mode in (\ref{linscal}) being eliminated by the
four diffeomorphism constraints coming from the $ti$ and $tt$
components of the equations in a canonical analysis splitting spacetime
into space + time.

Additional non-trivial solutions to both the semi-classical tensor and scalar
equations appear at $K^2 \sim M_{Pl}^2$ (for $\alpha_{_R}$ and $\beta_{_R}$
of order unity). However, these solutions are outside the range of
applicability of the semi-classical effective theory, which requires
$K^2 \ll M_{Pl}^2$, and hence should be discarded. Physically, this
must be the case since there is no independent length scale against
which the wavelength of the metric perturbations of a conformally
invariant quantum field can be compared in flat spacetime other than
the Planck length. The situation will be quite different for cosmological
spacetimes where the horizon scale plays an important role. From
(\ref{vartracflat}) or (\ref{linscal}) it is clear that the Planck scale
solutions in the scalar or trace sector involve $\delta R \neq 0$. These
Planck scale solutions will be present in the linear response equations
around arbitrary spacetimes as well. Hence we shall find it convenient
to set $\delta R = 0$, in order to exclude these Planck scale solutions
in the effective field theory description, and focus on the remaining
solutions (if any) that are determined by the low energy curvature or
horizon scale in cosmology, which is assumed many orders of magnitude
greater than the Planck scale.

\section{Linear Response of Conformal Matter in Cosmology}

In this section some general properties of conformally invariant fields in
conformally flat spacetimes are briefly reviewed along with some previous
computations of the perturbed stress-energy tensor for these fields in these
spacetimes.  Then the specific form of the linear response equations in a
general spatially flat Friedman-Robertson-Walker (FRW) cosmological spacetime
is derived and the infrared physical content of the equations is shown to reside
purely in the $tt$ component.

A conformally flat spacetime is one for which the metric is
\be
g_{ab}(x) = \Omega^2(x) \eta_{ab} \label{conformal}
\ee
with $\eta_{ab}$ the Minkowski metric. The function $\Omega(x)$ is called
the conformal factor. Conformally invariant fields are ones whose classical
action is invariant under conformal transformations when the field is
multiplied by some power of the conformal factor. If the spacetime under
consideration is conformally flat and the matter fields are conformally
invariant, then the state of the fields that is most often considered is
called the conformal ``vacuum'' \cite{BirDav}. It is obtained by mapping
the Minkowski vacuum and all of its Green's functions in flat spacetime
to the conformally flat spacetime using the same transformation of the
fields which keeps the action invariant under conformal transformations.
In this special state the expectation value of the stress tensor of the
conformally invariant fields is completely determined by its trace anomaly
(\ref{tranom}) and can be written in terms of local curvature invariants
and their derivatives in the geometry specified by (\ref{conformal})
in the form \cite{BirDav,BroCas},
\be
\lag T_{ab}\rag_{_{R \,, conf}} = -2 b'\
^{(3)\!}H_{ab} -\frac{b}{9}\,B_{ab}\,,
\label{Tconf}
\ee
where
\be
^{(3)\!}H_{ab} \equiv - R_a^{\ c}R_{cb} + \frac{2}{3}\, R\,R_{ab}
+ \frac{1}{2}\, R_{cd}R^{cd}\,g_{ab}  - \frac{1}{4}\, R^2\, g_{ab}
- 2\,C_{acbd}\,R^{cd}  \,,
\label{Hthree}
\ee
and $B_{ab}$ is defined by (\ref{H1def}).  Of course, just as there
are many states in flat space that are of interest besides the Minkowski
vacuum, the conformal state is simply one choice of
allowed state among many, even if we restrict ourselves to the
subclass of states which are spatially homogeneous and isotropic,
consistent with the symmetries of (\ref{conformal}). Applying the
term, ``vacuum" to the conformal state is also somewhat misleading,
since $\lag T_{ab}\rag_{_{R \,, conf}} \neq 0$.

Some time ago several authors considered the linear variation of the
expectation value (\ref{Tconf}) for conformal matter fields by
conformally transforming the linear variation (\ref{varflat}) in flat
space \cite{HorWal,Starob}. It was assumed in that work that the fields
are in the conformal vacuum state and the variation considered was of
the first kind,  namely it was assumed that the quantum state of the
fields follows the metric variation in a prescribed way, with no other
state dependent variations considered. The results of Refs. \cite{HorWal,Starob}
are used in this section to write the specific form of the linear response
equations in a spatially flat FRW spacetime, with the addition of the
state dependent terms which are omitted in \cite{HorWal,Starob}, but
certainly allowed on general grounds \cite{BirDav}.

The line elements for spatially flat Friedman-Robertson-Walker (FRW)
spacetimes can be written in the alternate forms,
\be
ds^2 = -dt^2 + a^2(t)\, d\vec x^2 = \Omega^2(\eta)
\left( -d\eta^2 + d\vec x^2 \right)\,,
\label{FRW}
\ee
where $\eta$ is the conformal time and $t$ is the comoving time. These
cosmological spacetimes are conformally flat with the conformal factor
$\Omega (\eta) = a(t)$, the FRW scale factor.

In the flat space version of the linear response equations it was useful
to perform a full Fourier transform. For a FRW spacetime it is useful
to Fourier transform only in space, but not in time. The inverse Fourier
transform can be used to undo the time part of the logarithmic kernel in
(\ref{varflat}) with the result,
\be K (\eta-\eta'; k; \mu) =
\int_{-\infty}^{\infty}\frac{d\omega}{2\pi}\, e^{-i\omega
 (\eta-\eta')} \,\ln\left[\frac{ -\omega^2 + k^2 - i\epsilon\,
   {\rm sgn}\,\omega} {\mu^2}\right]\,.
\label{Kt}
\ee
This is a distribution, which may be defined by its action upon integration
against a suitably smooth, restricted class of test functions $f(\eta')$
for which $\int d\eta' K(\eta-\eta') f(\eta')$ exists. Because of the
$-i\epsilon\, {\rm sgn}\,\omega$ prescription, $K$ has support only
for times $\eta'$ earlier than the time $\eta$, consistent with causal linear
response. At $\eta = \eta'$ the distribution (\ref{Kt}) is singular and
requires regularization. The properties of this distribution and two
possible regularizations are studied in detail in Appendix A.

In terms of this distribution the linearized perturbation of $\lag
T_{ab} \rag_{_R}$ for a conformal field in the conformal vacuum around a
conformally flat spacetime with metric \eqref{FRW} is \cite{HorWal,Starob}
\be
\delta \lag T_{ab} (\eta ; {\vec k}) \rag_{_R} = -\,\frac{b\hspace{.1cm}}{\Omega^2}
\int_{\eta_0}^{\eta} \, d\eta' K(\eta-\eta'; k ; \mu^2)\ \delta
\bar A_{ab} (\eta'; {\vec k}) + L_{ab} + \frac{1\ }{\Omega^2}\
\delta \lag \bar T_{ab} (\eta ; {\vec k}) \rag_{_R}\,.
\label{conflat}
\ee
Here $L_{ab}$ is the local contribution,
\be
L_{ab} = -\alpha_{_R} (\mu^2) \,\delta A_{ab}
- \left(\beta_{_R} + \frac{b}{9}\right)\, \delta B_{ab}
- 2 b'\ \delta\, ^{(3)\hspace{-.1cm}}H_{ab}
-\frac{8b\hspace{.1cm}}{\Omega^2} \ \partial_c\partial_d
\left[(\ln \Omega)\ \delta\bar C_{a\ b}^{\ c\ d}\right]\,,
\label{Lab}
\ee
and the final term in (\ref{conflat}) is the variation of the stress tensor
arising from the variation of the state away from the conformal ``vacuum,"
which was omitted in \cite{HorWal,Starob}, but which is present in the
more general context of state dependent variations \cite{BirDav}. Except for
the lower limit of the integral in (\ref{conflat}) introduced to allow for
the possibility of starting the linear response system at some finite
initial conformal time $\eta_0$, and the addition of the variation of
the final state dependent term in (\ref{conflat}), these are the results
for the variation of the stress tensor of conformal matter around a
conformally flat space as given by the authors of Refs. \cite{HorWal,Starob}.

In (\ref{conflat}) both $\delta\bar A_{ab}$ and
$\delta\bar C_{a\ b}^{\ c\ d}$ are evaluated on the linearized perturbation,
\be
\bar h_{ab} \equiv \Omega^{-2} h_{ab}\,.
\label{barh}
\ee
from {\it flat} spacetime, with $h_{ab}$ the full linearized metric perturbation
about the conformally flat spacetime $g_{ab} = \Omega^2 \eta_{ab}$ of (\ref{FRW}).
The linearized variation around flat space of $^{(C)\!}H_{ab} = -A_{ab}$ is given
in momentum space by Eqs. (\ref{projs})-(\ref{wre}) of the previous section,
while $\delta \bar C_{a\ b}^{\ c\ d}$ is given by the completely traceless part
of
\be
\delta \bar R_{a\ b}^{\ c\ d} =
\frac{1}{2} \left( \partial_b\partial^c \bar h_a^{\ d}
+ \partial_a\partial^d \bar h_b^{\ c} - \partial_a\partial_b \bar h^{cd}
- \partial^c\partial^d \bar h_{ab}\right)
\ee
in the usual way, all derivatives and indices referring to flat Minkowski
coordinates.

As in the corresponding expression for flat spacetime (\ref{varflat}), the
$\alpha_{_R}(\mu^2) \delta A_{ab}$ term in (\ref{Lab}) comes from varying
the local $C_{abcd}C^{abcd}$ Weyl invariant term in the effective action,
while the $\beta_{_R}\delta B_{ab}$ term is a finite renormalization of the
local $R^2$ term in the effective action, which we have chosen to include in
the variation of $\lag T_{ab} \rag_{_R}$. The $b\ \delta B_{ab}$ and
$b'\ \delta\, ^{(3)\hspace{-.1cm}}H_{ab}$ terms are the variations of the
background geometric terms in (\ref{Tconf}), while the
$\partial_c\partial_d[\ln\Omega\,\delta\bar C_{a\ b}^{\ c\ d}]$ term has been
obtained by the authors of Refs. \cite{HorWal,Starob} in different ways, and
is necessary for consistency. This term will be shown in the next section
to be a necessary consequence of the covariant effective action functional
determined by the trace anomaly (\ref{tranom}).

Since $\delta A_{ab}$, $\delta \bar A_{ab}$ and the last term in (\ref{Lab})
involving the Weyl tensor variation are traceless, the trace of the linear
response equations for conformal matter perturbed away from a FRW background
for which $C_{abcd} = 0$ simplifies considerably. In fact, for variations
of the first kind, the trace is determined completely by the trace anomaly
(\ref{tranom}), which is given in terms of local curvature terms and is
independent of the quantum state of the field. Hence the trace of the
variation is given by the local geometric equation,
\be
-\delta R = 8\pi G_{_N} \delta \lag T^a_{\ a} \rag_{_R}
= 16\pi G_{_N} \left\{ -2b' \left(R^a_{\ c}\, \delta R^c_{\ a} - \frac{1}{3} R \,\delta R\right)
+ \left( 3\beta_{_R} + \frac{b}{3}\right) \delta (\sq R) \right\}\,.
\label{scalR}
\ee
By assumption, in the semi-classical approximation we require the Ricci curvature
and its variations to be much smaller than the Planck scale.  Thus all the terms on the right
side of (\ref{scalR}) are small compared to the left side. Hence we may restrict ourselves
to only those solutions of the full linear response equations for which
\be
\delta R = 0\,.
\label{delR0}
\ee
The only solutions eliminated by this condition are Planck scale solutions,
such as the ones we have already encountered in flat spacetime, with
$k^2 \sim M_{Pl}^2$, which lie outside the range of validity of the
semi-classical approximation under consideration. Henceforth we will
restrict ourselves therefore to the tracefree or non-conformal solutions
of the linear response Eqs. (\ref{linresgen}) around FRW spacetimes. Earlier
authors have studied the conformal variations of the metric and stress tensor
for conformally invariant fields in de Sitter space in more detail, and found no
interesting gauge invariant modes other than
those expected on the Planck scale \cite{IsaRog}. The condition (\ref{delR0})
eliminates these trace or conformal Planck scale solutions from the start, and
simplifies the analysis of the remaining tracefree, non-conformal perturbations.
This condition also eliminates the class of quantum inflationary solutions
considered in Refs. \cite{StarobN,AzumWad2}.

With $\delta R = 0$, we are restricted to the sector of tracefree metric
perturbations (in the four dimensional sense).
Further, in expanding about a FRW background one can decompose the metric
perturbations into scalar, vector and tensor perturbations with respect to
the conformally flat three dimensional metric $g_{ij} = a^2 \eta_{ij}$,
and focus on the scalar sector (in the three dimensional sense). The
vector and tensor perturbations for conformal matter around a conformally flat
spacetime are not expected to show any non-trivial degrees of freedom
in linear response other than those in the standard classical theory.

The metric perturbations which are scalar with respect to the background three-metric
can be parameterized in terms of four functions, $({\cal A, B, C, E})$, of the form
\cite{Bard,Stew},
\bes\bea
h_{tt}&=&-2{\cal A} \\
h_{tj}&=&a \partial_j {\cal B} \rightarrow ia k_j {\cal B} \\
h_{ij}&=& 2a^2 \left[\eta_{ij}\, {\cal C} + \left( \frac{\eta_{ij}}{3}\,k^2
- k_ik_j\right) {\cal E}\right]\, \, .
\eea
\label{linmet}
\ees
\vspace{-.6cm}

\noindent
in momentum space. As is well known, and elaborated further in Sec. VI, only two
linear combinations of these four functions are gauge invariant. Since we have also
required (\ref{delR0}), this means that there remains only one dynamical gauge invariant
metric function to be determined by linear response in this scalar sector. The information
about this remaining metric degree of freedom is contained completely in the $tt$
component of the full linear response equations.

To show this, let the background semi-classical Einstein Eqs. (\ref{scE}) be written in the form,
\be
{\cal T}^a_{\ b} \equiv  \lag T^a_{\ b}\rag_{_R}
-\frac{1}{8\pi G_{_N}}\left(R^a_{\ b} - \frac{R}{2}\, \delta^a_{\ b} + \Lambda\, \delta^a_{\ b}
\right) = 0\,.
\label{backg}
\ee
Their covariant conservation,
\be
\nabla_a {\cal T}^a_{\ b} = \partial_a  {\cal T}^a_{\ b} + \Gamma^a_{\ ac} {\cal T}^c_{\ b}
- \Gamma^c_{\ ab} {\cal T}^a_{\ c} = 0\,,
\label{cons}
\ee
and first variation imply
\bes\bea
&&\partial_t(\delta {\cal T}^t_{\ t}) + \partial_i (\delta {\cal T}^i_{\ t})
+ \Gamma^j_{\ jt} (\delta {\cal T}^t_{\ t})- \Gamma^i_{\ jt} (\delta {\cal T}^j_{\ i}) = 0 \,,
\label{delTa}\\
&&\partial_t (\delta {\cal T}^t_{\ i}) + \partial_j (\delta {\cal T}^j_{\ i}) +
\Gamma^j_{\ jt} (\delta {\cal T}^t_{\ i}) -\Gamma^j_{\ it} (\delta {\cal T}^t_{\ j})
- \Gamma^t_{\ ij} (\delta {\cal T}^j_{\ t})= 0\,,
\label{delTb}
\eea
\label{delT}
\ees
\vspace{-.6cm}

\noindent for the time and space components respectively.  Here
the background equations (\ref{backg}) have been used along with the spatial homogeneity
and isotropy of the background.  This leaves the Christoffel symbols shown in (\ref{delT})
as the only remaining non-zero ones. In coordinates (\ref{FRW}) the
non-vanishing Christoffel symbols are
\bes\bea
&&\Gamma^i_{\ jt} = \Gamma^i_{\ tj} = \frac{\dot a}{a}\ \delta^i_{\ j}\,,\\
&&\Gamma^t_{\ ij} = \frac{\dot a}{a}\ \eta_{ij}\,,
\eea
\label{Chris}
\ees
\vspace{-.6cm}

\noindent where the overdot denotes differentiation with respect to the comoving
proper time variable $t$. Note that because of the self-consistent
background equations (\ref{backg}), one may equally well consider the variations,
$\delta {\cal T}^a_{\ b}$, $\delta {\cal T}_{ab}$ or $\delta {\cal T}^{ab}$,
freely raising and lowering indices with the background metric.

Using relations (\ref{Chris}) in (\ref{delTa}), and imposing
the time-time component of the linear response equations,
\be
\delta {\cal T}^t_{\ t} = 0
\label{delTtt}
\ee
and the total trace of the linear response equations,
\be
\delta {\cal T}^a_{\ a} = \delta {\cal T}^t_{\ t} + \delta {\cal T}^i_{\ i} = 0\,,
\label{traceT}
\ee
equivalent to (\ref{scalR}), gives
\be
\partial_i (\delta {\cal T}^i_{\ t}) = 0\,.
\label{delTitd}
\ee
For scalar perturbations $\delta {\cal T}^i_{\ t} = g^{ij}\partial_j {\cal P}$
for some scalar function $\cal P$, there being no way for a spatial vector to
appear in the scalar sector other than by differentiation with respect to $x^i$.
Then (\ref{delTitd}) implies that $\stackrel{\rightarrow}{\nabla}\!\!^2\, {\cal P} = 0$
or ${\cal P}= 0$ up to a possible constant. Since a constant in $\cal P$ is
eliminated in $\delta {\cal T}^i_{\ t} = g^{ij}\partial_j {\cal P}$,
(\ref{delTitd}) implies
\be
\delta {\cal T}^i_{\ t} = 0\,,
\label{delTit}
\ee
or in other words, the time-space components of the linear response equations
around a general FRW space are automatically satisfied for scalar metric
perturbations of the form (\ref{linmet}), provided the self-consistent
background equation (\ref{backg}), the time-time component (\ref{delTtt}),
and the trace equation (\ref{traceT}), and are all satisfied.

To complete the proof one then notes from (\ref{delTtt}) and (\ref{traceT}),
that the spatial trace $\delta {\cal T}^i_{\ i}$ vanishes, leaving only a
possible traceless part, in $\delta {\cal T}^i_{\ j}$ which for scalar perturbations
can be expressed in the form,
\be
\delta {\cal T}^i_{\ j} = \left( \frac{1}{3}\delta^i_{\,j} - g^{il} \partial_l
\frac{1}{\stackrel{\rightarrow}{\nabla}\!\!^2}\partial_j\right) {\cal Q}
\ee
for some scalar function $\cal Q$. Substituting this form into
(\ref{delTb}), and using (\ref{delTtt}), and (\ref{delTit}) gives
$\partial_i {\cal Q} = 0$, which by an argument identical to that for the
time-space component implies
\be
\delta {\cal T}^i_{\ j} = 0\,.
\label{delTij}
\ee
Thus, (\ref{delTit}) and (\ref{delTij}) show that the sufficient conditions for
{\it all} of the semi-classical linear response equations to be satisfied around
a general FRW background for scalar perturbations of the metric (\ref{linmet}) are:
\begin{itemize}
\item (i) the background is self-consistent, so that Eq. (\ref{backg}) is
satisfied;
\item (ii) the time-time component (\ref{delTtt}) is imposed;
\item (iii) the trace equation (\ref{traceT}) is imposed.
\end{itemize}
Since the trace equation can be imposed by (\ref{scalR}), for non-Planck scale
fluctuations, the $tt$ component contains the only remaining gauge
invariant dynamical information about scalar perturbations parameterized
by (\ref{linmet}) around a self-consistent solution of the semi-classical
Einstein equations (\ref{backg}). Being able to focus solely on the
time-time component will simplify our analysis considerably.

Since the variation of $\lag T_{ab}\rag$ of conformal matter in a conformally
flat spacetime is determined by the form of the trace anomaly (\ref{tranom}),
another route to the linear response equations in this case is afforded by the
effective action and stress tensor corresponding to the anomaly, which
has been derived in several previous works.  In the next section this alternative
derivation from the effective action and stress tensor of the anomaly is given and
compared with the above derivation based on Refs. \cite{HorWal,Starob}.

\section{Anomaly Action and Stress Tensor}

For massless matter or radiation fields, there is no intrinsic length scale
associated with their quantum fluctuations, and the conformal variation of the
one-loop effective action in a general curved background is determined completely
by the trace anomaly. Although the anomaly determines uniquely only the
part of the effective action which responds to conformal variations, and
leaves the remaining conformally invariant part undetermined, it has been
argued elsewhere that the anomaly action contains the only effective field theory
corrections to Einstein's theory which can be relevant in the infrared, {\it i.e.}
at macroscopic distance scales much greater than the Planck scale \cite{AntMot,MazMot}.
The availability of the alternative formulation of linear response for
conformal matter around conformally flat backgrounds reviewed in
the previous section allows this argument to be tested in FRW spacetimes
by comparing directly the energy-momentum tensor and linear response equations
derived from the anomaly action with the previous exact formulation.

The general form of the effective action which records the effects of the
trace anomaly is non-local in the metric. However in several earlier works
a local form was obtained by the introduction of two scalar auxiliary
fields, $\varphi$ and $\psi$ \cite{MotVau,RN}. This auxiliary field effective
action is of the form,
\be
S_{anom} = b'\, S^{(E)}_{anom}[g; \varphi] + b\, S^{(F)}_{anom}[g; \varphi, \psi]\,,
\label{Sanom}
\ee
with
\bes
\bea
&& \hspace{-1cm}S^{(E)}_{anom}[g; \varphi] \equiv \frac{1}{2} \int\,d^4x\,\sqrt{-g}\
\left\{-\left(\sq \varphi\right)^2 + 2\left(R^{ab} - \frac{R}{3}g^{ab}\right)
(\nabla_a \varphi) (\nabla_b \varphi) + \left(E - \frac{2}{3} \sq R\right) \varphi\right\}\\
&&\hspace{-1cm} S^{(F)}_{anom}[g; \varphi, \psi] \equiv \,\int\,d^4x\,\sqrt{-g}\
\left\{ -\left(\sq \varphi\right)
\left(\sq \psi\right) + 2\left(R^{ab} - \frac{R}{3}g^{ab}\right)(\nabla_a \varphi)
(\nabla_b \psi)\right.\nonumber\\
&& \hskip 6cm + \left.\frac{1}{2} F \varphi +
\frac{1}{2} \left(E - \frac{2}{3} \sq R\right) \psi \right\}\,.
\eea\label{SEF}
\ees
\vspace{-.6cm}

\noindent
By varying this coordinate invariant effective action functional with
respect to the spacetime metric, we obtain the stress-energy tensor,
\be
T^{\rm anom}_{ab} = b' E_{ab} + b F_{ab}
\label{Tanom}
\ee
where the two separately conserved tensors are give by Eqs.
(3.41) and (3.42) of \cite{MotVau}, namely,
\bea
E_{ab} &=&-2\, (\nabla_{(a}\varphi) (\nabla_{b)} \sq \varphi)
+ 2\,\nabla^c \left[(\nabla_c \varphi)(\nabla_a\nabla_b\varphi)\right]
- \frac{2}{3}\, \nabla_a\nabla_b\left(\nabla\varphi\right)^2\nn
&&
+ \frac{2}{3}\,R_{ab}\, (\nabla\varphi)^2
- 4\, R^c\,_{\! (a}(\nabla_{b)} \varphi) (\nabla_c \varphi)
+ \frac{2}{3}\,R \,(\nabla_a \varphi) (\nabla_b \varphi)\nonumber\\
&& + \frac{1}{6}\, g_{ab}\, \left\{-3\, (\sq\varphi)^2
+ \sq \left(\nabla\varphi)^2\right)
+ 2\left( 3R^{cd} - R g^{cd} \right) (\nabla_c \varphi)(\nabla_d
\varphi)\right\}\nonumber\\
&&
\hspace{-1cm} - \frac{2}{3}\, \nabla_a\nabla_b \sq \varphi
- 4\, C_{a\ b}^{\ c\ d}\, \nabla_c \nabla_d \varphi
- 4\, R^c\,_{\!(a} \nabla_{b)} \nabla_c\varphi
+ \frac{8}{3}\, R_{ab}\, \sq \varphi
+ \frac{4}{3}\, R\, \nabla_a\nabla_b\varphi \nonumber\\
&&
- \frac{2}{3} \left(\nabla_{(a}R\right) \nabla_{b)}\varphi
+ \frac{1}{3}\, g_{ab}\, \left\{ 2\, \sq^2 \varphi
+ 6\,R^{cd} \,\nabla_c\nabla_d\varphi
- 4\, R\, \sq \varphi
+ (\nabla^c R)\nabla_c\varphi\right\}\,,
\label{Eab}
\eea
and
\bea
&& F_{ab} = -2\, (\nabla_{(a}\varphi) (\nabla_{b)} \sq \psi)
-2\, (\nabla_{(a}\psi) (\nabla_{b)} \sq \varphi)
+ 2\,\nabla^c \left[(\nabla_c \varphi)(\nabla_a\nabla_b\psi)
+ (\nabla_c \psi)(\nabla_a\nabla_b\varphi)\right]
\nonumber\\
&&
\hspace{.5cm} - \frac{4}{3}\, \nabla_a\nabla_b\left[(\nabla_c \varphi)
(\nabla^c\psi)\right]
+ \frac{4}{3}\,R_{ab}\, (\nabla_c \varphi)(\nabla^c \psi)
- 4\, R^c\,_{\! (a}\left[(\nabla_{b)} \varphi) (\nabla_c \psi)
+ (\nabla_{b)} \psi) (\nabla_c \varphi)\right]\nonumber\\
&&
\hspace{1.5cm} + \frac{4}{3}\,R \,(\nabla_{(a} \varphi) (\nabla_{b)} \psi)
+ \frac{1}{3}\, g_{ab}\, \Big\{-3\, (\sq\varphi)(\sq\psi)
+ \sq \left[(\nabla_c\varphi)(\nabla^c\psi)\right]
\nonumber\\
&&
\hspace{1.5cm} \left. + 2\, \left( 3R^{cd} - R g^{cd} \right) (\nabla_c
\varphi)(\nabla_d \psi)\right\}- 4\, \nabla_c\nabla_d\left( C_{(a\ b)}^{\ \ c\ \ d}
\varphi \right)  - 2\, C_{a\ b}^{\ c\ d} R_{cd} \varphi \nonumber\\
&&
\hspace{.5cm} - \frac{2}{3}\, \nabla_a\nabla_b \sq \psi
- 4\, C_{a\ b}^{\ c\ d}\, \nabla_c \nabla_d \psi
- 4\, R^c\,_{\!(a} \nabla_{b)} \nabla_c\psi
+ \frac{8}{3}\, R_{ab}\, \sq \psi
+ \frac{4}{3}\, R\, \nabla_a\nabla_b\psi \nonumber\\
&&
\hspace{1cm} - \frac{2}{3} \left(\nabla_{(a}R\right) \nabla_{b)}\psi +
\frac{1}{3}\, g_{ab}\,  \left\{ 2\, \sq^2 \psi + 6\,R^{cd} \,\nabla_c\nabla_d\psi
- 4\, R\, \sq \psi + (\nabla^c R)(\nabla_c\psi)\right\}\,,
\label{Fab}
\eea
where $(\nabla\varphi)^2 = (\nabla_a \varphi)(\nabla^a \varphi)$.
By varying \eqref{Sanom}  with respect to the local auxiliary fields one obtains their
equations of motion,
\bes
\bea
&&\Delta_4 \varphi = \frac{E}{2} - \frac{\sq R}{3}\,,
\label{phieom}\\
&&\Delta_4 \psi = \frac{F}{2} = \frac{1}{2} C_{abcd}C^{abcd}\,,
\label{psieom}
\eea
\label{auxeom}
\ees
\vspace{-.6cm}

\noindent
which are linear in $\varphi$ and $\psi$. Here
\be
\Delta_4 \equiv \sq^2 + 2 R^{ab}\nabla_a\nabla_b - \frac{2}{3}R \sq +
\frac{1}{3} (\nabla^a R)\nabla_a \,.
\label{Delfdef}
\ee
Since the terms quadratic in the auxiliary fields in (\ref{SEF}) are conformally
invariant, they lead to only traceless terms in the corresponding stress tensors
(\ref{Eab}) and (\ref{Fab}). Therefore the terms in the trace are linear in $\varphi$
and $\psi$, and in fact are proportional to (\ref{auxeom}), yielding
the local geometrical traces,
\bes
\bea
E^a_{\ a} &=& 2 \Delta_4 \varphi = E - \frac{2}{3} \sq R\,, \label{Etr}\\
F^a_{\ a} &=& 2 \Delta_4 \psi = F = C_{abcd}C^{abcd}\,,
\label{Ftr}
\eea
\label{EFtraces}
\ees
\vspace{-.6cm}

\noindent Thus the {\it classical} trace of the stress tensor following
from the local effective action (\ref{SEF}) reproduces the quantum
trace anomaly~\eqref{tranom}.

Moreover, it was shown in Refs. \cite{AMM,MazMot,AMMDE} by considering its conformal
variation under $g_{ab} \rightarrow \bar g_{ab} = e^{-2\sigma} g_{ab}$ that
\be
T_{ab}^{WZ}[\bar g;\sigma] \equiv
-\frac{2}{\sqrt{-g}} \frac{\delta}{\delta g^{ab}}\,
\Gamma^{WZ} [g; \sigma]= 2\, ^{(3)\!}\bar H_{ab}
+ \frac{1}{9}\, ^{(1)}\bar H_{ab}\,,
\label{H13}
\ee
where $T_{ab}^{WZ}[g;\sigma]$ is the stress tensor of the Wess-Zumino effective
action,
\be
\Gamma^{WZ}[g; \sigma] = \int\,d^4x\sqrt{-g} \, \left[2 \sigma\Delta_4\sigma
+ \left(E - \frac{2}{3} \sq R \right)\sigma \right] =
- S^{(E)}_{anom} [g; \varphi =-2\sigma]
\ee
obtained by varying the metric $g_{ab}$, keeping $\sigma$ fixed. Considering
equations (\ref{Sanom})-(\ref{Eab}), dropping the overbar on (\ref{H13}) and
using the notation of (\ref{Hdef}) one finds that
\be
E_{ab}\big\vert_{\varphi = 2 \ln \Omega}  = -T_{ab}^{WZ}\left[g;\sigma =
-\frac{\varphi}{2}\right]
= -2 \,^{(3)\hspace{-.1cm}}H_{ab} + \frac{1}{9} B_{ab}\,,
\label{Especial}
\ee
in the full metric $g_{ab}$, provided
\be
g_{ab} = e^{\varphi} \eta_{ab} = \Omega^2 \eta_{ab}
\label{ggbar}
\ee
is conformally flat, with the auxiliary field $\varphi = 2 \ln \Omega$. It is
easily checked from the conformal variation,
\be
E - \frac{2}{3} \sq R = 4 \Delta_4 \ln \Omega = 2 \Delta_4 \varphi
\ee
in the full metric, that $\varphi = 2 \ln \Omega$ satisfies (\ref{Etr}). These
relations may be proven directly as well, by the use of formulae for the conformal
variations of the Ricci tensor, such as
\bes\bea
&&R_{ab} = -\nabla_a\nabla_b \varphi - \frac{1}{2} (\nabla_a \varphi)(\nabla_b\varphi)
+ \frac{g_{ab}}{2}
\left[ - \sq \varphi  + (\nabla\varphi)^2\right]\,.\\
&&R = -3 \sq\varphi + \frac{3}{2}\, (\nabla \varphi)^2
\eea\ees
in the full metric $g_{ab} = e^{\varphi} \eta_{ab}$.

Thus, provided that the auxiliary field is fixed in terms of the geometric conformal
factor via $\varphi = 2 \ln \Omega$, the tensor $E_{ab}$ reduces to the purely local
combination of geometric tensors given by (\ref{Especial}). Taking the second auxiliary
field $\psi = 0$, and the Weyl tensor $C_{a\ b}^{\ c\ d} = 0$, gives
\be
T_{ab}^{anom} = b' E_{ab} = -2b'\, ^{(3)\hspace{-.1cm}}H_{ab} + \frac{b'}{9}\, B_{ab}
\label{Tanomconf}
\ee
for the full contribution of the anomaly effective action (\ref{Sanom}) to the expectation
value of the energy-momentum tensor of conformal matter in an arbitrary conformally flat
spacetime. We note that for a conformally flat spacetime of the form (\ref{FRW}) and
$\varphi(t) = 2 \ln a(t) + c_0$, the tensor $E^a_{\ b}$ derived from the auxiliary field
effective action gives rise to precisely the geometric tensor $^{(3)}H^a_{\ b}$ discussed
in Ref. \cite{MazMot}, which is not derivable from a local geometric action and for
that reason was called ``accidentally conserved" in \cite{BirDav}. The actual origin
of this tensor is the form of the anomaly action (\ref{SEF}).

Eq. (\ref{Tanomconf}) differs from the known exact result (\ref{Tconf}) only by the
addition of a $b'' \sq R$ term to the anomaly and corresponding local $R^2$ term to the
effective action with $b'' = 2(b+b')/3$, the same choice of $b''$ expected from the flat
space result: {\it cf.} Eqs. (\ref{tranom})-(\ref{vartracflat}). Since such local terms
with arbitrary coefficients should be added to the effective action in any case, we
conclude that the action (\ref{Sanom}) plus these additional local terms gives the
correct general form for the stress tensor of conformal matter in its conformal
``vacuum" state in an arbitrary conformally flat spacetime.

To study linear response next we wish to vary the anomaly stress tensor (\ref{Tanom})
away from the conformally flat background. Since the relations (\ref{Especial}) and
(\ref{Tanomconf}) hold for arbitrary conformal transformations, it follows that they
hold also for linearized conformal transformations about a conformally flat FRW spacetime.
Hence it is clear that in the variation of (\ref{Tanomconf}) we must expect to obtain
local geometrical terms which are just the variations of the local geometrical terms
in (\ref{Tanomconf}).  By adding the local $F=C_{abcd}C^{abcd}$ and $R^2$ terms to the
effective action, it is clear that one should expect the variation of the local
geometrical tensors $A_{ab}$ and $B_{ab}$ which arise from them. Thus the origin of
all the local terms in (\ref{conflat}) and (\ref{Lab}) is clear, except perhaps for the last term
in (\ref{Lab}), whose origin we now discuss.

Inspection of the $\varphi$ dependent terms in (\ref{Fab}) shows that
\be
\delta F_{ab}\Big\vert_{\varphi = 2 \ln \Omega,\,\psi=0} =
- 8\, \nabla_c\nabla_d \left[(\ln \Omega)\, \delta C_{(a\ b)}^{\ \ c\ d}\right]
-4 (\ln \Omega)\, \delta C_{(a\ b)}^{\ \ c\ d}\,R_{cd}
\label{varF}
\ee
around any conformally flat spacetime with $C_{a\ b}^{\ c\ d} = 0$. Since the
Lagrangian density $\sqrt{-g} F \varphi$ from which this term is derived is
conformally invariant under $g_{ab} = \Omega^2 \eta_{ab} \rightarrow \eta_{ab}$,
$\delta C_{(a\ b)}^{\ \ c\ d}$ and the covariant derivatives in (\ref{varF})
may be replaced by their values, $\delta \bar C_{(a\ b)}^{\ \ c\ d}$ and
$\partial_c\partial_d$ on the perturbation (\ref{linmet}) in flat spacetime
by a simple overall scaling with $\Omega^{-2}$. That is, one may check
explicitly that in general
\be
2\, \nabla_c\nabla_d \left[\varphi \,C_{(a\ b)}^{\ \ c\ d}\right]
+ \varphi\, C_{(a\ b)}^{\ \ c\ d}\,R_{cd} = \frac{1\ }{\Omega^2}
\left\{2\,\overline\nabla_c\overline\nabla_d \left[\varphi \,
\overline C_{(a\ b)}^{\ \ c\ d}\right]
+ \varphi\, \overline C_{(a\ b)}^{\ \ c\ d}\,\overline R_{cd}\right\}
\label{conftens}
\ee
for any two metrics conformally related by $g_{ab} = \Omega^2 \bar g_{ab}$. Varying
this relation keeping $\varphi$ and $\Omega$ fixed, with $\delta g_{ab} =
\Omega^2 \delta \bar g_{ab}$, and taking $\bar g_{ab} = \eta_{ab}$ to be flat gives
upon making use of (\ref{varF}),
\be
\delta F_{ab}\Big\vert_{\varphi = 2 \ln \Omega,\,\psi=0} =
-\frac{8\ }{\Omega^2}\ \partial_c\partial_d \left[(\ln \Omega)
\,\overline C_{(a\ b)}^{\ \ c\ d}\right]\,.
\label{confvar}
\ee
Hence the last local term in (\ref{Lab}) of the variation of the stress tensor
expectation value of conformal matter in a conformally flat spacetime may be
regarded as a direct result of the anomaly effective action (\ref{Sanom})-(\ref{SEF})
in the auxiliary field formalism. Since the $A_{ab}$ tensor is just proportional
to (\ref{conftens}) with $\varphi$ replaced by a constant, we also deduce that
\be
A^a_{\ b} = \frac{1\,}{\Omega^4}\, \bar A^a_{\ b}\,,
\label{raiseA}
\ee
upon raising one index, for the two metrics related by (\ref{ggbar}).

Thus the anomaly action and stress tensor reproduces all the local terms
denoted by $L_{ab}$ in (\ref{Lab}) in the exact result (\ref{conflat}).
However it does not reproduce the non-local term
in the variation (\ref{conflat}), involving the logarithmic distribution $K$.
This term comes from the part of the one-loop effective action which is
invariant under conformal transformations $g_{ab} \rightarrow \Omega^2 g_{ab}$,
and does not contribute to the anomaly or the anomaly action. It may be argued
from the addition of the local $\alpha_{_R}(\mu^2) C_{abcd}C^{abcd}$ term,
renormalized at an arbitrary mass scale $\mu$ that a non-local logarithmic
term involving $K$ with the correct coefficient must be present in the effective
action in order for the scale $\mu^2$ to drop out of the low energy dynamics
according to (\ref{muindep}). Whether the non-local term is essential or
not in perturbations around de Sitter space will be studied in detail in Sec. VII.

Through the anomaly effective action and its auxiliary field
description, it is possible to express the variation of the stress tensor
in a completely local form, and to gain the advantage of examining a
wider class of state variations parameterized by the auxiliary fields
as well. By specializing to de Sitter space in the next section it is shown
that the independent variation of the metric and auxiliary fields of
the anomaly stress tensor (\ref{Tanom}) exhibits specific additional state
dependent horizon scale degrees of freedom which are not present in the
classical Einstein theory.

\section{Linear Response of Conformal Matter in de Sitter Spacetime}

In this section we specialize the discussion to de Sitter spacetime,
and derive the linear response equations in the anomaly
action auxiliary field formulation, comparing it in detail to
the general expression for the linearized stress tensor
variation in (\ref{conflat}).

In FRW coordinates with flat spatial sections (\ref{FRW}) the scale factor
of de Sitter spacetime is
\be
a(t) = e^{Ht} = -\frac{1}{H\eta} = \Omega(\eta)\,,\qquad \eta \in (-\infty, 0)\,.
\label{RWdS}
\ee
Although the coordinates (\ref{FRW}) cover only half of the fully
analytically extended de Sitter spacetime, they exhibit the spatial
homogeneity and isotropy and standard FRW form explicitly, and are
convenient for admitting the standard Fourier mode decomposition in the
flat spatial coordinate $\vec x$.  This allows the linear response
equations to be expressed as ordinary differential equations in the
time variable $t$ (or $\eta$), for each spatial Fourier mode
separately.

In the maximally symmetric de Sitter spacetime,
\bes\bea
&&R^{a\  b}_{\ c\ d} = H^2 \left(g^{ab}\,g_{cd} -
\delta^a_{\ d}\,\delta^b_{\ c}\right)\,, \\
&&R^a_{\ b} = 3H^2\, \delta^a_{\ b}\,,\qquad H = \frac{\dot a}{a}\,,\\
&&R=12H^2\,, \qquad \nabla_a R = 0 = \sq R\,, \\
&&E_{ab} = -2 \ ^{(3)\hspace{-.1cm}}H_{ab}
= 6 H^4\,g_{ab}\,, \qquad E = 24H^4\,, \\
&&C^{a\  b}_{\ c\ d} = 0\,,\qquad A_{ab}= B_{ab} = 0\,,\qquad F=0\,.
\eea
\label{Rieds}
\ees
\vspace{-.6cm}

\noindent
Because of the $O(4,1)$ maximal symmetry, with $10$ Killing generators, the
conformal ``vacuum" of a conformal field in  de Sitter spacetime also is
maximally symmetric. The linear response approach requires a self-consistent
solution of equations (\ref{scE}), around which we perturb the metric and
stress tensor together. Hence the choice of this maximally symmetric
state, known as the Bunch-Davies (BD) state is the natural one about which
to consider perturbations. From the point of view of a free falling observer
this conformal ``vacuum" is in fact a thermally populated state with the
temperature $T_{_H} = H/2\pi$ \cite{BirDav}. In this state $O(4,1)$ de Sitter symmetry
implies that $\lag T^a_{\ b}\rag$ is also proportional to $\delta^a_{\ b}$.
The self-consistent value of the scalar curvature $R$ including the
quantum contribution from $\langle T^a_{\ b}\rangle$ is given by the trace
equation,
\bes\bea
-R + 4 \Lambda &=& 8 \pi G_{_N} b' E = \frac{4 \pi G_{_N}b'}{3} R^2\,,
\qquad {\rm or}\\
H^2 &=& \frac{\Lambda}{3} \left[ 1 - \frac{16\pi G_{_N} b'\Lambda}{3}
+ \dots \right]\,,
\eea
\label{selfcons}
\ees
\vspace{-.6cm}

\noindent
in an expansion around the classical de Sitter solution with
$8\pi G_{_N}\Lambda |b'|\ll 1$ \cite{WadAzu}. Thus, in this limit the stress
tensor source of the semi-classical Einstein's equations (\ref{scE}) in the BD
state is a small finite correction to the classical cosmological term in the
self-consistent de Sitter solution. The fractional size of this correction is
$\hbar G_{_N}\Lambda/c^3 \ll 1$, if the de Sitter horizon scale $H^{-1}$ is
much greater than the Planck scale $L_{Pl} \equiv \sqrt{ \frac{\hbar G}{c^3}}
= 1.616 \times 10^{-33}$ cm.

Given the smallness of the quantum correction in a self-consistent solution
of (\ref{scE}), at first sight it would appear unlikely for the quantum fields
to have any significant effects at the macroscopic scales of cosmology. However,
these effects will become apparent only when both the geometry and quantum state
of the matter are not fixed but allowed to vary dynamically. Since a quantum
state specified over a macroscopic region of space is highly non-local,
variations of the state are independent of the magnitude of local curvature
variations. Although there is a very small coupling $L_{Pl}^2 H^2 \ll 1$
controlling the loop expansion around the classical de Sitter background, the
expansion in this coupling may be non-uniform or even infrared divergent due
to state dependent variations on the horizon scale $H^{-1}$.

In the auxiliary field effective action approach, information about the
quantum state of the underlying conformal matter fields is contained in
the particular solution of the auxiliary field equations (\ref{auxeom}).
In de Sitter spacetime the conformal differential operator $\Delta_4$
of (\ref{Delfdef}) factorizes,
\be
\Delta_4\vert_{dS} = -\sq (-\sq + 2 H^2)\,.
\ee
and the equation satisfied by $\varphi$ is inhomogeneous whereas
the equation for $\psi$ is homogeneous.  Thus it is consistent to set $\psi = 0$
in order to obtain a de Sitter invariant state. Taking $\psi \neq 0$ will not
lead to a de Sitter invariant stress tensor. In any spatially homogeneous
and isotropic state $\varphi$ can be a function only of time. With $\varphi
= \bar\varphi (t)$ one easily finds that the general homogeneous
solution of either of Eqs. (\ref{auxeom}) is the linear combination,
$c_0 + c_1 a^{-1} + c_2 a^{-2} + c_3 a^{-3}$. All of these behaviors
except the first lead to de Sitter non-invariant stress tensors, with
the constant $c_0$ dropping out of the stress tensor entirely. Hence
to obtain a de Sitter invariant state all the coefficients of
these homogeneous solutions of Eqs. (\ref{auxeom}) must be set to zero.
The remaining inhomogeneous solution to (\ref{phieom}) is easily
found to be $2Ht$. If the solutions
\bes\bea
&& \bar\varphi = 2 \ln \Omega = 2Ht\\
&& \bar\psi = 0\,,
\eea
\label{auxBD}
\ees
\vspace{-.6cm}

\noindent
are substituted into  the stress tensors (\ref{Eab}) and (\ref{Fab}) the result is
\be
T^a_{\ b} [\bar\varphi, \bar\psi=0]_{dS} = b' E^a_{\ b}[g^{dS}; \bar\varphi] =
6\, b'\, H^4\, \delta^a_{\ b}\,
\label{TabBD}
\ee
which is exactly the value of the renormalized stress tensor of a conformally
invariant field of any spin in the BD state, found either by direct
computation or the methods of \cite{BroCas}.  Thus the solutions \eqref{auxBD}
correspond to the Bunch-Davies state.

With the self-consistent BD de Sitter solution (\ref{FRW}), (\ref{RWdS}),
(\ref{selfcons}), (\ref{auxBD}), and (\ref{TabBD})  one may consider the
linear response variation,
\bea
\delta R^a_{\ b} -\frac{\delta R}{2} \delta^a_{\ b} &=& 8\pi G_{_N}\,
\delta\, \left\{(T^a_{\ b})^{loc} + (T^a_{\ b})^{anom}\right\}  \nn
&=&  8\pi G_{_N} \left\{ - \alpha_{_R}\,\delta A^a_{\ b} - \beta_{_R}\,
\delta B^a_{\ b}
+ b'\, \delta E^a_{\ b} + b\, \delta F^a_{\ b}\right\}\,,
\label{linaux}
\eea
where all terms to linear order in $\delta g_{ab} = h_{ab}$ and in the variations
of the auxiliary fields,
\bes
\bea
&& \delta \varphi \equiv \varphi - \bar\varphi \equiv \phi\\
&& \delta\psi \equiv \psi - \bar\psi = \psi \,,
\eea
\label{auxvar}
\ees
\vspace{-.6cm}

\noindent
are to be retained. All indices are raised and lowered by the background de Sitter
metric $g_{ab}$ at linear order in the perturbations.

For the first local tensor $A^a_{\ b}$ it is simplest to vary the alternative
form,
\be
A^a_{\ b} = - 4 C^a_{\ cbd}R^{cd} - 2 \sq R^a_{\ b}
+ \frac{2}{3} \nabla^a\nabla_b R
+ \frac{1}{3}\delta^a_{\ b} \sq R + 4 R^a_{\ c}R^c_{\ b}
- \frac{4}{3}R^a_{\ b} R
- \delta^a_{\ b} R^c_{\ d}R^d_{\ c} + \frac{1}{3} \delta^a_{\ b} R^2\,,
\ee
which follows from the Bianchi identities. The variations are simplified by
using the symmetry properties of de Sitter space catalogued in (\ref{Rieds}),
and by imposing the $\delta R = 0$ condition discussed in the previous section.
It is easily demonstrated that this implies that $\delta (\nabla^a\nabla_b R) = 0$
and $\delta (\sq R) = 0$, as well, while $\delta (\sq R^a_{\ b}) =
\sq \delta R^a_{\ b}$. Hence,
\be
\delta A^a_{\ b}\Big\vert_{dS, \, \delta R = 0} =
2 \left( - \sq + \frac{R}{3}\right) \delta R^a_{\ b}\,.
\label{varA}
\ee
For the second local tensor $B^a_{\ b}$ one finds,
\be
\delta B^a_{\ b}\Big\vert_{dS, \, \delta R = 0} = - 2R\, \delta R^a_{\ b}\,.
\label{varB}
\ee

The variation of the stress tensors obtained from the anomaly action,
$E^a_{\ b}$ and $F^a_{\ b}$ depend on both the variations of the metric
and the variations of the auxiliary fields. It is shown in Appendix B that
\be
\delta F^t_{\ t}\Big\vert_{dS, \, \delta R = 0} = 2Ht\ \delta A^t_{\ t}
- \frac{2}{3}\frac{\stackrel{\rightarrow}{\nabla}\!\!^2\,}{a^2}\,
\left[\partial_t^2 + H\partial_t
- \frac{\stackrel{\rightarrow}{\nabla}\!\!^2\,}{a^2}\right]\psi
\label{varFa}
\ee
From (\ref{Tanomconf}) one expects the variation of $E^a_{\ b}$ to
contain the local geometric term,
\be
-2 \delta\, ^{(3)\!}H^a_{\ b} + \frac{1}{9}\, \delta B^a_{\ b}
= -\frac{R}{3}\, \delta R^a_{\ b} - \frac{2R}{9}\, \delta R^a_{\ b}
= - \frac{20H^2}{3}\,\delta R^a_{\ b} \,.
\ee
This has been verified by direct calculation using the algebraic manipulation
programs Mathematica and MathTensor. The variation of the auxiliary
field equations, (\ref{auxeom}) gives
\bes\bea
&& \delta (\sq^2 \varphi) - \frac{R}{6} \delta (\sq \varphi) =
\left(\sq - \frac{R}{6}\right) \delta (\sq \varphi)=
- 2 (\nabla_a\nabla^b \bar\varphi) \delta R^a_{\ b}
\label{varyphi}\\
&& \delta (\sq^2 \psi) - \frac{R}{6} \delta (\sq \psi) =
\left(\sq - \frac{R}{6}\right) \sq\psi = 0\,.  \label{varypsi}
\eea
\label{varyaux}
\ees
\vspace{-.6cm}

\noindent
Both this variation and that of the  additional terms in
$\delta E^t_{\ t}$ dependent upon $\delta \varphi = \phi$ can
be simplified somewhat by a convenient choice of gauge. Consider
\be
\delta (\sq \varphi) = \delta (g^{ab}\nabla_a \nabla_b \varphi) =
-h^{ab} \nabla_a\nabla_b\bar\varphi
- g^{ab} (\delta \Gamma^c_{\ ab})\nabla_c \bar\varphi + \sq \phi\,,
\label{varboxphi}
\ee
where
\bes\bea
&&\delta \Gamma^c_{\ ab} = \frac{1}{2} \left( -\nabla^ch_{ab}
+ \nabla_a h^c_{\ b} + \nabla_b h^c_{\ a}\right)\,,\\
&&g^{ab}\delta \Gamma^c_{\ ab} = \nabla_a h^{ac} - \frac{1}{2} \nabla^c h\,,
\eea\ees
\vspace{-.6cm}

\noindent
is the variation of the Christoffel symbol and its trace (with
$h \equiv g^{ab}h_{ab}$). The fact that $\bar\varphi$ is a linear function
of $t$ allows the variation to be taken inside one wave operator, {\it i.e.}
$\delta (\sq^2 \varphi) = \sq \delta(\sq \varphi)$ as in (\ref{varyphi}), but
from (\ref{varboxphi}), $\delta (\sq \varphi) \neq \sq \phi$ in general.
However the additional terms in (\ref{varboxphi}) may be set to zero
by making the {\it gauge choice},
\bes
\bea
h^{ab} \nabla_a \nabla_b \bar\varphi &=& 0\,\\
g^{bc} \delta\Gamma^a_{\ bc} \nabla_a\bar\varphi &=& \left(\nabla_b h^{ab} -
\frac{1}{2}\nabla^a h\right)\nabla_a\bar\varphi = 0\,.
\eea
\label{gaugec}
\ees
\vspace{-.6cm}

\noindent
Because of the simple form of $\bar\varphi$ from (\ref{auxBD}),
these two gauge conditions are also equivalent to
\bes
\bea
&&h^{ij}g_{ij} = h^a_{\ a} + h_{tt} = 0\,\\
&&\nabla^i h_{ti} = g^{ij} \nabla_i h_{tj} = \frac{1}{2} \partial_t h_{tt}\,.
\label{gaugeb}
\eea
\ees
\vspace{-.6cm}

\noindent
or in terms of the decomposition (\ref{linmet}),
\bes
\bea
-\stackrel{\rightarrow}{\nabla}\!\!^2\, {\cal B} &=&
a (\dot {\cal A} + 6 H {\cal A})\,,\\
{\cal C} &=& 0\,.
\eea
\label{gaugecond}
\ees
\vspace{-.6cm}

\noindent
With this choice of gauge the variation in (\ref{varboxphi}),
$\delta (\sq \varphi) = \sq \phi$, and the auxiliary field
Eqs. (\ref{varyaux}) can be written in Fourier space in the simple form,
\bes\bea
&&\left(\frac{d^2}{dt^2} + 5 H\frac{d}{dt} + 6H^2 + k^2 e^{-2Ht} \right) v
 = 0\,,
\label{veq}\\
&&\left(\frac{d^2}{dt^2} + 5 H\frac{d}{dt} + 6H^2 + k^2 e^{-2 Ht} \right)
\left(w - 2h_{tt} \right) = 0\,,
\label{ueq}
\eea
\label{auxlineqs}
\ees
\vspace{-.6cm}

\noindent
with
\bes\bea
H^2 v &\equiv & \left(\frac{d^2}{dt^2} + H \frac{d}{dt} + k^2 e^{-2Ht}\right) \psi
= \frac{1\,}{a^2}\,\left(\frac{d^2}{d\eta^2} + k^2\right) \psi
\,. \label{vdef} \\
H^2 w &\equiv & \left(\frac{d^2}{dt^2} + H \frac{d}{dt} + k^2 e^{-2Ht}\right) \phi
= \frac{1\,}{a^2}\,\left(\frac{d^2}{d\eta^2} + k^2\right) \phi\,.  \label{udef}
\eea
\label{uvdef}
\ees
\vspace{-.6cm}

\noindent
In terms of these quantities, with the condition $\delta R = 0$ and the
gauge choice (\ref{gaugec}), the full variation of $E^t_{\ t}$ is
\bea
\delta E^t_{\ t} &=&   -\frac{20H^2}{3} \,\delta R^t_{\ t}
-\frac{2}{3} \frac{\stackrel{\rightarrow}{\nabla}\!\!^2\,}{a^2}
\left[\left(\partial_t^2  + H\partial_t
- \frac{\stackrel{\rightarrow}{\nabla}\!\!^2\,}{a^2}\right) \phi
- 2 H^2 h_{tt}\right]\nn
&=&  -\frac{20H^2}{3} \,\delta R^t_{\ t} -\frac{2H^2}{3\,a^2}\,
\stackrel{\rightarrow}{\nabla}\!\!^2\, (w - 2h_{tt})\,,
\label{varEt}
\eea
while that for $F^t_{\ t}$ in (\ref{varFa}) becomes
\be
\delta F^t_{\ t}\Big\vert_{dS, \, \delta R = 0} = 4Ht\ \left(-\sq + 4H^2\right)
\delta R^t_{\ t}
- \frac{2H^2}{3a^2}\stackrel{\rightarrow}{\nabla}\!\!^2\, v \,.
\label{varFt}
\ee
The gauge choice (\ref{gaugec}) is useful for simplifying the scalar auxiliary
field contributions to (\ref{varEt}) and (\ref{varFt}), while the local
geometric terms involving $\delta R^t_{\ t}$ in both expressions are independent
of this gauge choice. In the next section it is shown that in fact the quantities
$w-2h_{tt}$ and $v$ are gauge invariant.

All the variations in (\ref{varA}), (\ref{varB}), (\ref{varEt}) and (\ref{varFt})
involve the local geometric variation $\delta R^t_{\ t}$. Thus the linear
response equation in de Sitter space derived from the auxiliary field effective
action takes the form,
\bea
\delta R^t_{\ t} &=& -\frac{\bar\alpha_{_R}}{H^2}\left(-\sq + 4 H^2\right)
\delta R^t_{\ t}
+ (\bar \beta_{_R} + 5\varepsilon')\delta R^t_{\ t} + \frac{\varepsilon t}{H}\,
\left(-\sq + 4H^2 \right) \delta R^t_{\ t} \nonumber \\
& & \ + \frac{\varepsilon' }{2a^2}\, \stackrel{\rightarrow}{\nabla}\!\!^2\,
(w - 2h_{tt}) - \frac{\varepsilon }{6a^2}\,
\stackrel{\rightarrow}{\nabla}\!\!^2\, v\,,
\label{tteq}
\eea
where the coupling constants have been written in terms of the dimensionless
small numbers,
\bes
\bea
\bar\alpha_{_R} \equiv && 16 \pi  G_{_N}H^2\alpha_{_R}\,,\\
\bar \beta_{_R} \equiv && 192 \pi  G_{_N}H^2\beta_{_R}\,\\
\varepsilon \equiv && 32 \pi G_{_N}H^2 b \,,\\
\varepsilon' \equiv && -\frac{32\pi }{3} G_{_N} H^2 b'\,.
\eea
\label{pert}
\ees
\vspace{-.6cm}

\noindent
Defining the dimensionless variable $q$ by
\be
q \equiv - \frac{2a^2}{H^2}\delta G^t_{\ t}= - \frac{2a^2}{H^2}
\left( \delta R^t_{\ t} - \frac{1}{2}\delta R\right)\,,
\label{qdef}
\ee
it is straightforward to show that under the condition $\delta R = 0$ and
with the gauge choice (\ref{gaugec}),
\bea
\left(-\sq + 4 H^2\right)\delta R^t_{\ t} &=&
-\left[\partial_t^2 + 7H\partial_t + 12H^2
- \frac{\stackrel{\rightarrow}{\nabla}\!\!^2\,}{a^2}\right]\frac{qH^2}{2a^2}\nn
&=& - \frac{H^2}{2a^2} \left[\partial_t^2 + 3H\partial_t + 2H^2
- \frac{\stackrel{\rightarrow}{\nabla}\!\!^2\,}{a^2}\right]q\,.
\label{commutea}
\eea
Thus, multiplying (\ref{tteq}) through by $-2a^2/H^2$ and Fourier transforming
the spatial variables gives
\be
(1 - \bar \beta_{_R} - 5\varepsilon') q = \varepsilon Ht\, {\cal D}q
- \bar\alpha_{_R}\,{\cal D}q
+ \varepsilon' \frac{k^2}{H^2} (w-2h_{tt}) - \frac{\varepsilon}{3} \frac{k^2}{H^2} v\,.
\label{qeq}
\ee
with
\be
{\cal D}q \equiv \frac{1}{H^2} \left(\frac{d^2}{dt^2} + 3H\frac{d}{dt} + 2 H^2
+ k^2 e^{-2Ht}\right) q\,.
\label{diffD}
\ee
Eqs. \eqref{auxlineqs} and \eqref{qeq}, with \eqref{uvdef}, \eqref{pert}, and \eqref{qdef} are
the linear response equations for de Sitter space derived from the anomaly action and stress tensor.
Since by assumption $\bar\beta_{_R}$
and $\varepsilon'$ are very small quantities they may be neglected with respect to
unity on the left side of (\ref{qeq}). Note that in the strict classical limit,
where all the small quantities in (\ref{pert}) first order in $\hbar$ are set
to zero, the classical constraint equation $q=0$ or $\delta R^t_{\ t} = 0$
is recovered from (\ref{qeq}). This is consistent with the absence of any
dynamics in the pure Einstein theory in the sector of scalar perturbations
without matter sources.

This form can be compared now to those that are obtained by the method of
Ref. \cite{HorWal,Starob} reviewed in the previous sections. From
(\ref{linresgen}), (\ref{conflat}) and (\ref{Lab}), upon raising one index
with the inverse metric $g^{ab} = \Omega^{-2} \bar g^{ab}$, one finds
\bea
\delta \left\{R^t_{\ t} - \frac{R}{2} \delta^t_{\ t} +
\Lambda \delta^t_{\ t}\right\}
&=& 8\pi G_{_N}\left\{ -\frac{b}{\Omega^4} \int_{-\eta_0}^{\eta} \,
d\eta' K(\eta-\eta'; k ; \mu^2)\ \delta
\bar A^t_{\ t} (\eta'; {\vec k}) \right. \nonumber \\
& &  \left. \;\;\;\;\;\;\;\;  + L^t_{\ t} + \frac{1\ }{\Omega^4}
\delta \lag \bar T^t_{\ t} \rag_{_R} \right\}\,,
\label{linHW}
\eea
with
\be
L^t_{\ t}= -\alpha_{_R} \,\delta A^t_{\ t} - \left(\beta_{_R} +
\frac{b}{9}\right)\, \delta B^t_{\ t}
- 2 b'\ \delta\, ^{(3)\!}H^t_{\ t} -\frac{8b}{\,\Omega^4} \,
\partial_i\partial_j
\left[(\ln \Omega)\, \delta C^{ti\ j}_{\ \ t}\right]\,.
\label{localtt}
\ee
Upon making use of (\ref{confvar}), (\ref{varA}), (\ref{commutea}),
(\ref{qdef}), (\ref{diffD}), (\ref{pert}) and (\ref{raiseA}),
Eq.\ (\ref{linHW}) becomes
\bea
&&\left(1 - \bar\beta_{_R} - \frac{5\varepsilon}{3}\right) q =
-\frac{2\Omega^2}{H^2}\, \delta \lag T^t_t\rag_{_R} \nonumber\\
&& \qquad = \varepsilon Ht\, {\cal D}q  - \bar\alpha_{_R}\,{\cal D}q
- \frac{\varepsilon}{2\Omega^2 } \int_{\eta_0}^{\eta}\, d\eta'\,
K(\eta -\eta'; k ;\mu) [\Omega^2\,{\cal D}q]_{\eta'}
- \frac{2\,}{\Omega^2} \delta \lag \bar T^t_{\ t} \rag_{_R}\,.
\label{qeqHW}
\eea
Comparing to (\ref{qeq}), it is apparent that the linear response equation
(\ref{qeqHW}) based on the methods of \cite{HorWal,Starob} differs from that
based on the anomaly action in four respects:
\begin{itemize}
\item The coefficient of $\delta B^t_{\ t}$ in (\ref{linHW}) and (\ref{localtt}),
$-\beta_{_R} - \frac{b}{9}$ is replaced by $- \beta_{_R} + \frac{b'}{9}$
in (\ref{linaux}) and (\ref{qeq}). This is an inessential difference, amounting
to a different coefficient of the finite local $R^2$ term in the action, already
discussed, and which, as can be seen from \eqref{pert}, is any case negligible.
\item The linear response form (\ref{linHW}) or (\ref{qeqHW}) contains the general
state dependent term, $\delta \lag \bar T^a_{\ b}\rag_{_R}$, whereas (\ref{qeq})
contains the specific $w-2h_{tt}$ and $v$ auxiliary field terms.
\item The auxiliary field action gives rise to additional equations of motion
(\ref{auxlineqs}) for the state dependent terms, absent in the
approach of \cite{HorWal,Starob}.
\item Linear response based on the anomaly action, (\ref{qeq}) lacks the term
with the non-local kernel $K$ in (\ref{linHW}) and (\ref{qeqHW}).
\end{itemize}

In \cite{HorWal,Starob}, the variation of the quantum state of the field
was (implicitly) constrained by the metric variation, and not allowed to vary
independently. This is equivalent to setting $v = w- 2h_{tt} = 0$ in the anomaly
auxiliary field approach. Since these fields obey the homogeneous equations
(\ref{auxlineqs}), the particular solution in which they both vanish is allowed.
However, the local auxiliary field approach allows for a wider class of variations
of the state than that of \cite{HorWal,Starob}. That the $w-2h_{tt}$ and $v$
auxiliary field terms in (\ref{linaux}) are the result of allowing the state
to vary over and above the local metric variation follows from the fact that
the freedom to vary the auxiliary fields amounts to freedom to vary the boundary
conditions on the Green's function of $\Delta_4$ entering the non-local form
of the trace anomaly effective action \cite{MotVau}. From (5.24) one
sees that these terms scale with the conformal factor exactly as expected for
the state dependent variation $\delta \lag \bar T^t_{\ t} \rag_{_R}$ in the
general form (\ref{qeqHW}). In (\ref{linaux}) these terms take a specific form
related to the variation of the scalar auxiliary fields which obey their own
independent equations of motion (\ref{ueq}) and (\ref{veq}). The important
feature of the auxiliary field formulation of the anomaly effective action
(\ref{Sanom})-(\ref{SEF}) is that these additional state dependent terms are
expressed in terms of local field variations that have their own independent
equations of motion (\ref{varyaux}). Although such state dependent
variations are certainly allowed on general grounds, the specific form
of these variations are determined by the local auxiliary field form of
the effective action.

\section{Gauge Invariant Variables and Action}

As already remarked in Sec. III the scalar sector perturbations involve
the four functions ${\cal A}, {\cal B}, {\cal C}, {\cal E}$, only two
linear combinations of which are gauge invariant.  In this section gauge
transformations are discussed and a set of gauge invariant variables are
constructed. The linear response equations around de Sitter space are written
in terms of these gauge invariant variables, and the gauge invariant quadratic
action functional corresponding to these equations is given.

The linearized coordinate (gauge) transformation of the metric perturbation is
\be
h_{ab} \rightarrow
h_{ab} + \nabla_a \xi_b + \nabla_b\xi_a
\; .
\label{hgauge}
\ee
Under this gauge transformation the scalar auxiliary fields transform as
\bes\bea
&&\varphi \rightarrow \varphi + \xi^a\nabla_a \varphi\,,\\
&&\psi \rightarrow \psi + \xi^a\nabla_a \psi \,.
\eea\label{phpsgauge}
\ees
\noindent
With the definitions
\bes
\bea
&&\xi^t = a T\,,\\
&&\xi^i = \eta^{ij}\partial_j L
\; .
\eea
\ees
\noindent appropriate for scalar perturbations, it is found that the linearized metric functions in
the decomposition (\ref{linmet}) transform as
\bes
\bea
&&{\cal A} \rightarrow {\cal A} + \dot{a}T + a \dot{T}\,,\\
&&{\cal B} \rightarrow {\cal B} + a\dot{L} - T\,,\\
&&{\cal C} \rightarrow {\cal C} +
\frac{1}{3} \stackrel{\rightarrow}{\nabla}\!\!^2\, L + \dot{a}T\,,\\
&&{\cal E} \rightarrow {\cal E} +  L
\; .
\eea
\label{gtrans}
\ees
\vspace{-.6cm}

\noindent
These are the linearized coordinate transformations possible in the scalar
sector~\cite{Bard,Stew}.  By considering the first order
variations of Eqs.~(\ref{phpsgauge}) and making use of the
notation introduced in Eq.~(\ref{auxvar}), one obtains the
transformation for both auxiliary field perturbations:
\bes
\bea
&&
\phi \rightarrow \phi + 2\dot{a}T\,, \label{phitrans}\\
&&\psi \rightarrow \psi\,.
\eea \ees
\vspace{-.6cm}

\noindent
Thus $\psi$ is already gauge invariant, as is the quantity $v$ defined
by (\ref{vdef}), while $\phi$ transforms due to the
non-zero $\bar\varphi = 2 Ht$ in the BD state. It is easily checked
that the from the transformation of the $\phi$ field (\ref{phitrans}) that
\be
\Phi \equiv \phi + 2 \dot{a}{\cal B}- 2 a \dot{a}\dot{\cal E}
\label{Phidef}
\ee
is gauge invariant. This is similar to the gauge invariant variable that
can be constructed from the scalar field in scalar inflaton models
of slow roll inflation \cite{MukFelBra}. Finally the metric variables,
\bes
\bea
&& \Upsilon_{\cal A}
\equiv {\cal A} + \partial_t(a {\cal B})- \partial_t(a^2 \partial_t{\cal E})\,,\\
&&\Upsilon_{\cal C} \equiv {\cal C}
- \frac{\stackrel{\rightarrow}{\nabla}\!\!^2{\cal E}}{3}
+ \dot{a}B - a \dot{a} \dot{\cal E}\,.
\eea
\label{defUpsAC}
\ees
\vspace{-.6cm}

\noindent
are invariant under the linearized gauge transformations (\ref{gtrans}). These
are the gauge invariant Bardeen-Stewart potentials denoted by $\Phi_A$ and
$\Phi_C$ in Refs. \cite{Bard,Stew}

The two quantities $\delta R$ and $q$ encountered in the linear response analysis
can be written in terms of the metric gauge invariant variables $\Upsilon_{\cal A}$
and $\Upsilon_{\cal C}$. Indeed, using the variation for $\delta R$ \cite{BarChr}
one finds
\bea
\delta R &=& - \sq h + \nabla_a\nabla_b h^{ab} - R^{ab}h_{ab} \nn
&=&  6 (\ddot\Upsilon_{\cal C} - H \dot\Upsilon_{\cal A})
+ 24 H  (\dot\Upsilon_{\cal C} - H \Upsilon_{\cal A})
- \frac{2}{a^2} \, \stackrel{\rightarrow}{\nabla}\!\!^2\,
\Upsilon_{\cal A} - \frac{4}{a^2} \stackrel{\rightarrow}{\nabla}\!\!^2\,
\Upsilon_{\cal C}\,,
\label{delRg}
\eea
Hence condition~(\ref{delR0}), $\delta R = 0$, is gauge invariant, and
provides one constraint between the two gauge invariant potentials
$\Upsilon_{\cal A}$ and $\Upsilon_{\cal C}$. Next one can verify that $q$,
defined in Equation~\eqref{qdef}, can be written in the form,
\be
q \equiv -\frac{2a^2}{H^2} \delta G^t_{\ t} = 12a^2
\left({\frac{1}{H}}\dot\Upsilon_{\cal C}- \Upsilon_{\cal A}\right)
-\frac{4}{H^2}
\, \stackrel{\rightarrow}{\nabla}\!\!^2\, \Upsilon_{\cal C}\,,
\label{qUps}
\ee
which is also gauge invariant.

In the linear response equations the quantity $w-2h_{tt}$ is also encountered
in the gauge defined by the conditions (\ref{gaugecond}). The gauge invariant
quantity which reduces to $w - 2 h_{tt}$ in this gauge is
\be
H^2 u
= \left[\frac{d^2}{dt^2}
+ H\frac{d}{dt} +\frac{k^2}{a^2}\right] \Phi + 6H \, \frac{d \Upsilon_{\cal C}}{dt}
 - 2 H \, \frac{d \Upsilon_{\cal A}}{dt} - 8 H^2 \Upsilon_{\cal A}
\; .
\label{potphi}
\ee
This can be seen by evaluating $\Upsilon_A$, $\Upsilon_C$, and $\Phi$ in the gauge
(\ref{gaugecond}) using the definitions \eqref{Phidef} and (\ref{defUpsAC}).

Then the linear response equations around de Sitter space can be written in
terms of the gauge invariant variables $u$, $v$, $\delta R$, and $q$.
With the condition $\delta R =0$ imposed, the equations for $u$ and $v$ are
\bes\bea
\left( \frac{d^2}{dt^2}+ 5H \frac{d}{dt}+ 6H^2 + \frac{k^2}{a^2}
\right)u &=& 0\,,\\
\left( \frac{d^2}{dt^2}+ 5H \frac{d}{dt}+ 6H^2 + \frac{k^2}{a^2}
\right) v &=& 0\,,
\eea
\label{auxginv}
\ees
while that for $q$ is
\bea
&& q = - \frac{2\Omega^2}{H^2} \delta \lag {T^t}_t\rag_{_R} \nonumber\\
&& \qquad =  \varepsilon Ht\, {\cal D}q  - \bar\alpha_{_R}\,{\cal D}q
- \frac{\varepsilon}{2\Omega^2 } \int_{\eta_0}^{\eta}\, d\eta'\,
K(\eta -\eta'; k ;\mu)
[\Omega^2\,{\cal D}q]_{\eta'} +  \varepsilon' \frac{k^2}{H^2} u
- \frac{\varepsilon}{3} \frac{k^2}{H^2} v\,.
\label{qginv}
\eea
Here the differential operator $\cal D$ is defined by (\ref{diffD}). In
(\ref{qginv}) the $\bar\beta_{_R}$, $\varepsilon$ and $\varepsilon'$ terms on
the left side of (\ref{qeq}) or (\ref{qeqHW}) which differ in the two linear
response approaches but which in any case are small compared with unity, have
been dropped. The non-local term from the analysis of \cite{HorWal,Starob} has
been retained and the specific state dependent terms from the auxiliary field
anomaly action have been written in terms of the gauge invariant variables
$u$ and $v$. In arriving at (\ref{qginv}) we have reconciled the exact but formal
calculations of \cite{HorWal,Starob} with those following from the
anomaly effective action by adding the non-local term from the former to the
specific state dependent variations determined by the auxiliary fields of the
latter approach. By the arguments of Sec. III, all the gauge invariant
information about scalar perturbations around the self-consistent de Sitter
invariant BD state and all components of the curvature tensor variations can
be obtained from the solution(s) of (\ref{auxginv})-(\ref{qginv}). This is the
final gauge invariant form of the linear response equations for the only
remaining metric degree of freedom in the tracefree but spatial scalar
sector of perturbations of de Sitter spacetime whose solutions we will
analyze in the next section.

If the classical Einstein-Hilbert action is added to the effective action of the anomaly, Eqs. (4.1) and (4.2),
and both are expanded to quadratic order about the self-consistent BD de Sitter solution given by (5.3), (5.5), and
(5.6), the resulting quadratic action can be expressed in terms of the gauge invariant variables $u$, $v$, $\Upsilon_A$,
and $\Upsilon_C$ in the relatively simple form,
\begin{eqnarray}
\qquad S^{(2)} &=& \left(1 - 5 \varepsilon'\right) S_G
+ b^\prime \int d^3 \vec{x}\, dt \,a^3 \, \left\{ - \frac{H^4 {u}^2}{2} +  \frac{H^2\,u\, \delta R}{3}  \right\} \nonumber \\
   & & \;\; + b \int d^3 \vec{x}\, dt \,a^3 \left\{ - H^4 u\,v + \frac{H^2 \,v\, \delta R}{3}
+ \frac{4 \ln a}{3\, a^4}  \left[ \vec\nabla^2 (\Upsilon_A -  \Upsilon_C )\right]^2\right\} \nonumber\hspace{4.1cm}  (6.13)
\end{eqnarray}
where
\begin{eqnarray}
S_G &=& \frac{1}{8 \pi G_N} \int d^3 \vec{x}\, dt \,a^3 \left[ - 3\, \left( \frac{\partial \Upsilon_C}{\partial t} \right)^2
+    6\,H\,\Upsilon_A\, \frac{\partial \Upsilon_C}{\partial t} +   \frac{2}{\,a^2}\ (\vec{\nabla}\Upsilon_A)\cdot (\vec{\nabla}\Upsilon_C)
+   \frac{(\vec{\nabla} \Upsilon_C)^2}{a^2} -3\,H^2\,\Upsilon^2_A  \right] \nonumber\quad (6.14)
 \end{eqnarray}
is the Einstein-Hilbert part of the action, and $\delta R$ is given by (6.8). Varying
(6.13) with respect to $\Phi, \psi, \Upsilon_A$ and $\Upsilon_C$,
and setting $\delta R= 0$ yields the gauge invariant linear response equations
(6.11) and (6.12), without the non-local contribution involving $K$, and without the
$\bar\alpha_{_R}$ and $\bar\beta_{_R}$ terms which would require adding
to (6.13) the contributions of the purely local $C^2$ and $R^2$ actions also
expanded to quadratic order in the perturbations about de Sitter space.

\section{Solutions to the Linear Response Equations}

\subsection{Solutions of the first kind: $u = v = 0$}
\label{sec:veq0}

Since $u$ and $v$ satisfy the same homogeneous equation, {\it cf.}
(\ref{auxginv}), we first consider the case $u = v = 0$. Eq. (\ref{qginv})
is then homogeneous in $q$ and includes only those variations of
$\lag T^a_{\ b}\rag_{_R}$ which are driven by the metric fluctuations,
{\it i.e.} variations of the first kind. The homogeneous equation possesses the
trivial solution $q=0$. This is the unique classical solution in pure classical
gravity, for when $\hbar =0$, all effects of the stress tensor
$\lag T^a_{\ b}\rag_{_R}$ and its quantum fluctuations are set to zero on the
right side of (\ref{qginv}), and there is no dynamics at all in the scalar sector.
The classical equation, $q=0$ is an equation of constraint.

Next it is straightforward to find non-trivial solutions of (\ref{qginv})
if $u= v = 0 $ and the non-local term involving $K$ is neglected.
One may then check the consistency of this local approximation by inserting
the previous solution into the integral and evaluating the non-local term.
Since at late times the $\varepsilon H t \, {\cal D}q$ term dominates over the
$\bar\alpha_R\,{\cal D} q$ term, which in any case just shifts the origin of
the time variable $t$, the $\bar\alpha_R\, {\cal D}q$ term can be neglected as well
without loss of generality.

If the non-local term in Eq. \eqref{qginv} is ignored and $u = v = 0
= \bar\alpha_R = 0$ then the linear response equation reduces to
\be
q = \varepsilon H t\, {\cal D} q
\label{LocLinResp}
\ee
The Fourier components with $k\neq 0$ are red-shifted away exponentially
rapidly by the de Sitter expansion, $k^2 e^{-2Ht} \rightarrow 0$. Hence
after a few expansion times this term may be neglected, effectively setting
$k=0$. Then,
\be
\frac{d^2q}{d t^2} + 3H \frac{dq}{d t} + 2 H^2 q
- \frac{q H}{\varepsilon t} \simeq 0\,.
\label{geqk0}
\ee
This equation describes an exponential growth on the time scale
$H /\sqrt{\varepsilon} \sim M_{Pl}$. Substituting the exponential form,
\be
q(t) = \exp \left(\int^{Ht} W(\tau) d \tau \right)\,
\label{Wdef}
\ee
one finds
\be
\frac{1}{H} \frac{d W}{dt} + W^2 + 3 W + 2 - \frac{1}{\varepsilon H t} = 0 \;.
\ee
Ignoring the $dW/dt$ term, the zeroth order WKB solution is
\be
W \simeq -\frac{3}{2} + \frac{1}{2} \sqrt{1 + \frac{4}{\varepsilon H t}}\,.
\label{Wsoln}
\ee
Starting from an initial time $t_0$ such that $Ht_0$ is of order unity,
the WKB approximation (\ref{Wsoln}) is valid up until the
turning point where $W$ vanishes. This turning point is at
$Ht_{max} = 1/2\varepsilon$ and this is where $q$ achieves the maximum value,
\be
q_{max} = q (t_{max}) \simeq \exp \left( \frac{\ln 2}{\varepsilon}
- \frac{2Ht_0} {\sqrt{\varepsilon}}\right)\,,\quad \varepsilon \ll 1\,.
\label{qmax}
\ee
Here $t_0$ is the initial time which is assumed to be much less than
$(H\varepsilon)^{-1}$. Thus if the non-local term is ignored, then the
perturbation becomes exponentially large in
$\varepsilon^{-1} \sim M_{Pl}^2/ H^2 \gg 1$. The additional factor of the
time $t$ reduces this growth at later times, and after a time of order
$\varepsilon^{-1}$ eventually turns off the exponential growth.

Eq. (\ref{LocLinResp}) was also solved numerically for $\varepsilon = 0.01$ and
the results are shown in Figure~\ref{fig:q.pdf}.
\begin{figure}
\vskip -0.2in \hskip -0.4in
\includegraphics[angle=90,width=3.4in,clip]
{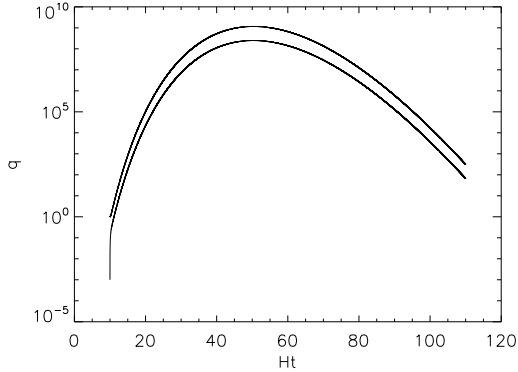}
\vskip -.2in
\caption{The plot shows solutions to the linear response equation \eqref{qginv}
in the approximation (7.1) when the non-local term is neglected, $u = v = 0$, $\varepsilon = 0.01$,
and $k = 100$.  The initial time is $Ht = 10$.  The solid curve is the
solution which begins with $q = 1$, $\dot{q} = 0$ and the dashed curve is the
solution which begins  with $q = 0$, $\dot{q} = 1$. The solutions grow
exponentially according to (\ref{Wdef}) and (\ref{Wsoln}) up until
$Ht_{max} \simeq  \frac{1}{2\varepsilon} = 50$.}
\label{fig:q}
\end{figure}
Note that the solutions grow rapidly for a long time before reaching a maximum
and finally decreasing again. The exponential growth is well described by the
analytic WKB approximation of (\ref{Wdef}) and (\ref{Wsoln}).

\begin{figure}
\vskip -0.2in \hskip -0.4in
\includegraphics[angle=90,width=3.4in,clip]
{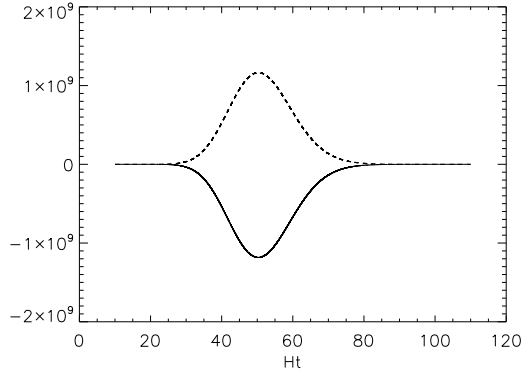}
\vskip -.2in \caption{Shown are the non-local and local terms for the case
$k = 100$. The dashed line corresponds to the local term
$\varepsilon Ht {\cal D} q = q$. The initial conditions for the solution
shown for $q$ are specified at $Ht = 10$.}
\label{fig:compare1}
\end{figure}

\begin{figure}
\vskip 0.2in \hskip -0.4in
\includegraphics[angle=90,width=3.4in,clip]
{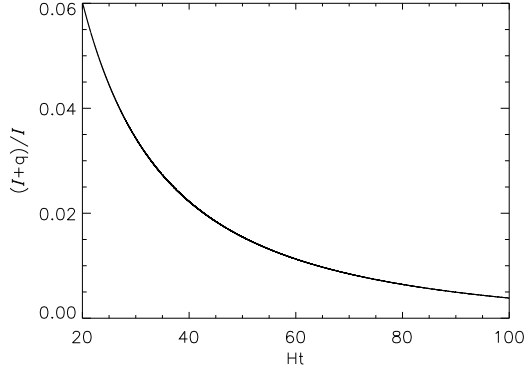}
\vskip -.2in \caption{Shown is the approximate cancelation which occurs when
the local and non-local terms, $\varepsilon Ht {\cal D} q = q $ and $I$
respectively are added together for the case $k = 100$.}
\label{fig:compare2}
\end{figure}

To test whether this general behavior survives in the full linear response
equation including the non-local term, one can first solve (\ref{LocLinResp})
and then substitute the solution into the non-local term to see if its
effects are significant or not. The non-local term in Eq. \eqref{qginv}
can be written in the form,
\be
I = -\frac{1}{2 \Omega^2} \int_{\eta_0}^\eta d \eta'
K(\eta-\eta';k;\mu) f(\eta') \,,
\label{Idef}
\ee
with $K$ given by (\ref{Kt}) and
\be
f(\eta) \equiv \varepsilon\Omega^2(\eta) {\cal D}q(\eta)\,.
\label{Fdef}
\ee
The behavior of the kernel $K$ is discussed in Appendix A.  For an initial
value formulation of linear response, the lower limit of the time integral
in (\ref{Idef}) should be taken as an arbitrary but finite $\eta_0$. As
discussed in Appendix A, a proper regularization of the non-local kernel $K$,
such as by means of a Pauli-Villars regulator mass, produces transient terms
which fall off at late times, and can be neglected. Then Eq.\ \eqref{Kpart}
can be used with the lower limit of the integral replaced by $\eta_0 \ll \eta$
to obtain the approximate form,
\be
I \simeq \frac{1}{\Omega^2(\eta)} \int_{\eta_0}^\eta d\eta'
{\rm ci}[k(\eta - \eta')] \frac{d f(\eta')}{d \eta'}
+ \ln\left(\frac{\mu}{k}\right) \frac{f(\eta)}{\Omega^2(\eta)}\,,
\label{I5}
\ee
For the solution of \eqref{LocLinResp},
\be
f(\eta) =  \frac{\Omega^2(\eta) q(\eta)}{\ln(\Omega(\eta))} \simeq
\frac{-1} {H^2\eta^2\ln(-H\eta)}
\exp \left( \int^{-\ln(-H\eta)} W(\tau) d \tau\right)\,.
\label{feta}
\ee
In Figs.~\ref{fig:compare1} and \ref{fig:compare2} the non-local and local
terms on the right hand side of the linear response equations are shown for
the case in which $q$ is the solution to the local linear response equations
displayed in Fig. \ref{fig:q}. It is clear from the Figures that for the
solutions shown the contribution of the local and non-local terms on the right
hand side of (\ref{qginv}) are not only comparable but tend to partially cancel.
Hence we conclude that it is {\it not} correct to ignore the non-local term
relative to the local terms in (\ref{LocLinResp}) for these homogeneous
solutions for which $u = v =0$.

It is possible to understand the behavior of the non-local term analytically
for the case that $f(\eta)$ is a rapidly increasing function, as it is for
(\ref{feta}). In this case the largest contribution comes from the region near
the endpoint of the integral (\ref{Idef}) as $\eta' \rightarrow \eta$. Consider
a time $\eta_1 < \eta$ close to the upper limit of the non-local integral, so
that the corresponding interval $\Delta \eta = \eta - \eta_1$ satisfies
\be
k \Delta \eta \ll 1 \,.
\label{etacond}
\ee
If this condition is satisfied then from (\ref{cosint}),
${\rm ci}[k(\eta-\eta')] \simeq
C + \ln[k(\eta-\eta')]$  and \eqref{I5} becomes
\bea
&&I \approx I_1 =
\frac{1}{ \Omega^2(\eta)} \left\{ \int_{\eta-\Delta \eta}^\eta d\eta'
\ln\left(k(\eta - \eta')\right) \frac{d f(\eta')}{d \eta'}
+ \left[C f(\eta) - C f(\eta - \Delta \eta)
+ f(\eta) \ln\left(\frac{\mu}{k}\right) \right] \right\}\nn
&& \quad \approx \frac{1}{\Omega^2(\eta)}\left\{\left[f(\eta)
- f(\eta - \Delta \eta)\right]
\left[\ln(k \Delta \eta) + C\right]
+ f(\eta) \ln\left(\frac{\mu}{k}\right) \right\} \nn
&& \quad \approx \left\{ \frac{1}{\Omega^2(\eta)}
\Delta \eta \frac{df(\eta)}{d\eta}
\left[\ln(k \Delta \eta) + C\right]
+ f(\eta) \ln\left(\frac{\mu}{k}\right) \right\}\,.
\label{Iapprox}
\eea
Now from (\ref{Wdef}), (\ref{Wsoln}) and (\ref{feta}), during the period of
exponential growth,
\be
\frac{df}{d\eta} \approx f\, \frac{W}{|\eta |} \approx \frac{f}{|\eta|
\sqrt{\varepsilon |\ln(-H\eta)|}}\,.
\label{fderiv}
\ee
Hence if
\be
\Delta \eta \lesssim |\eta| \sqrt{\varepsilon |\ln(-H\eta)|} \,,
\label{deleta}
\ee
for which (\ref{etacond}) is well satisfied, with $\varepsilon \ll 1$, then
\bea
&& I \approx I_1 \stackrel{<}{_\sim} \frac{f(\eta)}{\Omega^2(\eta)}
\ln(e^C\mu\Delta \eta)\nn
&& \quad \stackrel{<}{_\sim} \varepsilon {\cal D} q \ln\left(\frac{\mu}{H}
e^{C-Ht} \sqrt{\varepsilon Ht}\right)\nn
&& \quad = -\varepsilon (Ht) {\cal D} q
+ \varepsilon {\cal D} q \,
\ln \left(\frac{\mu}{H} e^C \sqrt{\varepsilon Ht}\right)\,.
\label{Iest}
\eea
The first term of (\ref{Iest}) exactly cancels the local term,
$\varepsilon (Ht) {\cal D} q$ on the right hand side of the linear response
equation (\ref{qginv}), thus giving an analytic approximation which reproduces
the cancelation found numerically, and exhibited in Fig. \ref{fig:compare2}.
The second term in (\ref{Iest}) gives an estimate for the remainder. Examination
of the steps leading to (\ref{Iest}) shows that this approximation is valid as
long as ${\cal D} q$ is a rapidly increasing function. In this case, the
non-local integral (\ref{Idef}) receives its dominant contribution from the
very short time ultraviolet region (\ref{deleta}) close to its upper endpoint.

From both this analytic estimate and the numerical studies we conclude
that for the homogeneous solutions of the linear response equation
(\ref{qginv}), the non-local term cannot be ignored. Thus the exponential
growth found in (\ref{Wdef})-(\ref{qmax}) by neglecting the non-local
integral in (\ref{qginv}) is not reliable. Actually this could have been
anticipated by returning to the origin of these terms in Section III. The
non-local term, and the local $\bar\alpha_{_R}\,{\cal D}q$ and
$Ht\, {\cal D}q$ terms, all of which involve ${\cal D}q$ have the same
origin in the logarithmic distribution in Fourier space. Eq. (\ref{muindep})
shows that the ${\cal D}q$ dependent terms should be considered together,
since only the sum is independent of the arbitrary renormalization scale $\mu$.
The $Ht\, {\cal D}q = \ln \Omega\, {\cal D}q $ term arises from the fact that
the logarithmic distribution is defined in flat conformal coordinates, in terms
of the conformal frequency and momentum $ K^a = (\omega, \vec k)$, whereas the
frequency and momentum relative to the physical line element (\ref{FRW})
is $K^a/\Omega$ (for slowly varying $\Omega$). Thus one should expect the
combination,
\be
\ln \left[\frac{-\omega^2 + k^2 -i \epsilon\, {\rm sgn}\, \omega}
{\mu^2} \right]   - 2 \ln \Omega =
\ln \left[\frac{-\omega^2 + k^2 -i \epsilon\, {\rm sgn}\, \omega}
{\mu^2 \Omega^2}\right] \,,
\label{logphys}
\ee
always to appear together. Heuristically, the local $\ln \Omega\, {\cal D}q$
term is just the term needed to insert $\Omega$ in the denominator of the
logarithm in (\ref{logphys}), and convert the conformal frequency and
wave-number to their values in the physical metric. Since it is known from
the linear response equations in flat space that the only non-trivial modes
occur at the Planck scale, these same Planck scale solutions should persist
in the de Sitter background when all three of the terms involving ${\cal D}q $
in (\ref{qginv}) are considered together and solved self-consistently.
For these homogeneous solutions of (\ref{qginv}) for which $u = v = 0$
the non-local integral $I$ cannot be neglected. This is the significance of
the integral being dominated by its extreme short distance logarithm near
its upper limit. To find these modes accurately one should solve the
full non-local Eq. (\ref{qginv}). However, these ultraviolet Planck scale
solutions should be excluded from consideration in de Sitter space in any
case for the same reason as in flat space, because they lie outside of
the range of validity of the low energy semi-classical description of gravity.
Therefore we do not pursue the solutions of the homogeneous Eq. (\ref{qginv})
further, and turn instead to the inhomogeneous solutions of the second kind.

\subsection{Solutions of the second kind: $u \ne 0$ or $v \neq 0$}
\label{sec:statechange}

Next consider the case in which the inhomogeneous gauge invariant state dependent
terms $u$ or $v$ are different from zero. The general solution of
(6.11) for either $u$ or $v$ is easily found. In either case
the solutions are
\be
u_{\pm} = v_{\pm} =  \frac{1}{H^2 a^2} \,\exp \left ( \pm \frac{ik}{Ha}\right)
e^{i \vec{k} \cdot \vec x}
= \eta^2 \,\exp (\mp i k\eta + i \vec k \cdot \vec x) \,.
\label{vsoln}
\ee
Note that this solution and the Eqs. (6.11) it satisfies involve
the cosmological horizon scale $H$, but not the Planck scale. Thus we
term these new solutions {\it cosmological horizon modes}. Taking {\it e.g.} $u=0$,
neglecting all the ${\cal D} q$ terms in (\ref{qginv}), we have
\be
q \simeq \frac{\varepsilon}{2}
\vec{k}^2 H^2 \eta^2 e^{\mp i k\eta + i \vec k \cdot \vec x}\,.
\label{inhomq}
\ee
This corresponds to a linearized stress tensor perturbation of
\be
\delta \lag T^{tt} \rag_{_R} = \frac{H^2\,q}{16 \pi G_{_N} a^2} =
b\,\frac{k^2 H^2}{a^4}\, e^{\mp i k\eta + i \vec{k} \cdot \vec x}\,.
\label{varTttq}
\ee
The solutions with $u$ of the form (\ref{vsoln}) give linearized
stress tensor perturbations similar to (\ref{varTttq}) with $b$ replaced
by $b'$.

We will now show that it is legitimate to neglect the terms involving
${\cal D} q$ in (\ref{qginv}) for these solutions. Note that
\be
{\cal D} q = \pm \frac{2 i k}{Ha} \, q = \pm 2 i \frac{k_{phys}}{H} \, q \,,
\label{Dqv}
\ee
where $k_{phys} = k/a = H |k \eta |$ is the redshifting physical
momentum of the mode. Recalling the definitions (\ref{pert}) we have
\be
\vert\bar\alpha_{_R}\,{\cal D} q \vert\sim
\vert\alpha_{_R}\, G_{_N} H k_{phys}\vert \,\vert q \vert \ll \vert q \vert \,,
\label{alphacond}
\ee
provided
\bes
\bea
&& G_{_N} H^2 \ll 1\,,\quad {\rm or} \quad H \ll M_{Pl}\,,\\
&& G_{_N} k_{phys}^2 \ll 1\,,\quad {\rm or} \quad k_{phys} \ll M_{Pl}\,,
\eea
\label{Hkcond}
\ees
\vspace{-.6cm}

\noindent
and $\vert \alpha_{_R} \vert $ is of order unity. Thus, from (\ref{alphacond})
the $\bar\alpha_{_R} {\cal D} q$ term in (\ref{qginv}) can be neglected.

For the non-local term note that (\ref{Dqv}) falls off with time, which
has the consequence that the function (\ref{Fdef}) appearing in the integral
also falls with time. In this case the nonlocal integral $I$ in Eq. \eqref{Idef}
receives contributions over the entire cosmological expansion time scale $H^{-1}$,
and remains bounded. This is in contrast to the non-local term evaluated on the
steeply rising function considered in the homogeneous case when $u = v =0$,
where (\ref{Idef}) was dominated by the short time behavior (\ref{deleta}) near its
upper endpoint. For the solution in (\ref{inhomq}), $I$ may even be calculated
analytically using the Pauli-Villars form of the kernel in Eq.\eqref{KPV},
for $M$ large and setting $M = \mu$ in the terms that remain after
this limit is taken. In this way one finds that all the terms in the non-local
integral are small and may be neglected, provided the conditions (\ref{Hkcond})
are satisfied, except for one possibly large logarithmic term which is of order,
\be
b G_{_N} H \frac{k}{a} \,\ln \left(\frac{\mu^2}{Hk}\right) \, q\,.
\ee
Because of (\ref{logphys}) with $\Omega = a$, this combines with the
remaining local term involving ${\cal D} q$, {\it i.e.}
\be
\varepsilon \,Ht\, {\cal D} q \sim b G_{_N} H \frac{k}{a}\, \ln(a)\, q
\ee
to give
\be
b G_{_N} H k_{phys}\, \ln \left(\frac{\mu^2}{Hk_{phys}}\right) \, q\,.
\ee
which is also of negligible magnitude with respect to $\vert q \vert$
provided $k_{phys} < \mu \lesssim M_{Pl}$, $b$ is of order unity, and
conditions (\ref{Hkcond}) are satisfied. Since these are necessary
conditions for the applicability of the semi-classical effective
theory (\ref{scE}) in the first place, we conclude that all
the ${\cal D}q$ terms in the linear response equation (\ref{qginv})
may be neglected compared to (\ref{inhomq}), and (\ref{inhomq})
is a non-vanishing solution to the {\it full} linear response equations,
to a very high degree of accuracy if the conditions (\ref{Hkcond}) are
satisfied.

The other components of the stress tensor perturbation can be found for
this solution from (\ref{varTttq}) by a general tensor decomposition for
scalar perturbations analogous to Eqs.  (\ref{linmet}) for the metric. That is,
the general perturbation of the stress tensor $\delta \lag T^{ab} \rag_{_R}$
in the scalar sector can be expressed in terms of $\delta \lag T^{tt} \rag_{_R}$
plus three additional functions. These three functions are determined
in terms of $\delta \lag T^{tt} \rag_{_R}$ by the conditions of
covariant conservation,
\be
\nabla_b\, \delta \lag T^{ab} \rag_{_R} = 0\,
\ee
for $a=t$ and $a=i$ (two conditions), plus the tracefree condition,
\be
\delta \lag T^a_{\ a} \rag_{_R} = 0\,,
\ee
imposed as a result of the $\delta R = 0$ condition. A straightforward
calculation using these conditions and the Christoffel coefficients
(\ref{Chris}) gives then
\begin{subequations}
\bea
&& \delta \lag T^{tt} \rag_{_R} = bH^2  \frac{k^2}{a^4} \,
e^{\mp i k\eta + i \vec k \cdot \vec x}
= - \frac{bH^2}{a^4}\, \vec \nabla^2_{\vec x}\,
e^{\mp i k\eta + i \vec k \cdot \vec x}\, ,\\
&&\delta \lag T^{ti} \rag_{_R} =
\pm b H^2\,\frac{k^i k}{a^5}\, e^{\mp i k\eta
+ i \vec k \cdot \vec x}
= \frac{bH^2}{a^4}\, \frac{\partial^2}{\partial x^i \partial t} \,
e^{\mp i k\eta + i \vec k \cdot \vec x}\, ,\\
&&\delta \lag T^{ij} \rag_{_R} = bH^2\, \frac{k^ik^j}{a^6}
\, e^{\mp i k\eta + i \vec k \cdot \vec x} =
- \frac{bH^2}{a^6}\, \frac{\partial^2}{\partial x^i \partial x^j}\,
e^{\mp i k\eta + i \vec k \cdot \vec x}\,.
\eea
\label{TkFRW}
\end{subequations}
\vspace{-.6cm}

\noindent
for the other components of the stress tensor variation for these modes
in the flat FRW coordinates of de Sitter space.

Thus, the auxiliary fields of the anomaly action yield the non-trivial
gauge invariant solutions (\ref{TkFRW}) for the stress tensor and corresponding
linearized Ricci tensor perturbations $\delta R^{ab}$. Being solutions of
(6.11) which itself is independent of the Planck scale, these solutions
vary instead on arbitrary scales determined by the wavevector $\vec k$, and are
therefore genuine low energy modes of the semi-classical effective theory. The
Newtonian gravitational constant $G_{_N}$ and the Planck scale enter
Eq. (6.12) only through the small coupling parameters $\varepsilon$
and $\varepsilon'$ between the auxiliary fields and the metric perturbation $q$.
Thus in the limit of either flat space, or arbitrarily weak coupling
$G_{_N}H^2 \rightarrow 0$ these modes decouple from the metric
perturbations at linear order.

The result (\ref{TkFRW}) would be obtained if the anomaly action (\ref{Sanom})
{\it alone} is used to generate the linear response Eqs. (\ref{qeq}), and the
non-local term is neglected completely. This demonstrates the relevance of the
anomaly action for describing physical infrared fluctuations in the effective
semi-classical theory of gravity, on macroscopic or cosmological scales
unrelated to the Planck scale.

\section{Cosmological Horizon Modes}

The modes found in Sec.\ref{sec:statechange} when the gauge invariant variables
$u$ and/or $v$ are nonzero can vary on any scale rather than the Planck
scale, which characterizes modes of the first kind found in Sec.\ref{sec:veq0} when
$u = v = 0$. The second set of modes arise naturally from the scalar auxiliary
fields which render the non-local anomaly action local, and are therefore
implied by the form of the trace anomaly at one-loop order. This is a non-trivial
result since these modes appear in the {\it tracefree} sector of the semi-classical
Einstein equations, with $\delta R=0$, and hence cannot be deduced from the
form of the trace anomaly itself, but only with the help of the covariant
action functional and the additional scalar degrees of freedom which the anomaly
implies. These additional modes, which couple to the scalar sector of metric
perturbations in a gauge invariant way, are due to a quantum effect
because the auxiliary scalar fields from which they arise are part of the
one-loop effective action for conformally invariant quantum fields,
rather than a classical action for an inflaton field usually
considered in inflationary models \cite{MukFelBra}.

In Section V it was shown that the background solutions $\bar\varphi = 2 \ln \Omega = 2Ht$
and $\psi = 0$  result in the stress-energy tensor for conformally invariant fields
in the Bunch-Davies state. The additional modes arising from perturbations of the
auxiliary fields $\varphi$ and $\psi$ from their background values correspond to
changes of state for the underlying conformal quantum fields from their de Sitter
invariant BD state. Some physical intuition about these modes and the changes of
state of the underlying quantum fields they correspond to may be gleaned from
the form of their stress tensor in (\ref{TkFRW}). If one averages this form over
the spatial direction of $\vec k$, a spatially homogeneous, isotropic stress
tensor is obtained with pressure $p = \rho/3$. Thus in FRW coordinates
this averaging describes incoherent or mixed state thermal perturbations
of the stress tensor which are just those of massless radiation, which
redshift with $a^{-4}$.

A second interpretation of this second set of modes emerges if we consider {\it coherent}
linear superpositions of different $\vec k$ solutions in different global coordinate
systems. In Ref. \cite{MotVau} a class of solutions to the auxiliary field equations
(\ref{auxeom}) were found for de Sitter space in {\it static} rather than cosmological
coordinates. In that background field calculations the spacetime geometry was held
fixed and there is no restriction on the state other than that required for
renormalization of the stress tensor for the quantum fields. In the linear response
equations the solutions to the perturbed equations (\ref{auxginv}) for the auxiliary
fields become source terms for the metric fluctuation $q$ in Eq. (\ref{qginv})
so their backreaction effects on the spacetime geometry are included to linear
order in fluctuations about the de Sitter background geometry. However, in either
case the equations (\ref{auxeom}) for the gauge invariant perturbations of the auxiliary fields,
$u$ and $v$, being linear, are the {\it same} equations whether or not the background
is varied. Thus they correspond to the same changes of state as in the background
field calculations of \cite{MotVau}, and can be compared directly with the fixed
de Sitter background solutions found for the auxiliary fields in the static coordinates
in \cite{MotVau}.

To make this comparison we first need to consider the relationship
between the flat FRW and static coordinate systems in de Sitter space.
The de Sitter line element in its static form,
\be
ds^2 = - (1-H^2r^2)d\hat t^2 + \frac{dr^2}{1-H^2r^2}
+ r^2 (d\theta^2 + \sin^2\theta d\phi^2)\,,
\label{static}
\ee
is identical to (\ref{FRW}) with the de Sitter scale factor (\ref{RWdS})
if the static time $\hat t$ and radius vector $\vec r$ are related to the flat FRW
coordinates  ($t, \vec x$) of (\ref{FRW}) by
\bes
\bea
&& r = |\vec x |\, e^{H t} \equiv \rho \, e^{H t}\,,\\
&& \hat t = t - \frac{1}{2H} \ln \left(1 - H^2 \rho^2 e ^{2Ht}\right)\,.
\eea
\ees
The inverse transformations are
\bes
\bea
\rho \equiv |\vec x|= \frac{r\,e^{-H \hat t}}{\sqrt{1-H^2r^2}}\,,\\
t = \hat t + \frac{1}{2H} \ln (1-H^2r^2)\,.
\eea
\ees
The Jacobian matrix of this $2 \times 2$ transformation is
\be
\left( \begin{array}{cc}
\frac{\partial \hat t}{\partial t} & \frac{\partial \hat t}{\partial \rho}\\
\frac{\partial r}{\partial t} & \frac{\partial r}{\partial \rho}
\end{array}\right) = \left(\begin{array}{cc}
\frac{1}{1-H^2r^2} & \frac{Hr^2}{\rho(1-H^2r^2)}\\
Hr & \frac{r}{\rho} \end{array}\right)
\label{Jacob}
\ee
Using these relations, one may express the action of the differential
operators in Eq. (6.11) in terms of the static coordinates
(\ref{static}) instead. We consider the case that the scalar auxiliary fields are
functions of $r$ only and focus attention on $v$. A short calculation using
(\ref{Jacob}) shows that
\be
- \frac{\vec \nabla^2_{\vec x}}{a^2}\, v(r) =
-\frac{1}{r^2} \frac{d}{dr} \left(r^2 \frac{dv}{dr}\right)\,.
\label{uinT}
\ee
while
\be
H^2 v(r) = \left(\frac{\partial^2}{\partial t^2} + H\frac{\partial}{\partial t}
- \frac{\vec \nabla^2_{\vec x}}{a^2}\right) \psi(r)
= - (1-H^2r^2)\, \frac{1}{r^2} \frac{d}{dr} \left(r^2 \frac{d\psi(r)}{dr}\right)\,,
\label{uop}
\ee
operating on functions $v = v(r)$ which are functions only of $r$ and not $\hat t$.

In \cite{MotVau} the static solutions to
\be
\Delta_4\, \psi(r) = 0
\label{psidel4}
\ee
are given. The four solutions are
\be
\psi = \frac{1}{Hr}\ln\left(\frac{1-Hr}{1+Hr}\right)\,,\quad
\ln\left(\frac{1-Hr}{1+Hr}\right)\,,
\quad 1\,, \quad \frac{1}{r}\,.
\label{psistatic}
\ee
The last of these is singular at the origin and so was not considered in \cite{MotVau}.
In any case this solution and the third constant solution to (\ref{psidel4})
give vanishing contribution to $v$ in (\ref{uop}) while from (\ref{uop}) the
first and second solutions give for $ v$,
\be
\frac{4}{1-H^2r^2}\,, \quad \frac{4}{H r}\,\frac{1+H^2r^2}{1-H^2r^2} \,,
\label{vstatic2}
\ee
respectively. The second gives a singular contribution to $v$ and the stress tensor
at $r=0$, and we consider only the first solution, taking
\be
v = \frac{4}{1-H^2r^2}\,.
\label{vstatic}
\ee
Then, from (\ref{varFt}), (\ref{tteq}), (\ref{uinT}) and (\ref{vstatic}), we obtain
\bes
\bea
&&\delta \lag T^{tt}\rag_{_R} = \frac{2bH^2}{3} \frac{\vec \nabla^2_{\vec x}}{a^2}\, v
= b \left\{ \frac{4H^4}{(1-H^2r^2)^2}
+ \frac{16H^6 r^2}{3(1-H^2r^2)^3}\right\}\,,\\
&& \delta R^t_{\ t} = 8\pi G_{_N} \delta \lag T^t_{\ t}\rag_{_R} = -\varepsilon H^2
\left\{ \frac{1}{(1-H^2r^2)^2} + \frac{4H^2 r^2}{3(1-H^2r^2)^3}\right\}\,.
\eea
\label{divTR}
\ees
\vspace{-.6cm}

\noindent
The solutions for $u$ are exactly analogous. This shows that a linear superposition
of solutions of the linear response equations of the second kind in static coordinates
can lead to gauge invariant perturbations which {\it diverge} on the de Sitter horizon.

To see what (\ref{divTR}) corresponds to in the static $(\hat t, r)$ coordinates,
we use the form of the other components in (\ref{TkFRW}) in FRW coordinates,
\bes
\bea
&&\delta \lag T^{ti}\rag_{_R} = \frac{b\,H^2}{a^2} \left( \frac{\partial}{\partial t}
+ 2H \right)\, \frac{\partial v}{\partial x^i}\,,\\
&&\delta \lag T^{ij}\rag_{_R} = - \frac{b\,H^2}{a^2} \frac{\partial^2\ v}{\partial x^i
\partial x^j}\,,
\eea
\ees
\noindent
and the transformation relation for tensors,
\be
T^{\hat t \hat t} = \left(\frac{\partial \hat t}{\partial t}\right)^2 T^{tt} +
2 \left(\frac{\partial \hat t}{\partial t}\right)
\left(\frac{\partial \hat t}{\partial x^i}\right)T^{ti}
+ \left(\frac{\partial \hat t}{\partial x^i}\right)
\left(\frac{\partial \hat t}{\partial x^j}\right) T^{ij}\,,
\ee
with (\ref{Jacob}), (\ref{uinT}), to obtain
\bea
&&\delta \lag T^{\hat t}_{\ \hat t}\rag_{_R} = -(1- H^2r^2)\,
\delta \lag T^{\hat t \hat t}\rag_{_R}\nn
&&\quad = b H^4 \left[ \frac{1}{r} \frac{d}{dr} \left( r \frac{dv}{dr}\right)\right]
+ \frac{b\,H^4}{(1-H^2r^2)} \left[ \frac{1}{r}\frac{dv}{dr}
- 5H^2 r \frac{dv}{dr}\right]\nn
&& \qquad = - \frac{16\,b\, H^4}{(1-H^2r^2)^2}\,.
\label{divstatT}
\eea
\vspace{-.6cm}

\noindent
Here use has been made of the following identities,
\bes
\bea
&&\frac{\partial \hat t}{\partial x^i} =
\frac{\partial \hat t}{\partial \rho} \frac{\partial \rho}{\partial x^i}
= \frac{\partial \hat t}{\partial \rho} \frac{x_i}{\rho}\,,\\
&& \frac{\partial}{\partial t} = \frac{\partial r}{\partial t}
\frac{\partial}{\partial r} + \frac{\partial \hat t}{\partial t}
\frac{\partial}{\partial \hat t} = Hr\,\frac{d}{dr}\,,\\
&& x_i \partial^i =  x^i \partial_i = \rho \frac{\partial}{\partial\rho} =
\rho \left( \frac{\partial r}{\partial \rho} \frac{\partial}{\partial r}
+ \frac{\partial \hat t}{\partial \rho} \frac{\partial}{\partial \hat t}\right)
= r \frac{d}{dr}\\
&&x_ix_j \partial^i\partial^j = (x^i\partial_i)(x^j\partial_j)
- x^i\partial_i = r \frac{d}{dr}\left(r \frac{d}{dr}\right)
- r\frac{d}{dr}
\eea
\ees
\noindent valid when operating on functions of $r$ only. Likewise we find
\be
\delta \lag T^r_{\ r}\rag_{_R} = \delta \lag T^{\theta}_{\ \theta}\rag_{_R} =
\delta \lag T^{\phi}_{\ \phi}\rag_{_R} = \frac{16\,b\, H^4}{3 (1-H^2r^2)^2}\,,
\label{statT}
\ee
corresponding to a perfect fluid with $p = \rho/3$, in {\it static} coordinates.
The contributions proportional to $b'$ instead of $b$ from the $u$ solutions
are also of the same form.

The form of the stress tensor (\ref{divstatT}), (\ref{statT}) is the form of a finite
temperature fluctuation away from the Hawking-de Sitter temperature $T_{_H} = H/2\pi$
of the Bunch-Davies state in static coordinates \cite{AndHisSam}. Since the equation
the solutions (\ref{psistatic}) satisfy is the same as (\ref{auxeom}), it follows
that there exist linear combinations of the solutions  (\ref{vsoln}) found in
Sec. \ref{sec:statechange} which give (\ref{vstatic}) and the diverging behavior
of the linearized stress tensor on the horizon, corresponding to this global
fluctuation in temperature over the volume enclosed by the de Sitter horizon.
Note that in static coordinates the stress tensor $p= \rho/3$ does not require
averaging over directions of $\vec k$, but a particular {\it coherent} linear
superposition over modes (\ref{vsoln}) with different $\vec k$ in order
to obtain a particular isotropic but spatially inhomogeneous solution of
(\ref{veq}), which selects a preferred origin and corresponding horizon
in static coordinates (\ref{static}). The fluctuations in Hawking-de Sitter
temperature thus preserve an $O(3)$ subgroup of the de Sitter isometry group
$O(4,1)$.

To follow the diverging behavior (\ref{divstatT}), (\ref{statT})  all the way to
the horizon one would clearly require a linear combination of the solutions (\ref{vsoln})
with large Fourier components. However once $8\pi G_{_N}$ times the perturbed stress
tensor in (\ref{divstatT}) becomes of the same order as the classical background Ricci
tensor $H^2$, the linear theory breaks down and non-linear backreaction effects must
be taken into account. This occurs at $r = H^{-1} - \Delta r$ near the horizon, where
\be
\Delta r \sim L_{Pl}\,,
\ee
or because of the line element (\ref{static}) at the proper distance from the horizon of
\be
\ell \sim \sqrt {\frac{L_{Pl}}{H}} \gg L_{Pl}\,.
\label{ellest}
\ee
Thus in the background field calculation, one finds that the stress-energy is relatively
small well inside the horizon. This correspond to a maximum $k_{phys} \sim 1/\ell \ll M_{Pl}$,
where the semi-classical description may still be trusted. At the distance (\ref{ellest})
from the horizon, the state dependent contribution to the stress-energy tensor becomes
comparable to the classical de Sitter background curvature, the linear approximation
breaks down, and non-linear backreaction effects may be expected. This is the same
estimate that was obtained in Refs. \cite{gstar}, where a matching of de Sitter space
within the horizon onto an exterior Schwarzschild spacetime was discussed.

\section{Summary and Conclusions}

In this paper the specific form of the linear response equations in a de Sitter space
background for semi-classical gravity coupled to conformally invariant matter fields
has been derived with the principal result being Eqs. (\ref{auxginv})-(\ref{qginv}).
In terms of a decomposition of the perturbations into scalar, vector, and tensor parts
with regard to spatial hypersurfaces \cite{Bard,Stew}, the analysis has been restricted
to the scalar sector in (\ref{linmet}), and it has been shown explicitly that the
resulting linear response equations can be written in terms of gauge invariant variables
through (\ref{defUpsAC})-(\ref{potphi}).  The quadratic action in terms of these gauge invariant
variables corresponding to the linear response equations is given by Eqs. (6.13) and (6.14).
 A general condition, $\delta R = 0$, has been
utilized which eliminates conformal or trace solutions which vary on the Planck
scale and are thus outside the range of validity of the semi-classical approximation.
This condition is approximate in generic spacetimes but is an exact restriction on
solutions in the maximally symmetric de Sitter spacetime. Remaining under
consideration are only those solutions of the linear response Eqs. (\ref{linresgen})
around de Sitter space which both have zero four dimensional trace, but which are
scalars with respect to the FRW spatial sections. Owing to the freedom to make
linearized coordinate transformations of the background de Sitter space, this leads
to one and only one remaining gauge invariant scalar metric degree of freedom, whose
dynamics is fully described by the $tt$ component of the perturbed semi-classical
Einstein equations in FRW coordinates (\ref{FRW}), namely (\ref{qginv}).

The importance of the trace anomaly and the auxiliary fields used to render it local
have been emphasized in our analysis. Among other things. the linear response analysis
presented here provides an important test for the auxiliary field action. If the auxiliary
field form of the anomaly effective action is used, then the linear response equations
consist entirely of local linear partial differential equations with state dependent
perturbations resulting from variations in the auxiliary fields.  In the linear response
equations derived from the exact effective action there are minor differences associated
with the values of renormalization parameters, along with two more significant differences.
One is that in the anomaly action approach the unspecified state dependent terms allowed
in the general approach take quite specific forms in terms of the scalar auxiliary fields,
which possess their own dynamical equations of motion. The second significant difference
is that the anomaly action takes no account of the conformally invariant term associated
with the non-local term with the kernel $K$.

The solutions to the linear response equations have been investigated in Sec. VII.
Our main results can be divided in two parts, depending upon whether the perturbations
of the stress tensor are driven by changes in the metric (solutions of the first kind),
or additional state dependent perturbations such as those generated by the auxiliary
fields of the anomaly action (solutions of the second kind) are considered. For
perturbations of the first kind, the local linear response equation has been solved
by first ignoring the non-local term involving $K$ and then evaluating the non-local
term using the solutions so obtained. It was found that the non-local term cannot
be neglected for this first class of solutions to the linear response equations,
and this is closely related to the fact that these modes are fundamentally short
distance Planck scale modes, analogous to those already found in flat spacetime,
which in any case lie outside the range of validity of the semi-classical effective
theory.

The only remaining solution to the exact linear response equations for perturbations
of the first kind is the constrained trivial solution $q = 0$. Since only this
trivial solution to the linear response equation survives when the Planck scale solutions
are eliminated, it means that our criterion for the validity of the semi-classical
approximation in de Sitter space is satisfied \cite{AndMolMot}. As originally formulated
this validity criterion applies only to gauge invariant fluctuations of the first kind,
which depend upon the two-point retarded correlation function for the stress tensor,
$\lag [T^a_{\ b}, T^{cd}]\rag$ and thus probe the response of the geometry to the
quantum fluctuations of the stress-energy tensor about its mean value.

For state dependent perturbations of the second kind the situation is much more
interesting. A new class of infrared modes has been found in the tracefree but
scalar sector, given by Eqs. (\ref{vsoln})-(\ref{varTttq}) and (\ref{TkFRW}),
which are associated with the scalar auxiliary fields of the anomaly effective
action. Since these modes vary not on the Planck scale but on arbitrarily large
scales which can be comparable to the cosmological horizon scale in de Sitter space,
we have termed them {\it cosmological horizon modes.} We emphasize that the existence
of these scalar modes is a necessary consequence of the one-loop effective action
and trace anomaly of quantum conformal matter or radiation. Thus the conformal
anomaly provides scalar degrees of freedom in cosmology without the {\it ad hoc}
introduction of an inflaton.

These scalar cosmological horizon modes satisfy the full linear response equations
to leading order in $G_{_N} H^2$. In particular, the non-local term is always
negligible for these modes, provided the semi-classical evolution is begun with
modes whose physical wavelength is much larger than the Planck length $L_{Pl}$.
Thus the simple analytic solutions to the equations resulting from the anomaly
action (\ref{vsoln}) give excellent approximations to the exact solutions of the
linear response equations around de Sitter space to first order in
$H^2 L_{Pl}^2 \ll 1$.

Although it is not the full quantum effective action obtained
by integrating out the conformal matter or radiation fields exactly, the difference
is a non-local term which is important for the Planck scale solutions of the first
kind only. Ignoring this non-local term, the auxiliary field action gives the correct
linear response equations valid on low energy scales far removed from the UV Planck
scale, {\it i.e.} in the infrared  limit where the semi-classical theory can be
trusted, and for the second class of state dependent solutions. This is consistent
with the general arguments that the anomaly action should capture the main infrared
or macroscopic quantum effects discussed in earlier work \cite{MazMot,MotVau}.
The fact that there is agreement with the local part of the perturbed stress-energy
tensor and the infrared relevant terms in a context where the results could be
compared to exact calculations from entirely different methods provides a good
indication of the usefulness of the anomaly action as a tool for investigating
infrared quantum effects due to conformally invariant fields in other contexts
where exact results may not be readily available.

In the FRW coordinates the cosmological horizon modes for fixed comoving wavenumber
$k$ redshift away at late times like $a^{-4}$. The linear homogeneous equation these modes
obey can also be considered in the static coordinates (\ref{static}) of de Sitter
spacetime. By recognizing the connection between these state dependent solutions to
the linear response equations and the solutions to the same auxiliary field equations
found from the background field analysis in de Sitter space in static coordinates
in Ref. \cite{MotVau}, we have found that the cosmological horizon modes
can describe state dependent changes on the horizon scale with a large
or even divergent stress tensor at $r = H^{-1}$ (with respect to an arbitrarily
chosen origin in de Sitter space). This may be understood as a thermal fluctuation
of temperature of the underlying conformal quantum field away from its value
in the Bunch-Davies state of $T_{_H} = H/2\pi$. If the temperature differs
from $T_{_H}$ by even a small amount, the corresponding stress tensor diverges
on the horizon \cite{AndHisSam}.

A diverging stress tensor on the cosmological horizon signals the breakdown of
the linear approximation and the necessity to consider non-linear backreaction
effects in the vicinity of the horizon. Thus our analysis of the cosmological
horizon modes and associated stress tensor perturbations in the static coordinates
suggests these infrared modes allow the temperature of the causal region interior
to the de Sitter horizon to fluctuate away from the BD value, leading to large
non-FRW fluctuations of the geometry in the vicinity of the cosmological horizon.

\vskip .7cm
\centerline{{\it Acknowledgments}}
\vskip .4cm

We, and especially E. M. are very grateful to many enlightening discussions with
A. Roura during the course of this work  We would also like to thank Bei-Lok Hu
for helpful conversations, and A. Starobinsky for reading a preliminary draft of
the manuscript and giving us his comments.  P.R.A. would like to thank J. Hartle for
a helpful conversation and access to some of his unpublished calculations. P.R.A.
acknowledges financial support from the Spanish Ministerio de Educaci\'on y Ciencia and
from the National Science Foundation under Grant Nos. PHY-0556292 and PHY-0856050.
Numerical computations were performed on the Wake Forest University DEAC Cluster with
support from an IBM SUR grant and the Wake Forest University IS Department. Computational
results were supported by storage hardware awarded to Wake Forest University through
an IBM SUR grant.

\appendix

\section{The Logarithmic Distribution in Flat Space}

There are several ways to work with the distribution $K$ defined by (\ref{Kt})
which lead to similar results. The distribution has the same form in a FRW
background in terms of the conformal time $\eta$ as it does in Minkowski
spacetime in terms of the usual time coordinate $t$. In terms of the latter
the distribution is
\be
K (t-t'; k; \mu) = \int_{-\infty}^{\infty}\frac{d\omega}{2\pi}\,
e^{-i\omega (t-t')} \,\ln\left[\frac{ -\omega^2 + k^2 - i\epsilon\,
{\rm sgn}\,\omega} {\mu^2}\right]\,.
\label{KtA}
\ee
As it stands this distribution is well-defined for $t \neq t'$, but
is undefined for $t=t'$. Two different approaches may be considered.
In one approach, which owes its origins to Hadamard \cite{Had,Schw}, an
appropriate strictly local distribution is added to (\ref{KtA})
to extend the definition to functions with non-vanishing support
at $t=t'$. We shall also describe a second approach, more akin to the
Pauli-Villars regulator method, which modifies the distribution
(\ref{KtA}) at large frequencies and allows for a smooth limit
at $t=t'$.

To begin, a mass $m$ can be introduced into the argument of the logarithm
in (\ref{KtA}) by replacing $k^2$ by $\omega^2_k = k^2 + m^2$.
Then differentiating with respect to $m^2$ gives:
\be
\frac{\partial}{\partial m^2} K (t-t'; \omega_k; \mu) =
\int_{-\infty}^{\infty} \frac{d\omega}{2\pi}\,e^{-i\omega (t-t')}
\frac{1}{-\omega^2 + {k}^2 + m^2 - i\epsilon\, {\rm sgn}\,\omega}
= \frac{\sin\left[\omega_k (t-t')\right]}{\omega_k}
\ \theta(t-t')\,.
\label{diffm}
\ee
This may be recognized as the spatial Fourier transform of the usual
retarded Green's function for a scalar field with mass $m$. For $t' > t$
the retarded Green's function vanishes and therefore is explicitly causal.
For $t=t'$ the distribution is also defined and vanishes at that point. Note
also that after differentiation with respect to $m^2$, there is no longer any $\mu$
dependence in (\ref{diffm}). The reason for this is that the $\mu$ dependence of
(\ref{KtA}) enters only through the purely local contribution,
\be
\mu \frac{d}{d\mu}\, K(t-t'; \omega_k ; \mu) =
- \int_{-\infty}^{\infty}\frac{d\omega}{\pi}\,
e^{-i\omega (t-t')} = - 2 \,\delta (t-t')\,,
\label{Kmu}
\ee
which is independent of $k$ or $m$.

If (\ref{diffm}) is integrated with respect to $dm^2 = d\omega^2_k =
2 \omega_k d \omega_k$ at fixed $k$, then one obtains
\be
K (t-t'; \omega_k; \mu) = K_{bare}(t-t'; \omega_k) + K_{local}(t-t'; \mu)\,,
\label{Kdecomp}
\ee
where
\be
K_{bare} (t-t'; \omega_k) = -\frac{2}{t-t'}\,
\cos \left[\omega_k (t-t')\right] \,
\theta(t-t')
\label{Kbare}
\ee
This determines the causal $K$ distribution for all $t' < t$, up to an
arbitrary local contribution $K_{local}(t-t';\mu)$, which is independent of
$\omega_k$ and satisfies (\ref{Kmu}). In the form (\ref{Kdecomp}) one can
set $m=0$ and $\omega_k = k$ to recover the distribution (\ref{Kt}).

The need for the local contribution $K_{local}$ in (\ref{Kdecomp}) appears
when $K$ is integrated against smooth test functions $f(t')$ which are
non-vanishing at $t'=t$, for in that case,
\be
\int^{t- \xi}_{-\infty}\, dt' K_{bare}(t-t'; k) f(t') =
-2 \int^{t- \xi}_{-\infty}\, dt' \,\frac{\cos [k (t-t')]}{t-t'} f(t')
\label{Kbareint}
\ee
diverges logarithmically at its upper endpoint as $\xi \rightarrow 0$.
This logarithmic divergence can be removed if the $\mu$ dependent
local distribution satisfying (\ref{Kmu}) is taken to be
\be
K_{local} (t-t' ; \mu) = -2 [\ln (\mu\xi) + C]\, \delta(t-t')
\label{Klocal}
\ee
and the limit $\xi \rightarrow 0$ is taken after summing $K_{bare} + K_{local}$.
Here $C = 0.577215...$ is Euler's constant which is a finite additional
term added for reason of normalization that will become clear shortly.
Then one can define the action of the distribution $K$ by the one
parameter sequence,
\be
K(t-t'; k ; \mu) \sim - 2\left\{
[\ln (\mu\xi) + C]\, \delta(t-t') + \frac{1}{t-t'}\,
\cos \left[k(t-t')\right]\,\theta(t-t'-\xi) \right\}_{\xi \rightarrow 0}\,,
\label{Kpf}
\ee
as $\xi \rightarrow 0$. The meaning of the symbol $\sim $ in (\ref{Kpf}) is
that acting upon an arbitrary smooth function $f(t')$,
\be
\int_{-\infty}^{t}\,dt'\, K(t-t'; k ; \mu)\, f(t')
\equiv  -2 \lim_{\xi \rightarrow 0}
\left\{ \left[ \ln (\mu \xi) + C\right] f(t) + \int_{-\infty}^{t-\xi} \,
\frac{dt'}{t-t'} \, \cos \left[k(t-t')\right]\, f(t') \right\}\,.
\label{Partf}
\ee
that is, the limit $\xi \rightarrow 0$ is to be taken after the integral over
$t'$ of a member of a class of suitably smooth, test functions $f(t')$ has been
computed, provided the integral in (\ref{Partf}) converges.

We note that the definition (\ref{Partf}) is not unique since an arbitrary
local distribution $K_{local} (t-t'; \mu)$ with the correct dimensions and
the property (\ref{Kmu}) could have been added to $K_{bare}$. The definition
(\ref{Partf}) is in fact an adaptation of Hadamard's {\it Partie finie}
definition of distributions of this kind \cite{Had}. Since
\be
\frac{d}{d \xi} \int_{-\infty}^{t-\xi} \, \frac{dt'}{t-t'} \,
\cos \left[k(t-t')\right]\,
f(t') = -\frac{\cos (k\xi)}{\xi}\, f(t-\xi) = - \frac{f(t)}{\xi} + f'(t)
+ {\cal O}(\xi)\,,
\ee
for test functions which are continuous and differentiable at $t'=t$,
it is clear that the integral's logarithmic dependence on $\xi$ is just cancelled
by the $ \ln (\mu \xi)$ term, and the limit $\xi \rightarrow 0$ of the sum in
(\ref{Partf}) is finite.

This finite limit in (\ref{Partf}) can be demonstrated directly by making use
of the cosine integral function,
\be
{\rm ci} (z)=  \int_{-\infty}^{-z} \,dx\, \frac{\cos x}{x} = C + \ln z
+ \sum_{n=1}^{\infty} (-)^n\frac{z^{2n}}{2n\cdot (2n)!}\,,\quad z>0\,,
\label{cosint}
\ee
to integrate (\ref{Partf}) by parts. Since
\be
\frac{d}{dt'} {\rm ci} \left[\omega_k(t-t')\right] =
- \frac{\cos\left[\omega_k(t-t')\right]}{t-t'}
\ee
we have
\bea
&&\int_{-\infty}^{t}\,dt'\, K(t-t'; k ; \mu)\, f(t') \nn
&& = -2 \lim_{\xi \rightarrow 0} \left\{ \left[ \ln (\mu \xi) + C\right] f(t)
-{\rm ci}(k\xi) f(t-\xi)
+ \int_{-\infty}^{t-\xi} dt' \,{\rm ci} \left[k(t-t')\right]\,
\frac{df}{dt'} \right\}\nn
&&= -2\int_{-\infty}^t\, dt'\, {\rm ci} \left[k(t-t')\right]\,
\frac{df}{dt'}
\, +\,  2\, f(t)\,\ln \left(\frac{k}{\mu}\right)\,,
\label{Kpart}
\eea
provided that $f(t')$ is differentiable, and
\[\lim_{t' \rightarrow -\infty} \{{\rm ci}\left[k(t-t')\right] f(t')\} = 0 \,. \]

The definition (\ref{Partf}) of the finite part of the logarithmically divergent
distribution is closely related to dimensional regularization and the subtraction
of simple poles at $n=4$, corresponding to logarithmic divergences of Feynman
integrals. As in dimensional regularization and subtraction, the introduction
of some large but otherwise arbitrary mass scale $\mu$ is required for dimensional
reasons. The remaining dependence on the arbitrary renormalization scale $\mu$
in the full linear response equations is removed by (\ref{muindep}), which amounts
to the freedom to shift the $\mu$ dependence into the unknown coefficient of
the local Weyl squared term in the full effective action and the coefficient
of the local tensor $A_{ab}$ in the semiclassical linear response equations.

It is instructive to apply (\ref{Partf}) or (\ref{Kpart}) to a simple example
of a suitable smooth test function, namely,
\be
f(t) = e^{\gamma t}\,,\qquad \gamma > 0\,,
\label{fexp}
\ee
for which the integral (\ref{Partf}) converges, and may be computed analytically.
Indeed,
\be
-2\int_{-\infty}^{t-\xi} \, \frac{dt'}{t-t'}\, \cos[\omega_k(t-t')]\,e^{\gamma t'}
= e^{\gamma t}\, {\rm Ei}\left[-(\gamma +i\omega_k)\xi\right] + e^{\gamma t}\,
{\rm Ei} \left[-(\gamma -i\omega_k)\xi\right]\,,
\label{regint}
\ee
where the exponential integral function for complex arguments, Ei$(z)$ is defined
by the analytic continuation of
\be
{\rm Ei}(z) =  \int_{-\infty}^z\,dx\,\frac{e^x}{x}  =  C + \ln(-z)
+ \sum_{n=1}^{\infty} \frac{z^n}{n \cdot n!}\,,
\ee
for negative real $z$. Using this in (\ref{regint}) and (\ref{Partf})
for $\xi \rightarrow 0$ gives
\be
\int_{-\infty}^t\, dt'\, K (t-t';k ;\mu)\, e^{\gamma t'} = e^{\gamma t} \,
\ln\left(\frac{\gamma^2 + {k}^2}{\mu^2}\right)\,,
\label{egH}
\ee
in which the dependence on Euler's constant has cancelled. The
argument of the logarithm is exactly $K^2 /\mu^2$ for the function (\ref{fexp}),
in which the frequency $\omega^2$ is replaced by $-\gamma^2$. This is the reason for
the introduction of Euler's constant into the definition (\ref{Kpf}).
In terms of a Fourier transform in time, this just amounts to the requirement
that  the $\mu$ introduced in (\ref{Klocal}) be identical to the $\mu$ in the
Fourier  transform (\ref{Kt}). The fact that (\ref{egH}) is strictly proportional
to $f(t)$ is a result of the simple Fourier transform of (\ref{fexp}),
and will not hold in the case of general $f(t)$.

Note that if $\gamma  > 0$, (\ref{egH}) possesses a smooth limit as $k \rightarrow 0$,
and indeed the finiteness of this limit can be demonstrated from (\ref{Kpart}) for a class
of test functions, satisfying
\be
\lim_{t' \rightarrow -\infty} \left\{\ln\left[\mu (t-t')\right] f(t') \right\}=  0\,,
\label{logcond}
\ee
which excludes $f(t) = const.$, {\it i.e.} $\gamma = 0$, in which case (\ref{egH})
diverges logarithmically for $k =0$. By repeating the steps leading from
(\ref{Partf}) to (\ref{Kpart}) one obtains for $k =0$,
\bea
&&\int_{-\infty}^t\, dt'\, K (t-t';k = 0 ;\mu)\, f(t') =
-2 \int_{-\infty}^t\, dt'\, \ln\left[\mu (t-t')\right] \, \frac{df}{dt'}
-2  C f(t)\nn
&& \quad = -2 \int_{-\infty}^t\, dt'\,  \ln\left[\mu e^C (t-t')\right] \,
\frac{df}{dt'}\,,
\label{Pfk0}
\eea
for differentiable functions satisfying (\ref{logcond}).

In this treatment of the logarithmic distribution $K$ it has been assumed that
the lower limit of the integration over $t'$ has been extended to $-\infty$.
However, for the application to linear response with a well defined initial
perturbation at a given finite time $t_0$, this definition requires modification.
The need for a modification can be seen from the integration by parts needed to
pass from (\ref{Partf}) to (\ref{Kpart}). If the lower limit of the integral is
replaced by $t_0$, there will be an additional contribution proportional to
${\rm ci} \left[k (t-t_0) \right]f(t_0)$. From the definition (\ref{cosint})
this vanishes as $t_0 \rightarrow -\infty$ provided that $f(t_0)/t_0$ vanishes
in this limit. However with $t_0$ fixed and finite, this surface contribution
diverges logarithmically as $t \rightarrow t_0$.

This problem with finite $t_0$ and indeed the previous divergence as
$t' \rightarrow t$ are encountered because the non-local kernel has been
treated in isolation from the other components of the linear response equation,
which must be solved self-consistently. In the self-consistent solution the
Fourier transform has zero support for $|\omega| \rightarrow \infty$ and the
full linear response equation contains no divergences for finite initial
conditions. Since the very large $|\omega|$ components will drop out in any
case, one may consider the associated distribution,
\be
K_{PV}(t-t'; \omega_k ; M) \equiv \int_{-\infty}^{\infty}\frac{d\omega}{2\pi}\,
e^{-i\omega (t-t')} \ln\left[\frac{ -\omega^2 + k^2 + m^2 - i\epsilon\, {\rm sgn}\,\omega}
{ -\omega^2 + {k}^2 + M^2 - i\epsilon\, {\rm sgn}\,\omega}\right]\,,
\label{defKPV}
\ee
in which the contributions at large $|\omega|$ are subtracted against an
equivalent distribution for a field with large Pauli-Villars mass $M$.
Comparison with (\ref{Kpf}) shows that when $M \gg \omega_k$ and $K$ is integrated
against smooth test functions $f(t')$ with bounded Fourier components possessing
vanishing support for $\omega \gg M$, the result should become indistinguishable
from that of the original distribution in (\ref{lintena}), with $M$ playing the
role of $\mu$ in (\ref{Kpf}).

The advantage of the explicitly subtracted Pauli-Villars definition (\ref{defKPV})
is that its ultra high frequency components are suppressed, with (\ref{defKPV})
vanishing for $|\omega| \gg M$. This leads to the integral being well
defined for all $t-t'$ with no purely local $\delta (t-t')$ contribution
of the form (\ref{Kmu}). Indeed the integration of (\ref{diffm}) between
$m$ and $M$ now gives
\be
K_{PV}(t-t'; \omega_k ; M) = -\frac{2}{t-t'} \left\{\cos [\omega_k (t-t')]
- \cos [\Omega_k (t-t')]\right\}\, \theta(t-t')\,,
\label{KPV}
\ee
with $\Omega_k \equiv \sqrt{k^2 + M^2}$, instead of (\ref{Kdecomp}).
Unlike (\ref{Kdecomp}) this function is strictly non-zero only for
$t' < t$, vanishing identically at $t'=t$. It is clear that if $M$ is
large, the second cosine in (\ref{KPV}) oscillates very rapidly and can
give a significant contribution only in the short time interval
$t - t' \lesssim 1/M$. Thus, the Pauli-Villars regulator plays a significant
role only in this very narrow interval of time where $t' \rightarrow t$,
making $K_{PV}$ essentially equivalent to $K_{bare}$ outside the region
$t - t' \lesssim 1/M$. Although there is no strictly local $\delta(t-t')$
contribution in the Pauli-Villars method, $K_{PV}$ has a {\it quasi-local}
contribution in the interval $t - t' \lesssim 1/M$. In this narrow
interval the role of the second contribution in (\ref{KPV}) is crucial since
it removes the logarithmic divergence of $K_{bare}$, and the difference,
$K_{PV}$ vanishes identically at $t=t'$. Hence the integral over test
functions $f(t')$ with non-vanishing support at $t'=t$ becomes well-defined.

For $K_{PV}$ there is no problem with an initial value formulation and we
may define the action of the retarded Pauli-Villars kernel on test functions
in the finite interval $[t_0, t]$ by
\bea
K_{PV}\circ f &\equiv& \int_{t_0}^t \,dt'\, K_{PV}(t-t'; k ; M)\, f(t')\nn
 &=& -2\,\int_{t_0}^t \,\frac{dt'}{t-t'}\,\left\{\cos [k (t-t')] -
\cos [\Omega_k (t-t')]\right\}\, f(t')\nn
&=& -2\,\int_{-\infty}^t \,\frac{dt'}{t-t'}\,\left\{\cos [k (t-t')] -
\cos [\Omega_k (t-t')]\right\}\, f(t')\nn
&& \qquad +2\,\int_{-\infty}^{t_0} \,\frac{dt'}{t-t'}\,\left\{\cos [k (t-t')]
- \cos [\Omega_k (t-t')]\right\}\, f(t')
\label{KPVf}
\eea
which is completely finite at both its upper and lower limits for suitably
smooth test functions $f(t)$. In the last form one may recognize the $t_0$
dependence as giving rise to a temporal transient that falls off for
$k (t-t_0) \gg 1$.

In order to compare the Pauli-Villars definition of the integral kernel (\ref{KPV})
with the finite part prescription (\ref{Partf}),  (\ref{KPV}) may be applied
to the same test function (\ref{fexp}). Taking first the infinite
range ($t_0 \rightarrow -\infty$), one finds
\be
\int_{-\infty}^t\, dt'\, K_{PV}(t-t';k;M) e^{\gamma t'} =  e^{\gamma t}
\ln\left(\frac{\gamma ^2 + {k}^2}{\gamma ^2 + {k}^2 + M^2}\right)\,.
\label{egKPV}
\ee
If $M^2 \gg \gamma ^2 + {k}^2$ this result can be expanded as
\be
\int_{-\infty}^t\, dt'\, K_{PV}(t-t';k; M) e^{\gamma t'} = e^{\gamma t}\left\{
\ln\left(\frac{\gamma ^2 + {k}^2}{M^2}\right) - \left(\frac{\gamma ^2 + {k}^2}{M^2}\right)
+ \frac{1}{2} \left(\frac{\gamma ^2 + {k}^2}{M^2}\right)^2 + \dots \right\}\,.
\label{egPV}
\ee
Thus in the limit of large $M$ the Pauli-Villars regularization
of the non-local term becomes equivalent to the definition (\ref{Partf}),
with the identification, $M \rightarrow \mu$, at least as far as the logarithmic
dependence upon $M$ is concerned.  This shows that
the arbitrary mass scale $\mu$ of (\ref{Partf}) is equivalent to an
ultraviolet cutoff scale $M$, and the additional terms of the expansion
in (\ref{egPV}) may be recognized in position space as an expansion of higher
derivative local terms $(-\sq/M^2)^n$ in ascending inverse powers of the UV
cutoff $M^2$, typical of a low energy expansion of an effective field theory.
These finite terms may be neglected for sufficiently large $M^2$, and
are absent entirely in the Hadamard finite part regularization (\ref{Partf}).

For finite $t_0$ the integral formula,
\be
2 \int_{-\infty}^{t_0}\, \frac{dt'}{t-t'} \, \cos\left[\omega_k(t-t')\right] \, e^{\gamma t'}
= - e^{\gamma t} \Big\{ {\rm Ei} \big[-(\gamma + i\omega_k)(t-t_0)\big] +
{\rm Ei} \big[-(\gamma - i\omega_k)(t-t_0)\big]\Big\}\,,
\ee
can be used in (\ref{KPVf}) to obtain
\bea
&&\int_{t_0}^t \,dt'\, K_{PV}(t-t'; k ; M)\, e^{\gamma t'}
= e^{\gamma t}\left\{ \ln\left(\frac{\gamma ^2 + {k}^2}{M^2}\right) \right.\nn
&& \qquad \bigg.- {\rm Ei} \left[-(\gamma + i k)(t-t_0)\right]
- {\rm Ei} \left[-(\gamma - i k)(t-t_0)\right]
+ 2\,{\rm ci}\big[M(t-t_0)\big] \bigg\} + \cdots
\label{initgam}
\eea
where all terms which vanish for $M \gg (\gamma, k)$ have been dropped. This expression
is completely finite as $t \rightarrow t_0$, and in fact vanishes in that limit. Hence
there are no unphysical divergences at the arbitrary initial time $t_0$. This is due to
the fact that unlike (\ref{KtA}) which requires a careful limiting procedure (\ref{Partf})
to remove its short time divergence as $t' \rightarrow t$, the very high frequency response
of (\ref{defKPV}) is suppressed at all times, {\it c.f.} (\ref{egKPV}) for $k \rightarrow \infty$.
Moreover, since Ei$(z) \rightarrow e^z/z$ for large $|z|$, the last three initial state
dependent terms in the bracket of (\ref{initgam}) become
\be
- \frac{2e^{-\gamma(t-t_0)}}{t-t_0} \,\frac{\left[ \gamma \cos \left(k(t-t_0)\right)
- k\sin\left(k(t-t_0)\right)\right]}
{\gamma^2 + k^2} + 2\, \frac{\sin \left(M(t-t_0)\right)}{M(t-t_0)}\,.
\label{tran}
\ee
The first term decays exponentially for $t-t_0 \gg \gamma^{-1}$, while the
latter falls only as a power times a rapidly oscillating function. Thus, at late times,
\be
\int_{t_0}^t \,dt'\, K_{PV}(t-t'; k ; M)\, e^{\gamma t'}
\rightarrow  e^{\gamma t} \ln\left(\frac{\gamma ^2 + {k}^2}{M^2}\right) \,,
\quad {\rm for} \  M \gg \gamma, k\,,
\label{PVgam}
\ee
which agrees with (\ref{egH}). Even for constant functions with $\gamma =0$, the initial
state dependent transient terms in (\ref{tran}) fall off linearly with large $t-t_0$, and
(\ref{PVgam}) approaches the Hadamard form (\ref{egH}) at late times, albeit more slowly.
This example shows that up to terms that vanish for large $M$, at late times or when
operating on continuous and differentiable functions $f(t)$ with bounded first derivative
at all times, the Pauli-Villars regularization of the logarithmic kernel becomes identical
to the Hadamard {\it Partie finie} definition, with $M \leftrightarrow \mu$.

In a different prescription, one could also consider the Hadamard definition of the
logarithmic kernel  for finite $t_0$ by replacing the lower time limit by $t_0$,
{\it after} the integration by parts in (\ref{Kpart}), {\it i.e.}
\be
\int_{t_0}^{t}\,dt'\, K(t-t'; k ; \mu)\, f(t')
= -2\int_{t_0}^t\, dt'\, {\rm ci} \left[k(t-t')\right]\, \frac{df}{dt'}
\, +\,  2\, f(t)\,\ln \left(\frac{k}{\mu}\right)\,.
\label{Kpart0}
\ee
Computing this for the same test function (\ref{fexp}) gives
\bea
&&\int_{t_0}^t \,dt'\, K(t-t'; k ; \mu)\, e^{\gamma t'}
= e^{\gamma t}\left\{ \ln\left(\frac{\gamma ^2 + {k}^2}{\mu^2}\right) \right.\nn
&& \quad \bigg.- {\rm Ei} \left[-(\gamma + i k)(t-t_0)\right]
- {\rm Ei} \left[-(\gamma - i k)(t-t_0)\right]
+ 2\,e^{-\gamma (t-t_0)}\,{\rm ci}\big[k(t-t_0)\big] \bigg\}\,.
\label{initgamH}
\eea
This is similar to the corresponding Pauli-Villars form (\ref{initgam}), with $\mu$
replacing $M$ and no $1/M$ terms, and is also finite as $t\rightarrow t_0$. It differs
from (\ref{initgam}) only in the last transient term, which falls exponentially for
$\gamma > 0$ in (\ref{initgamH}) but is given by the last term of (\ref{tran}) in
the Pauli-Villars method.

Unlike (\ref{egH}) for $k = 0$ and $\gamma = 0$, the result (\ref{initgam}) remains
finite, becoming
\be
\int_{t_0}^t \,dt'\, K_{PV}(t-t'; k = 0; M)
= - 2 \ln\left[M(t-t_0)\right] + 2\,{\rm ci}\big[M(t-t_0)\big] - 2 C\,.
\label{initgam0}
\ee
For times $t-t_0 \gg 1/M$, the logarithm dominates and (\ref{initgam0}) grows without
bound. Hence the logarithmic divergence of (\ref{Partf}) or (\ref{initgam0}) for
$k = \gamma = 0$ at all times is replaced by a logarithmic growth in time for
the Pauli-Villars definition of the distribution (\ref{defKPV}).

To see how the logarithmic kernel behaves in position space, one can return to the
unregulated form of the distribution (\ref{Kbare}) which is valid either if $t' < t$,
or if $K$ is integrated against test functions which vanish at $t=t'$.
Its form in position space is found to be
\be
K_{bare}(t, \vec x ) = \int \frac{d^3 \vec{k}}{(2\pi)^3}\, e^{i k \cdot \vec x}
\left\{- \frac{2}{t} \ \theta(t)\ \cos (kt)\right\}\,.
\ee
The integral over $\vec{k}$ can be performed by going to polar coordinates,
and integrating over the angles, with the result,
\bea
K_{bare}(t, \vec x ) &&= - \frac{2\theta(t)}{(2\pi)^2\,t}\ \int_0^{\infty}
k^2 dk \left(\frac{e^{ikr} - e^{-ikr}}{ikr}\right)\, \cos(kt)\nonumber\\
&&= \frac{1}{4\pi^2} \frac{\theta (t)}{tr} \frac{\partial}{\partial r}
\int_0^{\infty}\,dk\, \left[e^{ik(t+r)} + e^{-ik(t+r)} + e^{ik(t-r)} + e^{-ik(t-r)}\right]
\nonumber\\
&&=  \frac{1}{2\pi} \frac{\theta (t)}{tr} \frac{\partial}{\partial r}
\left[ \delta (t+r) + \delta (t-r)\right]\,, \quad t > 0, \ \ r > 0\,.
\label{derivdel}
\eea
Because of $\theta (t)$, the argument of the first delta function cannot vanish,
except possibly at $t=0$ and $r=0$, where the bare distribution $K_{bare}$ is
undefined. Hence the first delta function in (\ref{derivdel}) can be dropped if one
integrates against test functions which vanish at $t=t'$ and $\vec x = \vec x'$.

Reinstating $t-t'$ and $\vec x - \vec x'$, the effect of the
distribution on a test function,
\be
K_{bare}\circ f \equiv \int dt' \int d^3 \vec x'\,
K_{bare}(t-t', \vec x - \vec x' )\, f(t', \vec x')\,.
\ee
can be computed by first going to polar coordinates centered at $\vec x$,
\be
\vec x' = \vec x + \hat n r\,,
\ee
for some unit vector $\hat{n}$, and then introducing the standard retarded and
advanced time coordinates with respect to $(t, \vec x)$,
\bes
\bea
&&u= t'- t - r \,,  \\
&&v= t'- t + r \,.
\eea
\ees

\vspace{-.6cm}
\noindent
In these coordinates
\bea
K_{bare}\circ f &&= \int_{-\infty}^{\infty} dt' \int_0^{\infty} r^2 dr
\int \frac{d\Omega_{\hat n}} {2\pi} \,  \frac{\theta(t-t')}{(t-t')\,r}\
\frac{\partial}{\partial r}\delta(t'-t + r)\ f \nn
&&=  \int \frac{d\Omega_{\hat n}}{4\pi} \int_{-\infty}^{-\epsilon} du \,
\int_{-\infty}^{\infty} dv\,
\left[\frac {u-v}{u+v} \right] \frac{d\delta (v)}{d v}\ f \nn
&&=  -\int \frac{d\Omega_{\hat n}}{4\pi} \int_{-\infty}^{-\epsilon} du \,
\frac{\partial}{\partial v}\left[\frac {u-v}{u+v}\, f\right]_{v=0} \nn
&&=  -\int \frac{d\Omega_{\hat n}}{4\pi} \int_{-\infty}^{-\epsilon} du \,
\left[ -\frac{2f}{u} + \frac{\partial f}{\partial v}\right]_{v=0} \,.
\label{Kbareuv}
\eea
The integral over $\int^{-\epsilon}_{-\infty} \frac{du}{u}f$ diverges logarithmically
as $\epsilon \rightarrow 0^+$. Thus it may be handled by the same {\it Partie finie}
prescription as in (\ref{Partf}). In order to determine the local contribution at
$t'=t$, $\vec x' = \vec x$, a similar method can be used as that leading to (\ref{Pfk0})
for fixed $k =0$. As in (\ref{logcond}) note that one may integrate (\ref{Kbareuv}) by
parts and drop the lower limit surface term for functions $f$ for which
\be
\lim_{u \rightarrow -\infty}\left\{\ln (-\mu u) f(u, v=0)\right\} = 0\,,
\label{ulogcond}
\ee
which excludes spacetime constant functions. Then we may take the limit
$\epsilon \rightarrow 0^+$ and obtain the finite result \cite{Met},
\be
K \circ f = -\int \frac{d\Omega_{\hat n}}{2\pi} \int_{-\infty}^0 du \,
\left[\ln\left(-u/\lambda\right) \frac{\partial f }{\partial u} +
\frac{1}{2} \frac{\partial f}{\partial v}\right]_{v=0}\,,
\label{distpos}
\ee
for some constant $\lambda$. Taking into account a relative normalization factor
of $-1/2\pi$, (\ref{distpos}) agrees with Eq. (20) of Ref. \cite{Hor}, where the
notation $H_{\lambda}$ was used for the distribution $K$.  See also Ref.\cite{jordan2}

In (\ref{distpos}), $f$ is to be viewed as a function of $u, v$ and $\hat n$ according to
\be
f(t', \vec x') = f \left(t + \frac{u+v}{2}, \vec x + \hat n \frac{v-u}{2} \right) \,.
\ee
If one takes the previous example of
\be
f(t', \vec x') = e^{\gamma t'}  = e^{\gamma t + \gamma (u+v)/2}\,,
\ee
and substitutes this into (\ref{distpos}), using $\int_0^{\infty} dx\, e^{-x} \ln x = -C$,
one finds $2 e^{\gamma t} \ln (\gamma/\mu)$, in agreement with (\ref{egH}) for $k = 0$,
provided
\be
\frac{1}{\lambda} = \frac{\mu}{2}e^{C - \frac{1}{2}}\,.
\ee
Hence,
\bea
K \circ f &\equiv& \int dt'\int d^3\vec x'\, K(t-t'; \vec x-\vec x'; \mu) f(t',\vec x') \nn
&& \qquad=  -\int \frac{d\Omega_{\hat n}}{4\pi} \int_{-\infty}^0 du \,
\left[\ln\left(\frac{\mu^2 e^{2C} u^2}{4e}\right) \frac{\partial f }{\partial u} +
\frac{\partial f}{\partial v}\right]_{v=0}\,,
\label{KcF}
\eea
for differentiable functions satisfying (\ref{ulogcond}).

Unlike (\ref{Partf}), the definition (\ref{distpos}) or (\ref{KcF}) is Lorentz invariant.
In the Pauli-Villars regulated distribution (\ref{defKPV}), which is also Lorentz
invariant, the time limits are arbitrary and the lower limit can be taken to be finite.
Hence condition (\ref{ulogcond}) may be relaxed. In the definition (\ref{KcF}) there is
no {\it a priori} justification for replacing the lower limit with a finite time boundary.
Hence, although equivalent when evaluated on functions possessing vanishing support as
$|\omega| \rightarrow \infty$, which is the physically relevant case, the Pauli-Villars
subtracted definition (\ref{defKPV}) is both well defined over a wider class
of test functions, and better suited to a finite initial value formulation of linear response.

\section{The Variations in de Sitter Space}

In a conformally flat spacetime with $C^a_{\ bcd} = 0$, and in the conformal ``vacuum"
state for which the background auxiliary fields have the values $\bar \psi = 0$,
$\bar\varphi = 2 \ln \Omega$, the variation of the $F^a_{\ b}$ tensor~\eqref{Fab}
is given by:
\bea
&&\delta F^a_{\ b} = -2 (\nabla^{(a} \bar\varphi)\,(\nabla_{b)} \sq \psi)
-2 (\nabla^{(a} \psi)\,(\nabla_{b)} \sq \bar\varphi)
+ 2 \nabla^c\left[(\nabla_c \psi) (\nabla^a\nabla_b\bar\varphi)
+  (\nabla_c \bar\varphi) (\nabla^a\nabla_b\psi) \right]
\nn
&&-\frac{4}{3}\, \nabla^a\nabla_b\left[(\nabla^c \bar\varphi) (\nabla_c\psi)\right]
+ \frac{4}{3}\, R^a_{\ b} (\nabla_c \bar\varphi) (\nabla^c \psi)
- 4 R^{c(a}\left[ (\nabla_{b)} \bar\varphi) (\nabla_c \psi)
+ (\nabla_{b)} \psi) (\nabla_c \bar\varphi)\right] \nn
&&+ \frac{4}{3} R (\nabla_{a)} \bar\varphi) (\nabla_{b)} \psi)
- \delta^a_{\ b} (\sq \bar\varphi)(\sq \psi)
+ \frac{1}{3} \delta^a_{\ b} \sq \left[(\nabla^c\bar\varphi)(\nabla_c\psi)\right]
+ 2 \delta^a_{\ b} \left(R^{cd}
- \frac{R}{3} g^{cd} \right) (\nabla_c \bar\varphi) (\nabla_d \psi)\nn
&& -4 \nabla_c\nabla_d\,(\delta C^{(ac\ d}_{\ \ \ b)}\,\bar\varphi)
-2 R_{cd}\,\bar\varphi\, \delta C^{\ c\ d}_{a\ b }
- \frac{2}{3} \nabla_a\nabla_b \sq\psi
- 4 R^{c (a}(\nabla_{b)}\nabla_c\psi) + \frac{8}{3}\,R^a_{\ b} \sq\psi
+ \frac{4}{3}\,R\, \nabla_a\nabla_b\psi\nn
&& - \frac{2}{3}\,(\nabla^{(a}R)(\nabla_{b)} \psi)
+ \frac{2}{3}\delta^a_{\ b}\sq^2\psi
+ 2 \delta^a_{\ b}\,R^{cd}\,\nabla_c\nabla_d\psi
- \frac{4}{3}\,\delta^a_{\ b}R\,\sq\psi
+ \frac{1}{3}\,\delta^a_{\ b}\,(\nabla^c R)(\nabla_c \psi)\,.
\label{delF}
\eea
\vspace{-.3cm}

\noindent
Specializing to de Sitter spacetime and using the form of $\bar\varphi = 2Ht$
for the BD state in de Sitter space, one finds
\bes
\bea
&&\sq\bar\varphi = -6H^2 = -\frac{R}{2}\,,\\
&&(\nabla\bar\varphi)^2 = g^{ab}(\nabla_a \bar\varphi) (\nabla_b \bar\varphi)
= -4H^2 = -\frac{R}{3}\,.
\eea
\label{phipsibac}
\ees
\vspace{-.6cm}

\noindent
Using (\ref{phipsibac}), the equation of motion (\ref{varypsi}) for $\psi$ and
the condition (\ref{delR0}) $\delta R =0$ one
finds that (\ref{delF}) becomes
\bea
&&\delta F^a_{\ b} = -2 (\nabla^{(a} \bar\varphi)\,(\nabla_{b)} \sq \psi)
+ 2 \nabla^c\left[(\nabla_c \psi) (\nabla^a\nabla_b\bar\varphi)
+  (\nabla_c \bar\varphi) (\nabla^a\nabla_b\psi) \right]
-\frac{4}{3}\, \nabla^a\nabla_b\left[(\nabla^c \bar\varphi) (\nabla_c\psi)\right] \nn
&& \qquad +  \frac{1}{3} \,\delta^a_{\ b} \sq \left[(\nabla^c\bar\varphi)(\nabla_c\psi)\right]
+ \frac{R}{6}\,\delta^a_{\ b}\, (\nabla^c\bar\varphi)(\nabla_c\psi)
- \frac{2R}{3} \, (\nabla^{(a} \bar\varphi)\,(\nabla_{b)}\psi)
+ \frac{4R}{9}\,\delta^a_{\ b}\sq\psi\nn
&& \qquad\qquad -4\nabla_c\nabla_d\,(\delta C^{(ac\ d}_{\ \ \ b)}\,\bar\varphi)
-\frac{2}{3} \nabla^a \nabla_b \sq\psi
+ \frac{R}{3}\, \nabla^a\nabla_b\psi\,.
\eea
Evaluating the $a=b=t$ component of this tensor in the flat FRW coordinates
(\ref{FRW}) of de Sitter space, using (\ref{Aabdef}) and (\ref{varA}) gives
\bea
\delta F^t_{\ t} &=& 2Ht\ \delta A^t_{\ t}
- \frac{2}{3}\frac{\stackrel{\rightarrow}{\nabla}\!\!^2\,}{a^2}\,
\left[\partial_t^2 + H\partial_t
- \frac{\stackrel{\rightarrow}{\nabla}\!\!^2\,}{a^2}\right]\psi\nn
&=& 4Ht\,\left( -\sq + 4H^2\right) \delta R^t_{\ t}
- \frac{2}{3}\frac{H^2}{a^2}\,\stackrel{\rightarrow}{\nabla}\!\!^2\, v\,,
\eea
which is (\ref{varFt}) of the text.

The variation of the $E^a_{\ b}$ tensor is more involved. Omitting the
detailed steps we find in the gauge (\ref{gaugec}),
\bea
\delta E^t_{\ t} &=&   -\frac{20H^2}{3} \,\delta R^t_{\ t}
-\frac{2}{3} \frac{\stackrel{\rightarrow}{\nabla}\!\!^2\,}{a^2}
\left[\left(\partial_t^2  + H\partial_t
- \frac{\stackrel{\rightarrow}{\nabla}\!\!^2\,}{a^2}\right) \phi
- 2 H^2 h_{tt}\right]\nn
&=&  -\frac{20H^2}{3} \,\delta R^t_{\ t}
-\frac{2H^2}{3\,a^2}\, \stackrel{\rightarrow}{\nabla}\!\!^2\, (w - 2h_{tt})\,.
\label{ett}
\eea
Here $w$ and $v$ are defined in Eq. (\ref{uvdef}). The algebraic manipulation
programs Mathematica and MathTensor were used to help obtain this result.

\end{document}